\documentclass[times,final,longtitle]{elsarticle}
\usepackage{jasr}
\usepackage{framed,multirow}

\usepackage{amssymb}
\usepackage{siunitx}
\usepackage{upgreek}
\usepackage{latexsym}
\usepackage{multirow}%
\usepackage[intlimits]{amsmath}%
\usepackage{amssymb,amsfonts}
\usepackage{graphicx}%
\usepackage{amsthm}%
\usepackage{mathtools}
\usepackage{mathrsfs}%
\usepackage[title]{appendix}%
\usepackage{xcolor}%
\usepackage{textcomp}%
\usepackage{manyfoot}%
\usepackage{booktabs}%
\usepackage{algorithm}%
\usepackage{algorithmicx}%
\usepackage{algpseudocode}%
\usepackage{listings}%
\usepackage{natbib} 
\usepackage{tabularx,longtable,multirow,subfigure,caption}
\usepackage{fncylab} 
\usepackage{fancyhdr} 
\usepackage{url} 
\usepackage{bm} 
\usepackage[english]{babel}
\usepackage{caption}
\usepackage{subfigure}
\usepackage{dsfont}
\usepackage{titlesec}
\usepackage{epstopdf} 
\usepackage{array}
\usepackage{latexsym}
\usepackage{nomencl}
\usepackage{soul}
\usepackage[title]{appendix}

\usepackage[switch]{lineno}

\usepackage{url}
\usepackage{xcolor}
\definecolor{newcolor}{rgb}{.8,.349,.1}

\usepackage[hidelinks]{hyperref}
\usepackage[capitalise]{cleveref}

\DeclareMathOperator{\dif}{d}

\DeclareMathOperator{\PSD}{PSD}
\DeclareMathOperator{\erf}{erf}
\DeclareMathOperator{\erfc}{erfc}

\newcommand{\derivative}[2]{\frac{\dif #1}{\dif #2}}
\newcommand{\infint}[1]{\int_{-\infty}^{\infty} #1}
\def\myQuad{\hskip18\fontdimen6\font}

\makeatletter
\ams@newcommand{\iiiiiint}{\DOTSI\protect\MultiIntegral{6}}
\renewcommand{\MultiIntegral}[1]{%
  \edef\ints@c{\noexpand\intop
    \ifnum#1=\z@\noexpand\intdots@\else\noexpand\intkern@\fi
    \ifnum#1>\tw@\noexpand\intop\noexpand\intkern@\fi
    \ifnum#1>\thr@@\noexpand\intop\noexpand\intkern@\fi
    \ifnum#1>4 \noexpand\intop\noexpand\intkern@\fi 
    \noexpand\intop
    \noexpand\ilimits@
  }%
  \futurelet\@let@token\ints@a
}
\makeatother

\journal{Advances in Space Research}

\begin{document}

\verso{Sabin-Viorel Anton \textit{etal}}

\begin{frontmatter}

\title{A Wave Scattering Approach to Modelling Surface Roughness in Orbital Aerodynamics}%

\author[1]{Sabin-Viorel \snm{Anton}\corref{cor1}}
\cortext[cor1]{Corresponding author}
\ead{s.v.anton@tudelft.nl}
\author[1]{Bernardo \snm{Sousa Alves}}
\ead{b.sousaalves@student.tudelft.nl}
\author[1]{Christian \snm{Siemes}}
\ead{c.siemes@tudelft.nl}
\author[1]{Jose \snm{van den IJssel}}
\ead{j.a.a.vandenijssel@tudelft.nl}
\author[1]{Pieter N. A. M. \snm{Visser}}
\ead{p.n.a.m.visser@tudelft.nl}


\affiliation[1]{organization={Delft University of Technology},
                addressline={Kluyverweg 1},
                city={Delft},
                postcode={2629 HS},
                country={The Netherlands}}



\begin{abstract}
\noindent The growing number of space objects in low-Earth orbit necessitates accurate orbit predictions to decrease the likelihood of operational disruptions. The challenges in accurately capturing how gas particles interact with the objects’ surfaces result in errors in aerodynamic coefficient modelling, directly affecting the accuracy of orbit predictions. Currently, gas-solid boundary interactions are accounted for by empirical models like those proposed by Sentman and Cercignani-Lampis-Lord. These models have one or two adjustable parameters, typically tuned based on orbital tracking and acceleration data. However, these models are inadequate in accurately representing crucial processes at the gas-solid interface such as multiple reflections, shadowing, and backscattering resulting from the roughness of real surfaces. We propose a new, physics-based gas-surface interaction model that leverages electromagnetic wave theory to incorporate macroscopic effects on the gas particle scattering distribution resulting from surface roughness. Besides better describing the physics of gas-surface interaction, this model’s parameters can be determined by combining ground measurements to characterise the surface roughness and molecular dynamics simulations to specify the atomic-scale interaction. The model is verified for the entire parameter range using a test-particle Monte Carlo approach on a simulated rough surface. In addition, we successfully replicate several experimental results available in literature on the scattering of Argon and Helium on smooth and rough Kapton and Aluminium surfaces. We conclude by demonstrating the model’s effect on the aerodynamic coefficients for simple shapes and comparing these results with those produced with the Sentman and Cercignani-Lampis-Lord models, thereby demonstrating that previously observed inconsistencies between these models and tracking data of spherical satellites can be explained by surface roughness.
\end{abstract}


\end{frontmatter}

\section{Introduction}
\label{sec:Introduction}

\noindent Aerodynamic drag is a primary perturbation force for resident space objects (RSO) located in the thermosphere at altitudes below 600 km \citep{Mehta2014}. Therefore, accurate drag modelling is crucial for reducing orbit prediction uncertainties and enhancing space situational awareness amid the increasing number of RSOs at these altitudes. Estimating the drag forces for objects with known position and velocity relies on four quantities: their aerodynamic coefficients (drag, lift, and sideslip), their area to mass ratio, the atmospheric neutral density, and wind \citep{Bernstein2022, Mehta2023, Siemes2023}. As the majority of space debris in low-Earth orbit (LEO) have unknown geometries, equivalent aerodynamic coefficients normalised to a spherical shape are derived from past tracking data \citep{McLaughlin2011}. However, such an analysis requires accurate values for the previously mentioned atmospheric properties. While atmospheric models that can provide these properties exist, e.g. NRLMSISE-00 \citep{Picone2002}, JB2008 \citep{Bowman2008}, and the Drag-Temperature Model (DTM) \citep{Bruinsma2015},  they lack consistency, producing systematic differences of up to 30 \% in neutral density values \citep{Bruinsma2023} at altitudes above 400 km. Several satellites are currently used for collecting new thermospheric measurements to improve these models: the Gravity Recovery and Climate Experiment (GRACE) and their Follow-On (GRACE-FO) satellites, the Swarm satellites, the Gravity Field and Steady-State Ocean Circulation Explorer (GOCE) satellite, and the Challenging Minisatellite Payload (CHAMP) satellite \citep{Siemes2023}. These satellites have been chosen because of their low altitude and the high quality of their GNSS tracking data and/or the presence of onboard accelerometers. However, accurately deriving neutral density and crosswind observations from acceleration data requires prior knowledge of the satellites' aerodynamic coefficients across all orientations. Numerous studies, including those by \citet{Bernstein2022}, \citet{Mehta2023}, and \citet{SIEMES2024}, have highlighted that the accuracy of density observations derived from acceleration data is significantly influenced by these coefficients. Particularly during solar maximum conditions, variations in the aerodynamic model can result in discrepancies up to 30\% according to these authors. In free molecular flow, the aerodynamic coefficients are solely determined by the scattering of atmospheric gas particles on the satellite's surface. Therefore, reducing  density observations' uncertainty is crucially dependent on an in-depth understanding of  the intricate scattering mechanisms \citep{Mehta2014}.

Many models have been developed to capture the complexity of gas-solid dynamics at a microscopic scale under free-molecular-flow conditions. Scattering kernels treat this problem as a boundary condition to the Boltzmann equation and employ an interpolation through one or two adjustable parameters between scattering with zero and full thermal accommodation to the surface temperature \citep{Mehta2014, Bernstein2022, Livadiotti2020, March2021}; thus offering empirical simplicity at the cost of physical accuracy. Physical GSI models, on the other hand, model specific microscopic interaction phenomena assuming short residence times on the surface, but often disregard others \citep{Murray2015, Murray2017, Livadiotti2020, Kleyn2003}. Molecular dynamics simulations can accurately model most GSI encountered in the thermosphere. However, they become computationally impractical when modelling the full morphology of rough surfaces \citep{Liang2018, Liang2021}. An extensive overview of existing GSI modelling approaches is given in Sec.~\ref{sec:Background}. The most common approach in orbital aerodynamics is to employ a one-parameter scattering kernel that assumes a fully diffuse reflection of incident gas molecules with an exit velocity according to the degree of accommodation to the surface temperature \citep{March2021}. Chosen for its simplicity and ease of fitting to in-orbit aerodynamic acceleration data, this kernel rapidly loses accuracy at altitudes above 400 km, where helium becomes the predominant species over atomic oxygen \citep{Mehta2014_2, Pardini2010}. Other scattering kernels assume a lobular, quasi-specular reflection of incident particles and have found great success in reproducing ground-based experimental results for noble gases such as helium on clean, microscopically smooth surfaces \citep{Cercignani1971}. However, when applied to RSOs in the altitude range of interest, they result in even larger deviations in drag coefficients compared to those observed for spherical satellites \citep{Pardini2010}. 

The gap between the observed aerodynamic behaviour of objects in orbit and the scattering lobes observed in ground experiments has led to speculation about various potential sources of error. These include inaccuracies in the geometric models of RSOs, imprecise values for the atmospheric composition and temperature, and sub-optimal selection of empirical parameters in the employed scattering kernels, or the possibility that the GSI modelling has overlooked a crucial process. This situation has spurred researchers to significantly refine aerodynamic algorithms in the last decade. Higher-fidelity geometry models have been developed for the accelerometer-carrying satellites \citep{March2019}, with a GSI analysis as the natural next step \citep{March2021}. Further improvements have been made in the selection of parameters to existing scattering kernel approaches based on known atmospheric trends to empirically account for the degradation of surfaces when exposed to atomic oxygen \citep{Pilinski2013}. Most of these approaches, however, have either had little success in eliminating the high-altitude uncertainties or require a high complexity in the form of many tunable parameters, thus prohibiting their use due to the scarcity of data in the thermosphere \citep{Bernstein2022}. This has led in recent years to a shift from physical to data-based modelling of RSO aerodynamics, such as the response surface method developed by \citet{Mehta2017} and \citet{Mehta2018}, to circumvent the need to model a seemingly too complex phenomenon. Such methods, alas, do not shed insight into the underlying dynamics and have a range of applicability limited to the datasets they were trained on. Parallel to this, \citet{Erofeev2012}, \citet{Erofeev2014-kk},  and \citet{Liu1979} have conducted a large number of experimental studies on Kapton and aluminium surfaces, which were degraded to resemble those in the space environment, in which both the angular scattering indicators and the momentum impinged on the surface samples were recorded. \citet{Erofeev2012} found very large levels of backscattering of the incidence gas particles for many different gas-material combinations, especially at near-parallel incidence angles. Furthermore, a narrowing of the scattering indicator was detected at normal incidence, where a more diffuse scattering according to the Knudsen cosine law was expected. \cite{Erofeev2012} established a link between these unexpected effects and the roughness in the sample detected through either an electron microscope or atomic force microscopy (AFM). Although the most pronounced effects were found for the atomic oxygen-bombarded Kapton film, they were present for all materials, including those previously assumed to be smooth, such as aluminium \citep{Erofeev2014-kk}. Similar results were reported as early as the 1970s \citep{Erofeev1971} and also in more recent studies \citep{Roman2023,Hellwig2019} analysing the scattering of OH radicals and carbon ions from rough surfaces. Given these corroborating findings,  we consider the full spectrum of roughness displayed by real surfaces to play a crucial, possibly primary, role in determining the scattering behaviour of gas particles under free molecular flow conditions. Furthermore, we highlight that current GSI models do not adequately incorporate these effects, a point further discussed in \cref{sec:Background}. As such, the omission of roughness effects could be the main reason for the disagreement between existing GSI models and the observed aerodynamic behaviour in the thermosphere.

The aim of this paper is to address the inconsistencies between in-situ RSO aerodynamic observations and the predictions of different popular GSI kernels through a new theoretical framework founded in electromagnetic wave scattering theory. This framework leverages the wave nature of gas particles to capture the influence of the full roughness spectrum on their scattering dynamics. More specifically, we propose a three-dimensional multireflection model based on the Kirchhoff approximation by \citet{Beckman1987-kr} to account for the variation in the local surface normal induced by the roughness from macroscopic to atomic scales. The model assumes isotropic surfaces and allows for their height profile to be defined statistically through an arbitrary number of adjustable parameters pertaining to a Gaussian mixture model with a data-fitted Gaussian autocorrelation function. This approach circumvents the computationally impractical task of inscribing the roughness properties into the RSO geometry as performed in several other studies \citep{Erofeev2012, Erofeev2014-kk}. It further offers robust, analytical expressions for the rescattering probabilities and aerodynamic shadowing and rescattering expressions at each particle reflection without loss of generality, in contrast to the numerically precomputed functional space expansions of check the bib file because O. and I. should not appear here: \citet{Aksenova2022}. In practical use, the model itself must be complemented by a local scattering function that describes the gas-surface dynamics at an atomic scale for a perfectly smooth surface. 

This paper starts with a comprehensive review of the various GSI phenomena relevant to orbital aerodynamics, existing modelling approaches and their shortcomings in \cref{sec:Background}, serving as a justification for the proposed model. This is followed by a detailed derivation of the model itself in \cref{sec:Model} and a description of its algorithmic procedure. The model is verified in \cref{sec:Results} for many levels of roughness and various values in the parameter space of the local scattering function against scattering indicators produced by a raytracing algorithm applied to a high-detail geometry of a rough surface sample. Furthermore, in the same section, the model's ability to accurately replicate experimental scattering data from previous studies by \citet{Erofeev2012} and \citet{Erofeev2014-kk} is demonstrated. The section concludes with an application of the model to replicate the aerodynamic coefficients of the Stella and Gridsphere spherical satellites under solar minimum and maximum conditions. The obtained drag coefficients as a function of altitude are then compared with those of \citet{Mehta2014_2}. Consequently, the paper outlines several conclusions, and an outlook for future work. 




\section{The Gas-Surface Interaction Problem and Existing Solution Approaches} \label{sec:Background}

\subsection{The Interaction of Gas Particles with Real Surfaces} \label{subsec:GSI_Phenomena}

\noindent The interaction between atmospheric gas particles and the surfaces of RSOs involves a complex set of physical and chemical phenomena that affect the aerodynamic forces impinging on the object. At the time of writing, existing literature does not offer a complete overview of these in one publication. It is, therefore, the aim of this section to provide the reader with a short summary of said phenomena and explain their significance qualitatively, to provide a justification for the choices made in the forthcoming model methodology in Sec. \ref{sec:Model}. 

\begin{figure}[h]
    \centering
    \includegraphics[width=0.6\linewidth]{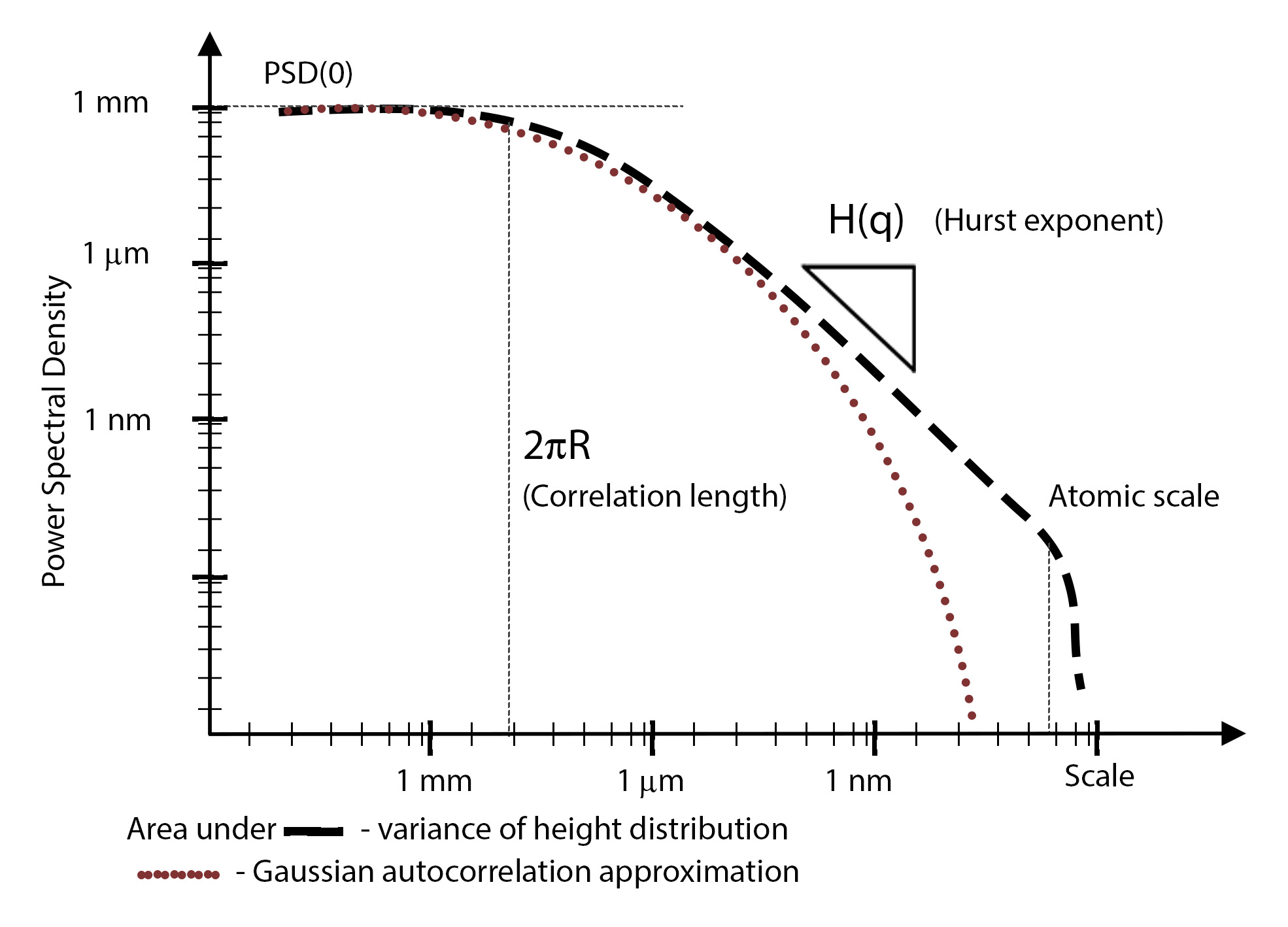}
    \caption{A sketch of the radially-averaged power spectral density of a typical engineering surface. Three regions are depicted: (left) a macroscopic roughness region characterised by uncorrelated features with large wavelengths, (centre) a microscopic fractal-like roughness region with a power-law behaviour defined by the Hurst exponent, (right) atomic scale roughness defined by individual atom corrugations. The symbol $R$ represents the correlation length, while $q$ is the length scale and $H(q)$ is the Hurst coefficient \citep{Karan2008}. The brown dotted curve correspond to a surface with an inverse squared exponential autocorrelation function. The dashed curve corresponds to a fractal surface.}
    \label{pp_1:fig:psd_rough_surfaces}
\end{figure}

\begin{figure}[h]
    \centering
    \includegraphics[width=0.99\linewidth]{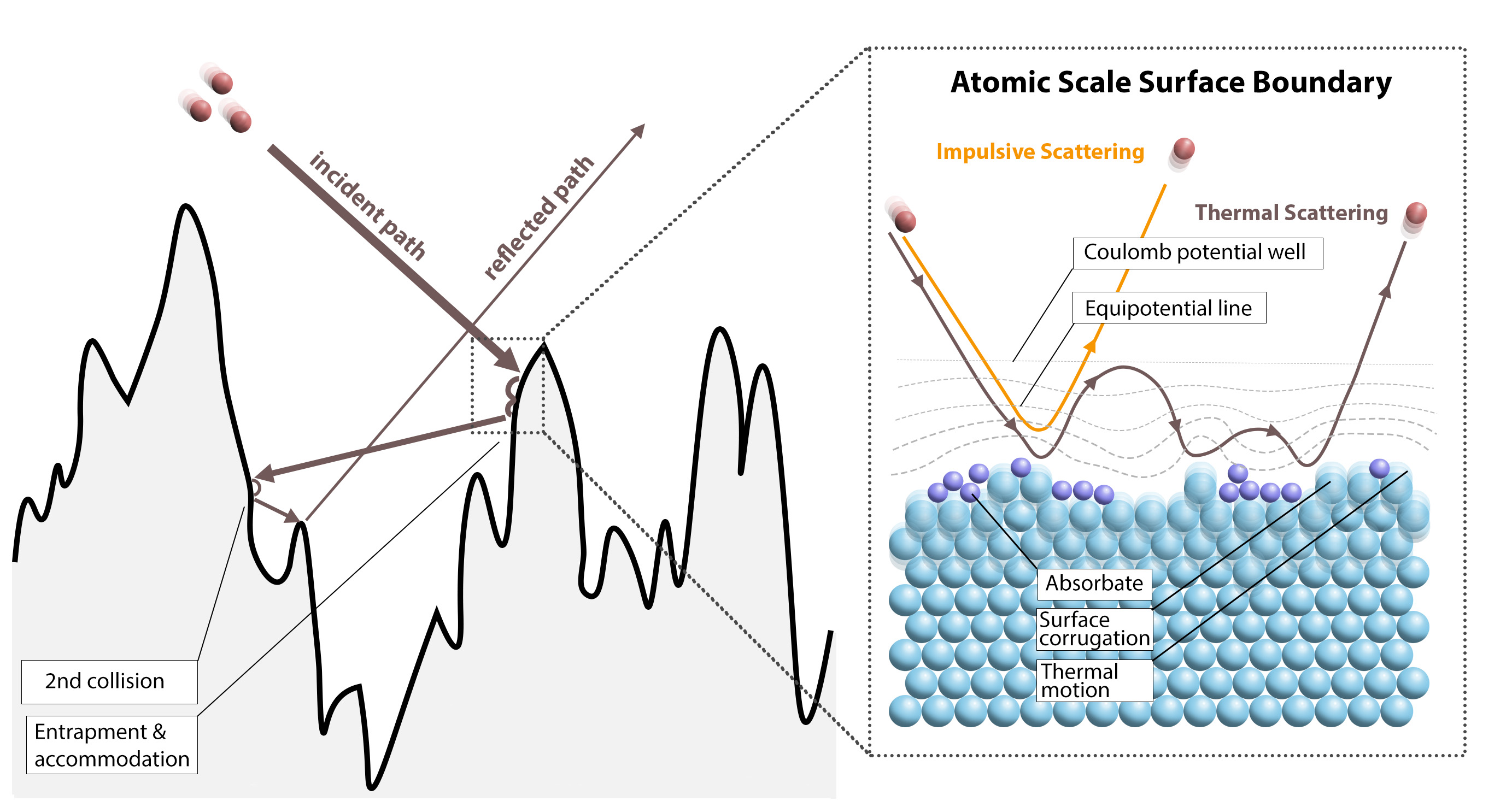}
    \caption{On the left: a sketch of the geometrically-rough profile of a typical engineering surface, and the multi-scattering of gas particles, with arrow thickness representing the particle velocity magnitude. On the right: a sketch of the impulsive scattering and scattering via thermal desorption typical for local interactions.}
    \label{pp_1:fig:gsi_phenomena}
\end{figure}

To determine how gas particles interact with the surface of an RSO, one must first understand the nature of real surfaces when exposed to the thermosphere environment. For brevity, this study shall limit itself to engineering surfaces, i.e. surfaces produced through manufacturing processes typically encountered in engineering and, more specifically, the space industry. Multiple studies \citep{Shu2023, Gong2016, Chen2013, Cutler2021, Karan2008} have measured the height profile of such surfaces, revealing a "cascade" of features, from large macroscopic scales (millimetres) to microscopic and atomic scales (nanometres). Furthermore, several of these investigations \citep{Karan2008, Cutler2021} analyse irregular surfaces with fractal characteristics, which can be attributed to the power law patterns observed in their power spectral densities (PSDs). A PSD for a standard engineering surface is sketched in Fig.~\ref{pp_1:fig:psd_rough_surfaces}. Three regions can be distinguished: a flat region on the left (at larger scales) in which the height features are uncorrelated, describing the macroscopic roughness of the surface; a microscopic power-law region of roughness with a slope in the log-log spectrum plot characterised by the Hurst exponent $H(q)$ defining the fractal dimension \citep{Karan2008}; and cut-off region (right) where the atomic scale is reached, defined by the lattice parameter for crystalline surfaces. Based on this description, two types of roughness can be identified. The first is a "geometric" roughness found at the macroscopic and microscopic scales, stemming, for example, from the manufacturing process and environmental exposure. This roughness is shown in Fig. ~\ref{pp_1:fig:gsi_phenomena} (left) and can be seen at scales that are much larger than the distance between two surface atoms (coined the "lattice parameter" for crystalline surfaces), and it is "continuum" in nature, i.e., it can be described by a continuous and differentiable height function $h : \mathbb{R}^2 \rightarrow \mathbb{R}$, $h = h(x, y)$ \citep{Karan2008}. The other type of roughness represents the morphology of the surface at an atomic level. At these scales, most surfaces display a periodic pattern of atoms, called a "lattice". Forces exerted by these atoms, i.e. strong and weak interactions, Coulomb (van der Waals) forces and so on, form a potential well above the surface, displayed in \cref{pp_1:fig:gsi_phenomena} (right) through the isopotential dotted lines. The shape of the lattice structure can disturb these lines, making them more irregular with decreasing height \citep{Rettner1991} as they begin to follow the shape of the surface itself. Randomness in the periodic structure of the lattice can be produced in several ways: through imperfections in the structure (i.e. dislocations and height-misplaced lattice atoms), the vertical Maxwellian thermal motion of the lattice atoms (shown in Fig. ~\ref{pp_1:fig:gsi_phenomena}), and the coverage with absorbates (shown in dark blue in the same figure). It should be mentioned that most nonmetallic surfaces may not follow the lattice structure shown in Fig.~\ref{pp_1:fig:gsi_phenomena} (right) and can become highly irregular. In this case, structural corrugations become the dominant source of roughness, and the isopotential lines follow suit. A final point to address, which drives the approach of gas-surface interaction modelling, is the transition point between the geometric roughness and the atomic-level corrugations, across all surface scales. While many studies \citep{Liang2018, Rettner1991, Karan2008, Shu2023, Gong2016, Chen2013, Cutler2021} suggest it is situated in the nanometre range, there is no consensus in the literature on a clear definition of this boundary. Therefore, for this study, the inflection point between the fractal region of the spectrum in Fig. ~\ref{pp_1:fig:psd_rough_surfaces} and the atomic scale region is defined as this boundary.

 Based on the aforementioned multiscale features of real surfaces, two types of GSI processes can be identified: local interactions, i.e. the momentum and energy exchanges between the surface atom lattice and the incident gas molecules depicted in the right part of Fig.~\ref{pp_1:fig:gsi_phenomena}, and geometric interactions, i.e. interactions of the gas particles with the geometric roughness of surfaces, indicated in the left panel of Fig.~\ref{pp_1:fig:gsi_phenomena}. 


\textbf{Local interactions} refer to the momentum and energy exchanges that occur at an atomic scale between the incident gas molecules, the atom lattice that forms the surface, and the absorbates that reside on it. The types of interactions that can take place at this scale are extensive, and can differ greatly with the type of gas and surface material being employed. For simplicity, these are classified into \textbf{impulsive scattering} and \textbf{thermal desorption} \citep{Murray2015, Murray2017}.

\begin{figure}[h]
    \centering
    \includegraphics[width=0.99\linewidth]{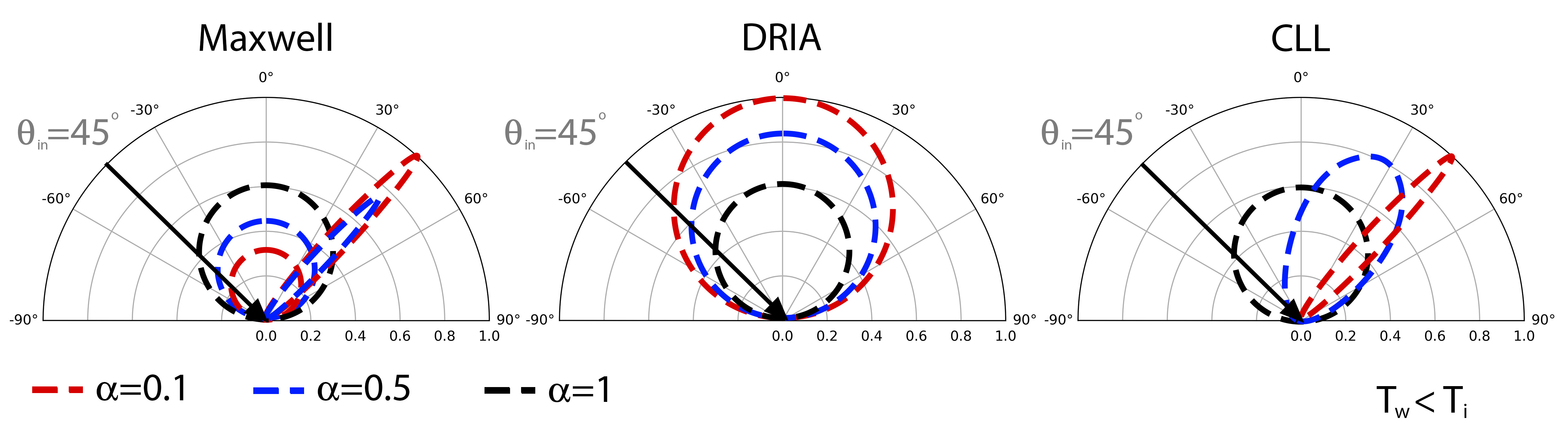}
    \caption{A sketch of the angular distributions of three scattering kernels for gas-surface interaction. The \textbf{Maxwell kernel} (left) employs a linear combination of fully specular and fully diffuse reflections. The \textbf{DRIA kernel} (center) also assumes fully diffuse scattering and adjusts the reflected particle momentum according to an energy accommodation coefficient. The \textbf{CLL kernel} (right) employs quasi-specular reflection behaviours.}
    \label{pp_1:fig:scattering_kernels}
\end{figure}

Impulsive scattering is characterised by a short time of flight of incident particles and is mainly influenced by the atomic morphology of the surface (see \cref{pp_1:fig:gsi_phenomena}, right, in yellow). For non-reactive interactions, we distinguish between the thermal and structural regime. In the thermal regime, incident gas particles have low kinetic energy and, therefore, do not penetrate deep into the van der Waals potential well, as depicted in \cref{pp_1:fig:gsi_phenomena} (right). As such, they only "see" a smooth surface, and their scattering behaviour is dictated only by the normal-oriented thermal motion of the surface atoms. This leads to partial accommodation of the momentum in the normal axis, and no accommodation in the tangential axis, resulting in a quasi-specular reflection \citep{Rettner1991}. In the structural regime, particles with larger kinetic energy are able to penetrate the potential well deep enough to be influenced by the surface atomic defects in addition to the atomic thermal motion. As such, the dependence of the scattering behaviour on surface temperature is decreased compared to the thermal regime, and the resulting quasi-specular angular lobes get oriented to the left of the specular direction (see \cref{pp_1:fig:scattering_kernels}, right sub-figure, red curve). This is a product of the expected thermal accommodation in the normal direction as well as a reflection in the tangential direction. For particles with high kinetic energy and large surface corrugations, a phenomenon called rainbow scattering is observed, where multiple quasi-specular lobular peaks are observed to the left of the specular direction as depicted \cref{pp_1:fig:scattering_kernels} \citep{Livadiotti2020}. Finally, the reactive processes equivalent of impulsive scattering are the Eley-Rideal and Hot Atom reactions \citep{Kleyn2003}, where a gas-phase atom abstracts an absorbed atom from the surface. 

The thermal desorption mechanism, on the other hand, is defined by long times of flight and accommodation of the particle momenta in the tangential and normal directions to the Maxwellian motions of the surface atoms (see \cref{pp_1:fig:gsi_phenomena}, right, in brown) \citep{Murray2015, Murray2017, Rettner1991}. Depending on the type of potential fields that act on the gas particles, these can be trapped on the surface through physisorption or chemisorption. In the case of physisorption, gas molecules may get trapped in the potential well formed by van der Waals forces (modelled through the Lennard-Jones potential) and form multiple weak absorbed layers above the surface. This is called non-activated trapping. If, however, gas molecules have an energy larger than a specific activation energy of about 0.5 eV \citep{Kleyn2003}, they undergo trapping by chemisorption (modelled through the Morse potential), and form strong, covalent bonds with the surface atoms in a monolayer. The probability of a gas molecule getting trapped onto the surface and scattering through thermal desorption is defined as the stick coefficient and is a function of surface temperature and absorbate coverage, as well as gas temperature and incident kinetic energy. Different stick coefficients are defined depending on the absorption type (physisorption or chemisorption). The type of scattering based on thermal desorption most often follows a Knudsen cosine law at the surface temperature $T_S$, resulting in a quasi-diffuse reflection. (see \cref{pp_1:fig:scattering_kernels}, DRIA, black curve). However, in the case of physisorption, preferential desorption might occur for near-parallel angles, resulting in a widened quasi-specular angular lobe. Conversely, for chemisorption, preferential desorption will occur for low reflection angles, resulting in a narrower angular lobe. The decoupling from the incidence conditions happens as the result of long trapping times for the gas particles. However, at high surface temperatures and dense adsorbate coverage, gas molecules might desorb early, before accommodating their tangential momentum. Hence, they form a preferential scattering lobe (as the one shown in \cref{pp_1:fig:scattering_kernels}, CLL, blue curve). The reactive interaction equivalent to thermal desorption is the Langmuir-Hinshelwood reaction \citep{Kleyn2003}. 

As the reader may conclude, there is a large variety of local gas-surface interaction phenomena that depend on the environmental conditions, gas, and surface properties. This diversity is partly responsible for the overall complexity of the gas-surface interaction problem, and the main reason why so many models have been developed to solve it (which shall be discussed in Sect.~\ref{subsec:Existing_Methods_GSI}). Nevertheless, the types of local scattering resulting from these phenomena can always be expressed as a linear combination of a symmetric, diffuse-like reflection and a quasi-specular term \citep{Murray2015, Murray2017}. The aforementioned stick coefficient, defining the fraction of particles undergoing diffuse scattering (i.e. thermal desorption), can be found experimentally or through molecular dynamics simulations. However, even with an informed choice of stick coefficient and both impulsive scattering and thermal desorption, as observed in ground experiments, thoroughly modelled, one cannot reconcile the resulting aerodynamic coefficients with those observed for satellites in the thermosphere at high altitudes \citep{Mehta2014_2, Mehta2014, Murray2017}. We believe this is due to neglecting the other very important type of gas-surface interactions, which result from the geometrical, fractal-like roughness of most real surfaces.

\textbf{Geometric interactions}, depicted in \cref{pp_1:fig:gsi_phenomena} (left), constitute single or multiple reflections of gas molecules, as a result of surface height variations that are much larger than its potential well depth. Therefore, such reflections may occur even when these molecules have completely "escaped" the potential well of the surface. If the surface geometry is very rough, these interactions can play a significant, sometimes even predominant role in influencing the angular scattering lobes. When combined with local interactions, they significantly enhance the total energy accommodation of gas particles departing from the surface. Similar to the rescattering of light photons off surfaces, these effects are especially prominent on surfaces that are extremely rough (e.g., due to deterioration in the space environment). Such surfaces include those compromised by atomic oxygen, like Kapton or other polyimide films, or those roughened by the physical impact of micrometeorites \citep{Banks2004}. Research on the scattering of hyperthermal flow gas against surfaces used in space engineering demonstrates that even minor levels of geometric roughness can significantly increase the diffusiveness in the angular scattering distribution's lobular patterns \citep{Erofeev2012, Erofeev2014-kk, Aksenova2022}. Furthermore, higher levels of geometric roughness, as seen on oxidised Kapton and aluminium surfaces, reveal backscattering as a distinct and highly impactful physical phenomenon at near-parallel angles. An unusual effect noted at normal incidence is the scattering lobe narrowing as surface roughness increases. These phenomena can noticeably increase the drag force on space objects with rough surfaces. In such scenarios, the scattering characteristics become less dependent on the incoming gas's properties and more on the geometric height profile of the surface itself.



\subsection{Existing Methods for Gas-Surface Interaction Modelling} \label{subsec:Existing_Methods_GSI}

\noindent Many methodologies have been developed in the past century to (partially) capture the gas-surface interaction processes outlined in Sect.~\ref{subsec:GSI_Phenomena}. These can be classified into three categories: \textbf{scattering kernels}, \textbf{physical models} and \textbf{molecular dynamics-informed models} \citep{Livadiotti2020}. 

\textbf{Scattering kernel models} are statistical kinetic models defined as boundary conditions to the Boltzmann transport equation in free-molecular flow. Mathematically, they relate the probability density function (PDF) of an incident gas molecule $f_i(\mathbf{x_i}, t_i, \mathbf{v_i})$ hitting the surface at position $\mathbf{x_i}$ and time $t_i$ with velocity $\mathbf{v_i}$ to that of the corresponding reflected molecule $f_r(\mathbf{x_i}, t_i, \mathbf{v_i})$, leaving the surface at position $\mathbf{x_r}$, time $t_r$ and velocity $\mathbf{v_r}$ as follows:
\begin{equation} \label{eq:KernelForm}
    \left( \mathbf{v_r} \cdot \mathbf{n} \right) f_r(\mathbf{x_r}, \mathbf{v_r}, t_r) = \int_{\mathbf{v_i} \cdot \mathbf{n} < 0} K\left(\mathbf{x_i} \rightarrow \mathbf{x_r}, t_i \rightarrow t_r,\mathbf{v_i} \rightarrow \mathbf{v_r}\right) \left( \mathbf{v_i} \cdot \mathbf{n} \right) f_i(\mathbf{x_i}, t_i, \mathbf{v_i}) \dif\mathbf{v_i}, \quad \mathbf{v_r} \cdot \mathbf{n} > 0
\end{equation}
where the subscripts $i$ and $r$ refer to the incident and reflected particles. The scattering kernel $K\left(\mathbf{x_i} \rightarrow \mathbf{x_r}, t_i \rightarrow t_r,\mathbf{v_i} \rightarrow \mathbf{v_r}\right)$ defines the mapping between the PDFs of incident and reflected gas particles, denoted by $f_i$ and $f_r$, respectively, and $\mathbf{n}$ is the outward pointing unit vector normal to the surface. Three mathematical conditions must be satisfied by every scattering kernel: the non-negativity condition, the normalisation condition, and the reciprocity condition \citep{Cercignani2001}. The non-negativity condition enforces a positive correlation between the reflected and incident gas particles' PDFs (i.e. larger incident velocities lead to larger reflected velocities):
\begin{equation} \label{eq:nonnegativity}
    K\left(\mathbf{x_i} \rightarrow \mathbf{x_r}, t_i \rightarrow t_r,\mathbf{v_i} \rightarrow \mathbf{v_r}\right) > 0
\end{equation}
The normalisation condition, on the other hand, is rooted in the requirements of any kernel from probability and statistics, i.e. that the integral of probabilities for all possible reflected gas states must be equal to one, i.e.
\begin{equation} \label{eq:normalisation}
    \int_{\mathbf{v_r} \cdot \mathbf{n}} K\left(\mathbf{x_i} \rightarrow \mathbf{x_r}, t_i \rightarrow t_r,\mathbf{v_i} \rightarrow \mathbf{v_r}\right) \dif \mathbf{v_r} = 1.
\end{equation}
The final condition of reciprocity concerns the reversibility of paths of the gas molecules, i.e. the incident path of a reflected molecule can be recovered using the kernel, given that the direction of its reflected path is inverted while the velocity magnitude $|\mathbf{v_r}|$ is kept constant. Physically, this conditions states that, in thermodynamic equilibrium, the number of molecules entering and leaving the surface remains constant (i.e. mass is conserved). This also implies a balance between absorption and desorption processes \citep{Cercignani1971}. Mathematically, this is enforced by
\begin{equation} \label{eq:reciprocity}
    K\left(\mathbf{x_i} \rightarrow \mathbf{x_r}, t_i \rightarrow t_r,\mathbf{v_i} \rightarrow \mathbf{v_r}\right) \left( \mathbf{v_i} \cdot \mathbf{n} \right) f_0(\mathbf{x_i}, t_i, \mathbf{v_i}, T_S) = K\left(\mathbf{x_r} \rightarrow \mathbf{x_i}, t_i \rightarrow t_r,-\mathbf{v_r} \rightarrow -\mathbf{v_i}\right) \left( \mathbf{v_r} \cdot \mathbf{n} \right) f_0(\mathbf{x_r}, t_r, \mathbf{v_r}, T_S),
\end{equation}
where $T_S$ is the temperature of the surface (wall), and $f_0(\mathbf{x}, t, \mathbf{v}, T_S) = \left(\frac{m}{2\pi k_B T_S}\right)^{\frac{3}{2}}\exp(-\frac{m v^2}{2 k_B T_S})$, with $v = \vert\mathbf{v}\vert$ is the three-dimensional Maxwellian probability distribution \citep{Maxwell1879}. In this expression, $k_B$ is the Boltzmann constant, $m$ is the molecular mass of the gas. These three mathematical conditions are necessary, but not sufficient in defining a scattering kernel, and further assumptions must be made about the gas particle behaviour to derive one \citep{Cercignani1971}. The reader should note that these conditions extend beyond scattering kernel-based methods and apply to every GSI model, including physical and data-based ones. Failure to do so can result in non-physical behaviour for a fraction or more of the parameter space \citep{Livadiotti2020, Liang2018, Mateljevic2009}. Furthermore, general assumptions made in literature for most scattering kernels are that the collisions are instantaneous, and the path travelled by the particles is negligible, i.e. $t_i = t_r$ and $\mathbf{x_i} = \mathbf{x_r}$. These need not necessarily exclude particle entrapment GSI processes that result in thermal desorption, as most of these occur on relatively small time scales (microseconds). Another common feature of these kernels is their reliance on one or two adimensional parameters describing the level of thermal or momentum accommodation of the particles. Historically, the first scattering kernel model was developed by \citet{Maxwell1879}, who described the behaviour of the reflected molecules as a linear combination of a fully specular reflection and a diffuse reflection, defined by one parameter only. This parameter is the fraction $\alpha$ of the gas molecules that is completely accommodated to the surface temperature $T_S$ and reflects diffusely with a cosine law, while the remaining fraction $1 - \alpha$ reflects specularly with its kinetic energy unchanged. The corresponding kernel takes the form
\begin{equation}
    K_{Maxwell}(\mathbf{v_i} \rightarrow \mathbf{v_r}) = \alpha \left(\frac{m}{2\pi k_B T_S}\right)^{\frac{3}{2}}\exp{\left[-\frac{m v^2}{2 k_B T_S}\right]} |\mathbf{v_r} \cdot \mathbf{n}| + (1 - \alpha) \delta(\mathbf{v_i} - \mathbf{v_{r_s}}),
\end{equation}
where $\delta(\mathbf{v_i} - \mathbf{v_{r_s}})$ is the Dirac delta function and $\mathbf{v_{r_s}}$ is the specular reflected velocity defined as $\mathbf{v_{r_s}} = \mathbf{v_i} - 2\left(\mathbf{v_i} \cdot \mathbf{n}\right) \mathbf{n}$. From a physical point of view, the Maxwell kernel could be interpreted as a simplified approach to modelling the impulsive scattering and thermal desorption processes mentioned in Sect.~\ref{subsec:Roughness_GSI}. The adjustable parameter $\alpha$ would, therefore, coincide with the stick coefficient. Under this interpretation, the kernel fails to capture the complex dynamics of impulsive scattering occurring in the structural and thermal regimes. Another simple scattering kernel that has found great success in orbital aerodynamics was proposed by \citet{Sentman1961FREEMF}, who assumed a fully diffuse reflection of particles with an incomplete thermal accommodation. Later on, the accommodation coefficient of this model was related to the atomic oxygen coverage of an RSO's surface \citep{Moe1972, Moe1993, Moe2005} and adapted into the fully empirical Diffuse Re-emission with Incomplete Accommodation (DRIA) scattering kernel 
\begin{equation}
    K_{DRIA}(\mathbf{v_i} \rightarrow \mathbf{v_r}) = \left(\frac{m}{2\pi k_B T_r}\right)^{\frac{3}{2}}\exp{\left[-\frac{m v^2}{2 k_B T_r}\right]} |\mathbf{v_r} \cdot \mathbf{n}|, \quad T_r = (1 - \alpha) \frac{|\mathbf{v_i}|^2}{3 R_{G}} + \alpha T_S,
\end{equation}
where the parameter $\alpha = \frac{T_r - T_S}{T_i - T_S}$ is the empirically found thermal (energy) accommodation coefficient. Furthermore, $T_r$ and $T_i$ are the reflected and incident gas temperatures, while $R_G = \frac{\mathcal{R}}{\mathcal{M}_G}$ is the ideal gas constant normalised by the molar mass of the gas species. The DRIA model has had great success in orbital aerodynamics at altitudes of approximately 200--400 km \citep{Mehta2014, Mehta2023, March2021} because of its one-parameter data-fitting simplicity and diffuse re-emission assumption. This assumption could capture, on the one hand, the thermal desorption processes occurring at a molecular level, which is the dominant reflection mode in an atomic oxygen-rich environment. On the other hand, it roughly resembles the diffusive and backscattering effects induced by geometric roughness described in Sect.~\ref{subsec:Roughness_GSI}. As such, this empirically-fitted kernel can be used in atomic-oxygen-rich conditions to describe both the local and geometric interactions with fair precision. At higher altitudes, however, where the presence of helium increases, the accuracy of the model is slowly lost. While the Maxwell and DRIA kernels are simple and, therefore, widely used as boundary conditions in Direct Simulation Monte Carlo (DSMC) simulations \citep{Livadiotti2020}, they fail to replicate the quasi-specular lobular scattering distributions observed in many ground experiments on clean surfaces \citep{Healy1967, Cercignani1971}. For this purpose, \citet{Cercignani1971} proposed a new scattering kernel that was later improved by \citet{Lord1995}. \citet{Cercignani1971} proposed using the eigenfunction spanning a four-dimensional space $S_4$, defined by the three Cartesian components of the reflected velocity vector $\mathbf{v_r}$ and an additional fictitious angle $\Psi$, which still satisfies Eqs.~(\ref{eq:nonnegativity}--\ref{eq:reciprocity}). This eigenfunction was then integrated over $\Psi$ to yield the kernel
\begin{equation}
    K_{CLL}(\mathbf{v_i} \rightarrow \mathbf{v_r}) = \frac{1}{\alpha_N (2 - \sigma_T)} \exp{\left[\frac{\alpha_N - 1}{\alpha_N}\left( v_{r_n}^2 + v_{i_n}^2\right) - \frac{(1 - \sigma_T)^2}{\sigma_T(2-\sigma_T)}\left( v_{r_t}^2 + v_{i_t}^2\right) + \frac{2(1-\sigma_t)}{\sigma_T(2-\sigma_T)}\left(\mathbf{v_{r_t}} \cdot \mathbf{v_{i_t}} \right)\right]} \times I_0\left( \frac{2\sqrt{1 - \alpha_N}}{\alpha_N} v_{i_n} v_{r_n}\right),
\end{equation}
where $I_0$ is the zeroth order modified Bessel function, $\mathbf{v_{i_n}} = \left(\mathbf{v_i} \cdot \mathbf{n}\right) \mathbf{n}$, $\mathbf{v_{r_n}} = \left(\mathbf{v_r} \cdot \mathbf{n}\right) \mathbf{n}$, $\mathbf{v_{i_t}} = \mathbf{v_i} - \left(\mathbf{v_i} \cdot \mathbf{n}\right) \mathbf{n}$, and $\mathbf{v_{r_t}} = \mathbf{v_r} - \left(\mathbf{v_r} \cdot \mathbf{n}\right) \mathbf{n}$ are the normal and tangential incident and reflected velocity vectors, respectively, which have the forms $\mathbf{v_{i_n}} = [0, 0, v_{i_n}]^T$, $\mathbf{v_{r_n}} = [0, 0, v_{r_n}]^T$, $\mathbf{v_{i_t}} = [v_{i_{t1}}, v_{i_{t2}}, 0]^T$, and $\mathbf{v_{r_t}} = [v_{r_{t1}}, v_{r_{t2}}, 0]^T$. A very important fact that makes the CLL kernel special compared to Maxwell or DRIA is that the adjustable parameters $\alpha_N$ and $\sigma_T$ have an actual physical meaning, describing the energy accommodation of the gas molecules in the normal direction and the momentum accommodation in the tangential direction, respectively. In fact, as stated in \cite{Cercignani1971}, a linear superposition of this kernel with different accommodation values can be used to describe any scattering shape resulting from thermal accommodation, with Eqs.~(\ref{eq:nonnegativity}~-~\ref{eq:reciprocity}) still being satisfied, since the proposed eigenfunction spans $S_4$:
\begin{equation}
    K_{CLL_{general}} = \sum_{k=0}^N c_k K_{CLL}(\mathbf{v_i} \rightarrow \mathbf{v_r}, \alpha_{n_k}, \sigma_{t_k}), \quad \sum_{k=0}^N c_k = 1, \quad c_k > 0 \quad \text{for all k}
\end{equation}
for an arbitrary number of eigenfunctions $N$. This can physically be interpreted as fractions $c_k$ accommodating the thermal motion of the surface atoms with different levels $\alpha_{N_k}$ and $\sigma_{T_k}$. Through this approach, impulsive scattering processes in a perfect thermal regime, as well as thermal desorption processes following particle entrapment, with complete or incomplete accommodation, can be modelled through an inspired choice of the parameters $\alpha_{N_k}$ and $\sigma_{T_k}$. In fact, such an approach with only two parameter sets has already been employed in studies by \citet{Murray2015} and \citet{Murray2017} to model the interaction of OH radicals with tenuous carbon surfaces with high accuracy. One drawback of the CLL model, besides the fitting requirement for the accommodation coefficients, is its limitation to thermal accommodation processes and inability to capture impulsive scattering processes in the structural regime, i.e. due to atomic-level corrugations. While best-fit parameters can be found to mimic scattering induced by corrugations in certain scenarios, these would lose their physical meaning and become empirical parameters only, as they would no longer describe an accommodation process. A sketch of the three main scattering kernels outlined above is given in \cref{pp_1:fig:scattering_kernels}, where the angular lobes have been scaled by the average reflected momentum of the gas particles. While scattering kernels display great versatility, being able to capture a large degree of local interactions and a specific case of geometric interaction, they are, in essence, incomplete models, due to the presence of adjustable parameters that are not linked to any specific physical property. For this reason, a different class of physics-based GSI models have also been developed, and are discussed next.

\begin{figure}[h]
    \centering
    \includegraphics[width=0.95\linewidth]{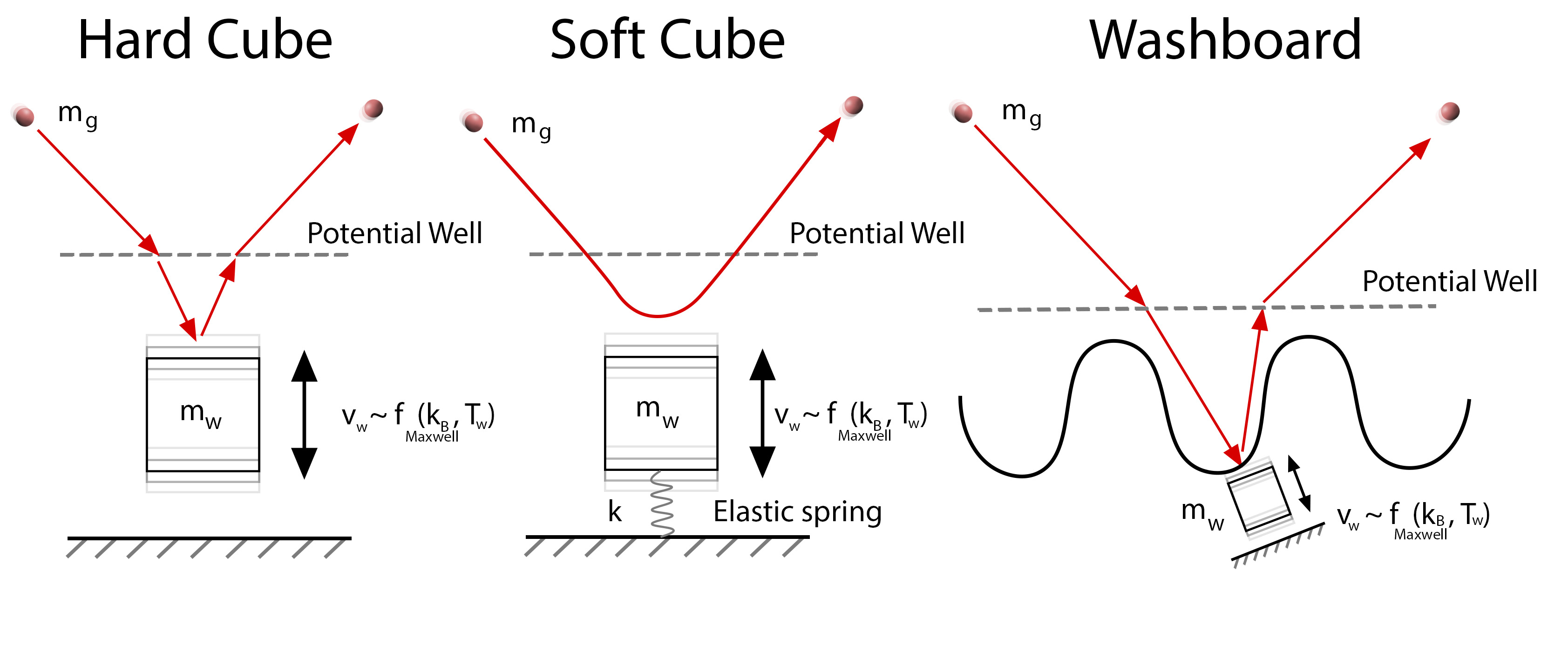}
    \caption{A sketch of three different physical gas-surface interaction models. The \textbf{Hard Cube model} (left) assumes rigid, elastic collisions between gas and surface atoms, the latter oscillating in the z-axis with a Maxwellian thermal velocity. The \textbf{Soft Cube model} further assumes linear spring connections between the surface atoms and non-negligible collision times. The \textbf{Washboard model} (right) further assigns a corrugated morphology to the surface. }
    \label{pp_1:fig:physical_GSI_models}
\end{figure}

\textbf{Physical GSI models} are complete, statistical kinetic models that approximate individual surface atoms with simple, linearly oscillating shapes surrounded by a potential well with a known depth. Inspired by the real physical GSI phenomena outlined in Sect.~\ref{subsec:GSI_Phenomena} and still bound to the mathematical constraints defined for scattering kernels (non-negativity, normalisation, and reciprocity), current physical GSI models are exclusively concerned with capturing impulsive scattering processes. Three popular models are depicted in Fig.~\ref{pp_1:fig:physical_GSI_models}: the hard cube model, the soft cube model, and the washboard model. The hard cube model proposed \citet{Goodman1965} and refined by \citet{Logan1966}, assumes a perfectly smooth surface composed of cubic, ideally elastic and rigid atoms oscillating according to a Maxwellian distribution with temperature $T_S$. Assuming instantaneous collisions of the gas particles, an impulsive-repulsive potential well is employed, resulting in momentum accommodation only in the normal direction \citep{Livadiotti2020}. The hard cube model can reasonably predict thermal-regime impulsive scattering but fails to capture the structural regime or any type of trapping and thermal desorption. These shortcomings are addressed by the soft cube model proposed  by \citet{Logan1968}, who assumed the atoms to be cubes connected by linear springs with each other. Furthermore, the impulsive-repulsive potential well was replaced by an exponential repulsive potential and a softer, square-well attractive potential, to model particle collisions with non-negligible collision times as well as nonactivated trapping. Hence, the stick coefficient of a surface can also be estimated with this model in the case of physisorption. A shortcoming of both the hard cube and soft cube models is the assumption of a perfectly smooth surface, which neglects the effect of atomic corrugations. To account for this, \citet{Tully1990} proposed a washboard model, which models these corrugations with either a 2D or 3D (periodic or random) height profile, and employs either the hard cube or soft cube procedure to handle particle-surface collisions. Several adaptations of the washboard model have been proposed to introduce 3D randomness in corrugations \citep{Mateljevic2009} and, eventually, multiple scattering \citep{Liang2018}. Even with the gradual increase in complexity, the aforementioned models are still limited to capturing local interactions only with slightly corrugated surfaces and fail to take into account the strong effects of the complex height profiles due to geometric roughness, which were observed in laboratory experiments by \cite{Erofeev1971},  \citet{Erofeev2012}, and \citet{Erofeev2014-kk}. This is primarily due to a lack of generality in the modelling of the surface geometry, combined with a failure to account for self-shadowing at near-parallel angles and for the variations with penetration depth of the rescattering function. 

\textbf{Data-informed GSI models} are entirely empirical approximators that are trained with either experimental data or molecular dynamics (MD) simulation outputs and are tasked with the prediction of scattering profiles. The most common training source is the latter, which is very successful in replicating the results of ground experiments on clean surfaces \citep{Livadiotti2020}. MD simulations used for gas-surface interaction capture the interactions between every atom in a surface sample, as well as those with the gas molecules, assuming various semi-empirical interaction potentials depending on the type of force being modelled (such as the Leenard-Jones or Morse potentials). The full spectrum of local interactions (reactive and nonreactive) can be described by this approach through an informed choice of these potentials. However, geometric interactions are impractical to model due to the high computational load that comes with simulating the full morphology of a surface from macroscopic to atomic scales \citep{Liang2018}. As such, data-informed models, while more accurate at an atomic level, suffer from the same limitations as physical GSI models when applied to orbital aerodynamics problems.  

\subsection{State of the Art in Modelling Roughness Effects} \label{subsec:Roughness_GSI}

\begin{figure}[h]
    \centering
    \includegraphics[width=0.99\linewidth]{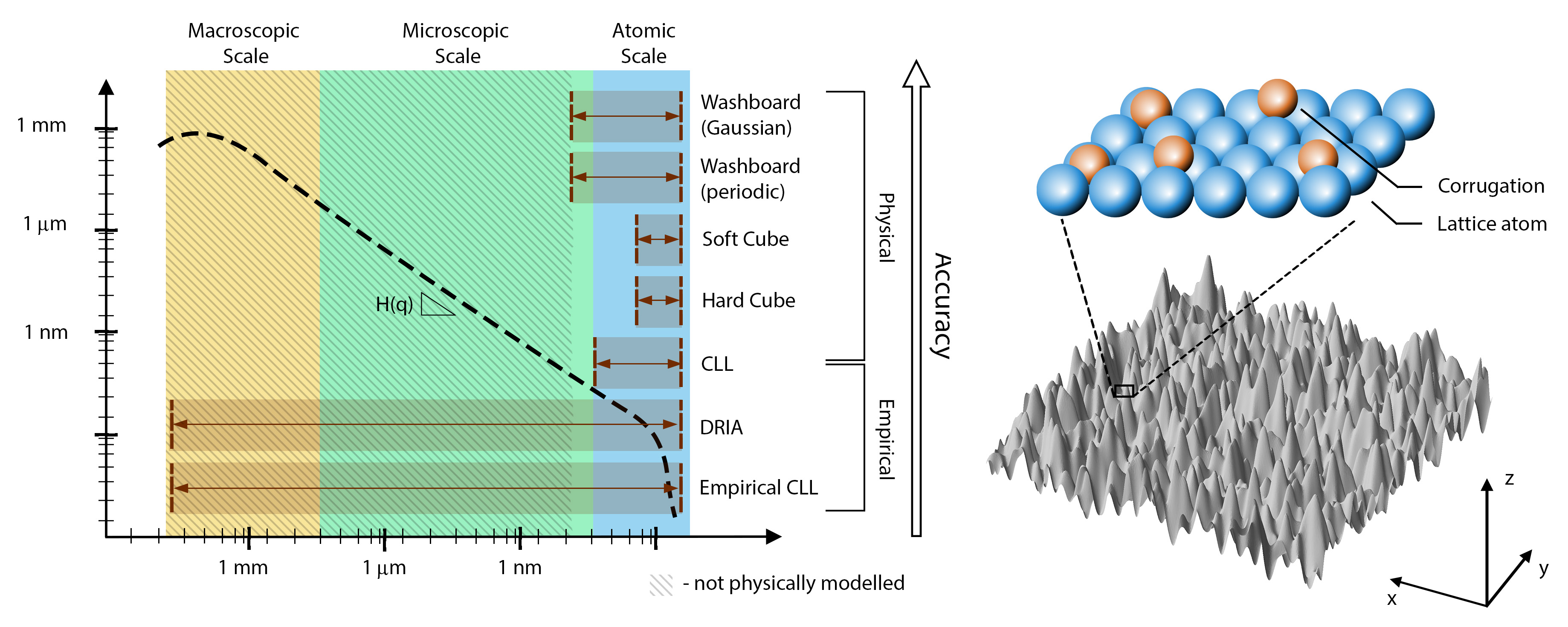}
    \caption{Left: The power spectral density of a general rough surface, and the scales included in each GSI model. Bottom-right: A realisation of Gaussian geometric \textbf{roughness} representative of real surfaces \citep{Shu2023, Erofeev2012}. Top-right: A sketch of atomic-level \textbf{corrugations} on a crystal lattice with a different atomic species.}
    \label{pp_1:fig:rough_surfaces_intro}
\end{figure}

\noindent Looking at the vast complexity of the gas-surface interaction problem and the diverse range of modelling approaches, an obvious conclusion is drawn: no unified model can be devised, which can describe the full spectrum of physical and chemical processes observed for all length scales. Furthermore, the current focus in literature is targeted towards understanding and capturing the atomic and microscopic scale interactions, while macroscopic effects induced by geometric roughness are mostly ignored. As Fig.~\ref{pp_1:fig:rough_surfaces_intro} shows, the most popular GSI models mentioned in this paper assume either an atomically smooth surface or only slight corrugations. While scattering kernels such as DRIA and CLL have been employed in several studies to model entire satellites with empirically-fitted parameters \citep{March2019, March2021, Mehta2014}, they achieved either poor or inconsistent accuracy. 

Currently, one of the most comprehensive GSI models developed is the washboard model by \citet{Liang2018}, which explicitly captures the influence of such corrugations in the structural regime of impulsive scattering. Another empirical approach for modelling nano-scale roughness has been proposed by \citet{Chen2024}, who linearly superimpose fully diffuse and quasi-specular reflections to capture the effects of slightly-rough surfaces. Nevertheless, accurate, physics-based analytical models that capture geometric interactions do not exist at the time of writing. Such an omission in the astrodynamics community is surprising, given the very significant roughness-induced effects observed by multiple studies \citep{Erofeev1971, Erofeev2012, Erofeev2014-kk, Roman2023, Liu1979}, and prompts the development of a new scale-dependent approach to modelling particle-surface dynamics in free-molecular flow. Hence, we believe that a combination of models tailored to specific length scales rather than a single overarching one is more effective at capturing these dynamics while maintaining computational efficiency. In this context, the wave scattering-based methodology proposed in this paper to describe geometric interactions serves as "an envelope" for a local scattering model capturing the multitude of previously mentioned local interaction processes. 

\section{Model Theory and Implementation} \label{sec:Model}

\subsection{Wave Approximation of Gas Particle Scattering} \label{subsec:Wave_Approx}

\noindent Let $P$ be a gas molecule travelling in vacuum with a constant velocity vector $\mathbf{v}$. Free molecular flow conditions are assumed (i.e. a high Knudsen number $K_n \gg 1$) and, as such, inter-molecular interactions can safely be neglected \citep{Vallado2007}. Furthermore, the effect of the electron orbitals on the atom nuclei is also neglected through the Born-Oppenheimer approximation \citep{Born1927}. In this context, the molecule can be interpreted as a quantum particle with a known velocity and probabilistic position. Under the quantum mechanical formalism, there is a scalar wave function $\Psi : \mathbb{R}^3 \times \mathbb{R} \rightarrow \mathbb{C}$,  $\Psi = \Psi(\mathbf{x}, t)$ whose squared modulus $|\Psi(\mathbf{x}, t)|^2$ defines the probability density function of the particle's position $\mathbf{x}$. This wave function satisfies the time-dependent Schr\"odinger equation
\begin{equation}
    i \hbar \frac{\partial \Psi(\mathbf{x}, t)}{\partial t} = \left[- \frac{\hbar}{2m}\nabla^2 + V(\mathbf{x}, t)  \right] \Psi(\mathbf{x}, t)
\end{equation}
with the Cartesian position vector $\mathbf{x} = \begin{bmatrix}
    x & y & z
\end{bmatrix}$, $t$ denoting time, $\hbar$ being the Planck constant, and $m$ the mass of the particle. Since the potential field $V(\mathbf{x}, t) = 0$ in a vacuum, the above equation reduces to the wave equation, which admits a plane wave solution of the form
\begin{equation} \label{eq:plane_wave}
    \Psi(\mathbf{x}, t) = \Psi_0 \exp\left[i\left(\mathbf{k} \cdot \mathbf{x} - \omega t\right)\right], \quad \mathbf{k} = \frac{2\pi}{\lambda} \overline{\mathbf{k}},
\end{equation}
where $\lambda$ is the de Broglie wavelength of the particle, which is $\lambda = \frac{\hbar}{mv}$ for non-relativistic particles, and $\overline{\mathbf{k}}$ is a unit vector representing the direction of the plane wave. Consider now a solid rough surface $S$, with a height profile described by a $C^{\infty}$ Lipschitz-continuous scalar function $\xi : \mathbb{R}^2 \rightarrow \mathbb{R}$, $\xi = \xi(x, y)$. Near $S$, the potential field $V(\mathbf{x}, t)$ is impossible to describe analytically as it is a complex superposition of surface atom potentials. To simplify the analysis assuming a classical mechanics regime, the surface is considered a perfectly rigid step potential barrier:
\begin{equation}
    V(\mathbf{x}, t) = \begin{cases}
        \infty, & z \leq \xi(x, y) \\
        0, & \text{otherwise}
    \end{cases}
\end{equation}
In effect, this ensures that no incoming particles can pass through the surface and, thus, act as their classical-mechanics counterparts, i.e. the power of the wave function is $|\Psi(z \leq \xi(x, y), t)|^2 = 0$ below the surface. The caveat to this simplification is the accompanying assumption of perfectly specular particle reflections of the surface, neglecting the local interactions discussed in \cref{subsec:Roughness_GSI}. However, these interactions can be easily reintroduced, which we will demonstrate in \cref{subsec:algorithm}. For the purposes of studying the averaged scattering behaviour, the time component of the Schr\"odinger equation is suppressed, leading to the partial differential equation system
\begin{equation} \label{eq:WaveEq}
     - \frac{\hbar^2}{2m} \nabla^2 \Psi(\mathbf{x}, t) = E \Psi(\mathbf{x}, t), \quad \Psi(z = \xi(x, y), t) = 0,
\end{equation}
where $E = \frac{\hbar^2 k^2}{2m}$ is the total energy of the system. The above boundary value problem is mathematically equivalent to the Helmholtz equation employed by \citet{Beckman1987-kr} in the scattering of horizontally-polarized electromagnetic waves from rough, two-dimensional surfaces. Hence, classic electromagnetic wave theory can be applied to predict the reflected wave functions of the gas particles and, thus, position and angular PDFs assuming local specular reflections.

\subsection{Statistical Surface Modelling}
\label{subsec:Surface_Modelling}

\noindent One of the most important aspects in solving the classical wave problem in \cref{eq:WaveEq} is the definition of the surface height profile $\xi(x, y)$. Since one is only interested in the averaged, macroscopic scattering off a surface, a statistical characterisation of the surface heights and slopes is sufficient. For the sake of simplicity, the surface is assumed to be isotropic and homogeneous. Under these constraints, the theory of Gaussian processes becomes a prime candidate approach, with the poly-Gaussian model developed by \citet{Litvak2012} being chosen as the starting point to characterise surfaces with arbitrary height PDF and autocorrelation functions. This approach was chosen for its ability to describe surfaces with highly non-Gaussian PDFs, using a minimum number of adjustable parameters. In this section, we extend the model to enable the characterisation of the surface slope PDF, which will become the centre point of the self-shadowing algorithm presented in \cref{subsec:shadowing_multireflection}. According to \citet{Litvak2012}, a surface defined as a two-dimensional stochastic process is described by
\begin{equation} \label{eq:PolyGaussian_Proc}
    \xi(x, y) = \sigma(x, y) \epsilon(x, y) + \mu(x, y), \quad P(\epsilon) = \frac{1}{\sqrt{2\pi}}\exp\left[ -\frac{\epsilon^2}{2} \right],
\end{equation}
where $P$ denotes probability, $\xi$ denotes the surface height function, and $\sigma, \mu, \epsilon :\mathbb{R}^2 \rightarrow \mathbb{R}$ are $C^{\infty}$ continuous and differentiable functions describing the variance, mean, and control stochastic processes of the surface with two-point autocorrelation functions $\mathcal{C}_{\sigma}, \mathcal{C}_{\mu}, \mathcal{C}_{\epsilon} : \mathbb{R} \rightarrow \left[ -1, 1\right]$ and $\epsilon$ following the standard normal distribution. If $\mu$ and $\sigma$ were deterministic functions, this would be reduced to a classic non-stationary Gaussian process \citep{MacKay2003-jc}. The novelty of Litvak's model is the introduction of a proxy stochastic process $\gamma:\mathbb{R}^2 \rightarrow \mathbb{R}$ with autocorrelation function $\mathcal{C}_{\gamma} : \mathbb{R} \rightarrow \left[ -1, 1\right]$ and, through the use of isotropy, redefining $\sigma$ and $\mu$ as nonlinear transformations $\sigma(\gamma(x, y))$ and $\mu(\gamma(x, y))$, thereby becoming deterministic. The choice of this process is arbitrary and will not influence the height PDF or correlation function $\mathcal{C}_{\xi}:\mathbb{R} \rightarrow [-1, 1]$ \citep{Litvak2012}. Under these assumptions, the characteristic function of the height profile is given by
\begin{equation} \label{eq:CharFunc}
    \chi_{\xi}(v) = 
    \infint{P(\xi)\exp\left[iv\xi\right]\dif\xi} =
    \infint{\infint{\infint{P(\sigma_i, \mu_i) P(\epsilon) \exp\left[i v \left(\sigma_i \epsilon + \mu_i \right) \right]\, \dif \sigma_i \dif \mu_i \dif \epsilon}}}, 
\end{equation}
where the joint probability of $\sigma_i$ and $\mu_i$ is given by
\begin{equation}
    P(\sigma_i, \mu_i) = \int_{-\infty}^{\infty}P(\sigma_i, \mu_i \, | \, \gamma) \, \dif F(\gamma) = \infint{\delta(\sigma(\gamma) - \sigma_i) \delta(\mu(\gamma) - \mu_i) P(\gamma) \, \dif \gamma}.
\end{equation}
Here, $F(\gamma)$ is the cumulative distribution function of $\gamma$ and $\delta(x)$ is the Dirac delta function. Substituting this expression into \cref{eq:CharFunc} and integrating with respect to $\mu_i$, $\sigma_i$, and $\epsilon$ leads to
\begin{eqnarray}
    \chi_{\xi}(v) & = & \infint{\infint{\infint{\left\{ \infint{\delta(\sigma(\gamma) - \sigma_i) \delta(\mu(\gamma) - \mu_i)  \frac{1}{\sqrt{2\pi}}\exp\left[ i v \left(\sigma_i \epsilon + \mu_i \right) - \frac{\epsilon^2}{2}\right] P(\gamma) \, \dif \gamma} \right\}  \, \dif \sigma_i \dif \mu_i \dif \epsilon}}} \nonumber \\
    & = & \infint{\infint{P(\gamma) \frac{1}{\sqrt{2\pi}}\exp \left[ i v \left(\sigma(\gamma) \epsilon + \mu(\gamma) \right) - \frac{\epsilon^2}{2}\right] \, \dif \epsilon \dif \gamma }} \nonumber \\
    & = & \infint{P(\gamma) \exp\left[ i v \mu(\gamma) - \frac{1}{2}v^2 \sigma^2(\gamma) \right] \, \dif \gamma}.
\end{eqnarray}
Taking the inverse Fourier transform of the above expression, one can observe that the PDF of the process $\xi(x, y)$ consists of a mixture of an infinity of Gaussian PDFs, which is indeed independent of the control process $\gamma(x, y)$: 
\begin{equation}
    P(\xi)=\mathcal{F}^{-1}\{\chi_\xi(v)\}=\infint{P(\gamma) \frac{1}{\sqrt{2\pi}\sigma(\gamma)}\exp\left[ -\frac{(\xi - \mu(\gamma))^2}{2\sigma^2(\gamma)} \right] \, \dif \gamma}
\end{equation}
Up to this point, we have outlined the derivation in \citep{Litvak2012}. Now, through a similar procedure, we proceed to extend it to describe the slope PDF of a surface, which is a heavily utilised feature in \cref{subsec:shadowing_multireflection} to characterise the multi-scattering effects of gas particles off very rough surfaces. Switching to polar coordinates $r$ and $\varphi$ through the assumption of isotropy, where $x = r \cos \varphi$ and $y = r \sin \varphi$, and assuming invariance in $\varphi$, it follows from \cref{eq:PolyGaussian_Proc} that the slopes of process $\xi$, i.e. $\Dot{\xi} = \derivative{\xi}{r} : \mathbb{R} \rightarrow \mathbb{R}$ are given by
\begin{equation}
    \Dot{\xi}(r) = \derivative{\sigma(r)}{r} \epsilon(r) + \sigma(r) \derivative{\epsilon(r)}{r} + \derivative{\mu(r)}{r} = \Dot{\sigma}(r) \epsilon(r) + \sigma(r) \Dot{\epsilon}(r) + \Dot{\mu}(r),
\end{equation}
where $\Dot{\sigma} = \derivative{\sigma(r)}{r}, \Dot{\mu} = \derivative{\mu(r)}{r}, \Dot{\epsilon} = \derivative{\epsilon(r)}{r}:\mathbb{R} \rightarrow \mathbb{R}$ are also $C^{\infty}$ continuous one-dimensional processes. Introducing the same control process $\gamma(r)$ and its derivative $\Dot{\gamma} = \derivative{\gamma}{r} :\mathbb{R} \rightarrow \mathbb{R}$, the slope process $\Dot{\xi}(r)$ is rewritten as
\begin{equation}
    \Dot{\xi}(r) = \Dot{\gamma}(r) \left[\derivative{\sigma}{\gamma} \epsilon(r) + \derivative{\mu}{\gamma}  \right] + \sigma(\gamma) \Dot{\epsilon}(r) = \Dot{\gamma}(r) \left[ \sigma_{\gamma} \epsilon(r) + \mu_{\gamma} \right] + \sigma(\gamma) \Dot{\epsilon}(r),
\end{equation}
where $\sigma_{\gamma} = \derivative{\sigma}{\gamma}$ and $\mu_{\gamma} = \derivative{\mu}{\gamma}$ are deterministic nonlinear transformations of the $\gamma$ process. To obtain the slope PDF, the characteristic function of $\Dot{\xi}$ is defined in a similar manner:
\begin{equation} \label{eq:Slope_Char}
    \chi_{\Dot{\xi}}(v) = \iiiiiint_{-\infty}^{\infty} P(\sigma_i, \Dot{\sigma}_i, \mu_i, \Dot{\mu}_i) P(\epsilon, \Dot{\epsilon}) \exp\left[i v \left( \Dot{\sigma}_i \epsilon + \sigma_i \Dot{\epsilon} + \Dot{\mu_i}\right) \right] \, \dif \sigma_i \dif \Dot{\sigma}_i \dif \mu_i \dif \Dot{\mu}_i \dif \epsilon \dif \Dot{\epsilon}
\end{equation}
To derive the joint PDF $P(\epsilon,\Dot{\epsilon})$, the relation between the processes $\epsilon$ and $\Dot{\epsilon}$ needs to be established first. Consider that $\epsilon_1$ and $\epsilon_2$ are two random variables of the same process $\epsilon$, with a distance of $r$ between them. Then, it is straightforward to find
\begin{equation}
    \Dot{\epsilon} = \lim_{r \rightarrow 0} \frac{\epsilon_2 - \epsilon_1}{r}.
\end{equation}
Thus, the processes $\epsilon$ and $\Dot{\epsilon}$ have a linear relation and, since $\epsilon$ follows a Gaussian distribution, $\Dot{\epsilon}$ does, too. Hence, their joint PDF can be written as
\begin{equation}
    P(\epsilon,\Dot\epsilon) = \frac{\exp\left( -\frac{1}{2} (\boldsymbol{e} - \boldsymbol{\mu})^\mathbf{T} \mathbf{\Sigma}^{-1} (\boldsymbol{e} - \boldsymbol{\mu}) \right)}{\sqrt{(2\pi)^2 \det(\mathbf{\Sigma})}}, \quad \text{with} \quad \boldsymbol{e} = \begin{bmatrix} \epsilon \\ \Dot\epsilon \end{bmatrix},
\end{equation}
where $\boldsymbol{\mu}$ and $\mathbf{\Sigma}$ are the mean and covariance matrix of vector $\boldsymbol{e}$, respectively \citep{Kac1939}, which will be derived in the next few steps. The mean of process $\Dot{\epsilon}$ is
\begin{equation}
    \boldsymbol{\mu} = E(\Dot{\epsilon})
    = E \left(\lim_{r \rightarrow 0} \frac{\epsilon_2 - \epsilon_1}{r}\right)
    = E \left(\lim_{r \rightarrow 0} \frac{\Dot\epsilon_2 - \Dot\epsilon_1}{1} \right)
    = E \left(\frac{0}{1} \right) = 0,
\end{equation}
where $E$ is the expectation operator. Here, L'H\^opital's rule was applied since $\lim_{r \rightarrow 0} \frac{\epsilon_2 - \epsilon_1}{r}$ is indeterminate as the mean of $\epsilon_1$ and $\epsilon_2$ is zero.
To obtain the covariance matrix, we first note the relation
\begin{equation}
    \begin{bmatrix} \epsilon \\ \Dot\epsilon \end{bmatrix} =
    \lim_{r \rightarrow 0} \begin{bmatrix} \epsilon_1 \\ \frac{\epsilon_2 - \epsilon_1}{r} \end{bmatrix} =
    \lim_{r \rightarrow 0}
    \begin{bmatrix} 1 & 0 \\ -\frac{1}{r} & \frac{1}{r} \end{bmatrix}
    \begin{bmatrix} \epsilon_1 \\ \epsilon_2 \end{bmatrix}
\end{equation}
and remind that process $\epsilon$ is describe by the auto-covariance function $\mathcal{C}_\epsilon(r)$ with variance $\mathcal{C}_\epsilon(0) = 1$. Thus, $\epsilon_1$ and $\epsilon_2$ have a unit variance and covariance $\mathcal{C}_\epsilon(r)$. Applying covariance propagation \citep[Chap.~2]{Koch1997} yields
\begin{equation}
    \mathbf{\Sigma} =
    \lim_{r \rightarrow 0}
    \begin{bmatrix} 1 & 0 \\ -\frac{1}{r} & \frac{1}{r} \end{bmatrix} 
    \begin{bmatrix} 1 & \mathcal{C}_\epsilon(r) \\ \mathcal{C}_\epsilon(r) & 1 \end{bmatrix}
    \begin{bmatrix} 1 & -\frac{1}{r} \\ 0 & \frac{1}{r} \end{bmatrix} =
    \lim_{r \rightarrow 0}
    \begin{bmatrix} 1 & \frac{\mathcal{C}(r)-1}{r} \\ \frac{\mathcal{C}(r)-1}{r} & \frac{2 - 2\mathcal{C}(r)}{r^2} \end{bmatrix} =
    \begin{bmatrix}
        1 & \left. -\frac{\dif \mathcal{C}_\epsilon}{\dif r} \right\vert_{r=0} \\
        \left. -\frac{\dif \mathcal{C}_\epsilon}{\dif r} \right\vert_{r=0} & -\left. \frac{\dif^2 \mathcal{C}_\epsilon}{\dif r^2} \right\vert_{r=0}
    \end{bmatrix},
\end{equation}
where L'H\^opital's rule was again applied to find the limits of the indeterminante forms. Assuming $\left.\frac{\dif \mathcal{C}_\epsilon}{\dif r}\right\vert_{r=0} = 0$, which is true for the auto-coriance function $\mathcal{C}_\epsilon$ defined later on, gives
\begin{equation}
    \mathbf{\Sigma} =
    \begin{bmatrix}
        1 & 0 \\
        0 & -\left. \frac{\dif^2 \mathcal{C}_\epsilon}{\dif r^2} \right\vert_{r=0}
    \end{bmatrix}.
\end{equation}
Inserting $\mathbf{\mu}$ and $\mathbf{\Sigma}$ into $P(\epsilon,\Dot\epsilon)$ leads to
\begin{equation}
    P(\epsilon, \Dot{\epsilon})
    = \frac{1}{2\pi \sqrt{\left.-\frac{\dif^2 \mathcal{C}_{\epsilon}}{\dif r^2}\right\vert_{r=0}}} \exp\left[- \frac{\epsilon^2}{2} - \frac{\Dot{\epsilon}^2}{-2\left.\frac{\dif^2 \mathcal{C}_{\epsilon}}{\dif r^2}\right\vert_{r=0}} \right]
    = P(\epsilon) P(\Dot{\epsilon}),
\end{equation}
which shows that the slope and height processes are independent of each other. The joint PDF of the processes $\sigma(\tau)$, $\Dot{\sigma}(\tau)$, $\mu(\tau)$, and $\Dot{\mu}(\tau)$ is given as
\begin{eqnarray}
    P(\sigma_i, \Dot{\sigma}_i, \mu_i, \Dot{\mu}_i) & = & \int_{-\infty}^{\infty}\int_{-\infty}^{\infty} P(\sigma_i, \Dot{\sigma}_i, \mu_i, \Dot{\mu}_i \, | \, \gamma, \Dot{\gamma}) \, \dif F(\gamma, \Dot{\gamma}) \nonumber \\
    & = & \infint{\infint{\delta(\sigma(\gamma) - \sigma_i) \, \delta(\Dot\sigma(\gamma,\Dot\gamma) - \Dot{\sigma}_i) \, \delta(\mu(\gamma) - \mu_i) \, \delta(\Dot\mu(\gamma,\Dot\gamma) - \Dot{\mu}_i) \, P(\gamma, \Dot{\gamma}) \, \dif \gamma \dif \Dot{\gamma}}}.
\end{eqnarray}
Substituting the two expressions above into \cref{eq:Slope_Char} results in the characteristic function
\begin{multline}
    \chi_{\Dot{\xi}}(v) = \iiiiiint_{-\infty}^{\infty}\left\{\infint{ \delta(\sigma(\gamma) - \sigma_i) \delta(\Dot{\sigma}(\gamma, \Dot{\gamma}) - \Dot{\sigma}_i) \delta(\mu(\gamma) - \mu_i) \delta(\Dot{\mu}(\gamma, \Dot{\gamma}) - \Dot{\mu}_i)}\right. \\ \times \Biggl.  \frac{1}{2\pi \sqrt{-\left.\frac{\dif^2 \mathcal{C}_{\epsilon}}{\dif r^2}\right|_{r=0}}} \exp\left[i v \left( \Dot{\sigma}_i \epsilon + \sigma_i \Dot{\epsilon} + \Dot{\mu_i}\right) - \frac{\epsilon^2}{2} - \frac{\Dot{\epsilon}^2}{-2\left.\frac{\dif^2 \mathcal{C}_{\epsilon}}{\dif r^2}\right|_{r=0}}\right] P(\gamma, \Dot{\gamma}) \, \dif \gamma \dif \Dot{\gamma}  \Biggr\}\, \dif \sigma_i \dif \Dot{\sigma}_i \dif \mu_i \dif \Dot{\mu}_i \dif \epsilon \dif \Dot{\epsilon}.
\end{multline}
Integrating the above equation with respect to $\sigma_i$, $\Dot{\sigma}_i$, $\mu_i$, and $\Dot{\mu}_i$ leads to
\begin{equation}
    \chi_{\Dot{\xi}}(v) = \infint{\infint{\infint{\infint{ \frac{1}{2\pi \sqrt{-\left.\frac{\dif^2 \mathcal{C}_{\epsilon}}{\dif r^2}\right|_{r=0}}} \exp\left[i v \left( \Dot{\gamma}\left(\sigma_{\gamma}(\gamma) \epsilon + \mu_{\gamma}(\gamma)\right) + \sigma(\gamma) \Dot{\epsilon}\right) - \frac{\epsilon^2}{2} - \frac{\Dot{\epsilon}^2}{-2\left.\frac{\dif^2 \mathcal{C}_{\epsilon}}{\dif r^2}\right|_{r=0}}\right] P(\gamma, \Dot{\gamma}) \, \dif \gamma \dif \Dot{\gamma} \dif \epsilon \dif \Dot{\epsilon}}}}}.
\end{equation}
Finally, integrating with respect to $\epsilon$ and $\Dot{\epsilon}$ gives the characteristic function of the slope PDF:
\begin{equation}
    \chi_{\Dot{\xi}}(v) = \infint{\infint{ P(\gamma, \Dot{\gamma}) \exp\left[ i v \mu_{\gamma}(\gamma) \Dot{\gamma} - \frac{v^2}{2}\left(\sigma_{\gamma}^2(\gamma)\Dot{\gamma}^2 - \left.\frac{\dif^2 \mathcal{C}_{\epsilon}}{\dif r^2}\right|_{r=0}\sigma^2(\gamma) \right) \right] \, \dif \gamma \dif \Dot{\gamma}}}
\end{equation}
Taking the inverse Fourier transform of this expression, one observes that the PDF of the slopes is also an infinite mixture of Gaussian PDFs with weights dictated by the joint probability of $\gamma$ and $\Dot{\gamma}$:
\begin{equation} \label{pp_1:eq:polyGaussian_height}
    P(\Dot{\xi}) = \mathcal{F}^{-1}\left\{ \chi_{\Dot{\xi}}(v)\right\}(\Dot{\xi}) = \infint{\infint{P(\gamma, \Dot{\gamma}) \frac{1}{\sqrt{2\pi\left( \sigma_{\gamma}^2(\gamma) \Dot{\gamma}^2 - \left.\frac{\dif^2 \mathcal{C}_{\epsilon}}{\dif r^2}\right\vert_{r=0}\right)}}\exp\left[ - \frac{\left( \Dot{\xi} - \mu_{\gamma} \Dot{\gamma}\right)^2}{2 \left( \sigma_{\gamma}^2(\gamma)\Dot{\gamma}^2 - \left.\frac{\dif^2 \mathcal{C}_{\epsilon}}{\dif r^2}\right\vert_{r=0}\right)}\right] \, \dif \Dot{\gamma} \dif \gamma}} 
\end{equation}
Unlike the surface height PDF $P(\xi)$, the slope PDF $P(\Dot{\xi})$ is dependent on the choice of the control process $\gamma$ or, more precisely, on the PDF of its radial derivative $\Dot{\gamma}$. Given that Gaussian processes are defined by their height PDF as well as all of their derivatives \citep[p.~540]{MacKay2003-jc}, this contradicts the claims of generality for poly-Gaussian models by \cite{Litvak2012}. However, in practice, this is a minor inconvenience, and only results in a y-axis-centred symmetry restriction on the stochastic radial derivative processes $\mu_{\gamma}(\gamma) \cdot \Dot{\gamma}$ and $\sigma_{\gamma}(\gamma) \cdot \Dot{\gamma}$ with respect to $r$, provided that $P(\gamma)$ is chosen as symmetric around $\gamma = 0$. For the purposes of this study, the $\gamma$ process was chosen as in \citep{Litvak2012} to follow the standard normal distribution
\begin{equation}
    P(\gamma) = \frac{1}{\sqrt{2\pi}}\exp\left[-\frac{\gamma^2}{2} \right]
\end{equation}
for simplicity. To further simplify the final form of the proposed model, the autocorrelation functions $\mathcal{C}_{\epsilon}$ and $\mathcal{C}_{\gamma}$ were defined as
\begin{equation}
    \mathcal{C}_{\epsilon}(r) = \exp\left[-\frac{r^2}{R^2} \right] \quad \text{and} \quad \mathcal{C}_{\gamma}(r) = \exp\left[-\frac{r^2}{R^2} \right],
\end{equation}
with the same autocorrelation length R. Then, the slope PDF takes the form
\begin{equation} \label{pp_1:eq:SlopePDF}
    P(\Dot{\xi}(r)) =  \infint{\infint{P(\gamma)P(\Dot{\gamma}) \frac{1}{\sqrt{2\pi\left( \sigma_{\gamma}^2(\gamma)\Dot{\gamma}^2 + 2 \left(\frac{\sigma(\gamma)}{R}\right)^2\right)}}\exp\left[ - \frac{\left( \Dot{\xi} - \mu_{\gamma} \Dot{\gamma}\right)^2}{2 \left( \sigma_{\gamma}^2(\gamma)\Dot{\gamma}^2 + 2\left(\frac{\sigma(\gamma)}{R}\right)^2\right)}\right] \, \dif \Dot{\gamma} \dif \gamma}},
\end{equation}
where
\begin{equation}
    P(\gamma) = \frac{1}{\sqrt{2\pi}} \exp\left[ - \frac{\gamma^2}{2}\right] 
\end{equation}
and
\begin{equation}
    P(\Dot{\gamma}) = \frac{R}{\sqrt{4\pi}}\exp\left[ -\frac{R^2\Dot{\gamma}^2}{4}\right].
\end{equation}
To make this model suitable for practical use in direct and inverse GSI problems, the functions $\sigma(\gamma)$, $\sigma_{\gamma}(\gamma)$, $\mu(\gamma)$, $\mu_{\gamma}(\gamma)$ need to be defined analytically using a compact set of parameters. For this purpose, the same approach as in \citep{Litvak2012} is employed, where $\sigma(\gamma)$ and $\mu(\gamma)$ are approximated by the physicist's Hermite polynomial expansions of order $N$:
\begin{equation}
    \sigma(\gamma) \approx \sum_{k=0}^{N}\sigma_k H_k(\gamma), \quad \mu(\gamma) \approx \sum_{k=0}^N \mu_k H_k(\gamma), \quad \sigma_k = \frac{1}{2^k k! \sqrt{\pi}}\infint{\sigma(\gamma) H_k(\gamma) \, \dif \gamma}, \quad \mu_k = \frac{1}{2^k k! \sqrt{\pi}}\infint{\mu(\gamma) H_k(\gamma) \, \dif \gamma} 
\end{equation}
Through this method, the function derivatives $\sigma_{\gamma}(\gamma)$ and $\mu_{\gamma}(\gamma)$ are straightforwardly computed as
\begin{equation}
    c_{\gamma}(\gamma) = \sum_{k=0}^N c_k \derivative{H_k(\gamma)}{\gamma} = \sum_{k=1}^N c_k k H_{k-1}(\gamma) \enspace \text{with} \enspace c \in \left\{ \sigma, \mu \right\},
\end{equation}
 where the constants $\sigma_k$ and $\mu_k$ become the free parameters of the surface model. Another important function to be defined is the height profile's autocorrelation function $\mathcal{C}_{\xi}(r)$, which is given in \citep{Litvak2012} as:
 \begin{equation}
    \mathcal{C}_{\xi}(r) = \frac{1}{\sigma_{\xi}}\left(\mathcal{R}_{\sigma}(r) \mathcal{C}_{\epsilon}(r) + \mathcal{R}_{\mu}(r) \right)
\end{equation}
where $\mathcal{R}_{\sigma}(r)$ and $\mathcal{R}_{\mu}(r)$ are the cross-correlation functions of the $\sigma(r)$ and $\mu(r)$ processes. Assuming without loss of generality that $\langle \mu(\gamma)\rangle = 0$, these two functions can be written in terms of $\sigma_k$ and $\mu_k$ as
\begin{eqnarray} \label{eq:CrossCorrelation}
    \mathcal{R}_{c}(r) & = & \infint{\infint{c(\gamma_1) c(\gamma_2) P(\gamma_1, \gamma_2 \, | \, r) \, \dif \gamma_1 \dif \gamma_2}} \nonumber \\
    & = & \infint{\infint{c(\gamma_1) c(\gamma_2) \left\{\sum_{k=0}^N H_k(\gamma_1) H_k(\gamma_2) \frac{c_{\gamma}^k(r)}{2^k k! \sqrt{\pi}}\frac{1}{2\pi}\exp\left[ - \frac{\gamma_1^2}{2} - \frac{\gamma_2^2}{2}\right]\right\} \, \dif \gamma_1 \dif \gamma_2}} \nonumber \\
    & = & \sum_{k=0}^N c_k^2 c_{\gamma}^k(r) 2^k k! \sqrt{\pi} \quad \text{with} \quad c \in \left\{ \sigma, \mu \right\}.
\end{eqnarray}
From \cref{eq:PolyGaussian_Proc} and \cref{eq:CrossCorrelation}, it is evident that the radially averaged autocorrelation function $\mathcal{C}_{\xi}(r)$ has the form
\begin{equation}
    \mathcal{C}_{\xi}(r) = \frac{1}{\sigma_{\xi}}\sum_{k=0}^N\left( \sigma_k^2 \mathcal{C}_{\epsilon}(r) + \mu_k^2\right) \mathcal{C}_{\gamma}^k(r) 2^k k! \sqrt{\pi}
\end{equation}
with
\begin{equation}
    \sigma_{\xi} = \infint{\left[ \sigma(\gamma)^2 + \mu(\gamma)^2 \right] P(\gamma) \, \dif \gamma} = \sum_{k=0}^N \left[\sigma_k^2 + \mu_k^2 \right] 2^k k! \sqrt{\pi}.
\end{equation}
Using this expression, we derive a relation between the radially averaged PSD of a real surface sample and the poly-Gaussian autocorrelation length $R$ of the processes . This is done by approximating $\mathcal{C}_{\xi}(r)$ with a Gaussian autocorrelation function, i.e. $\mathcal{C}_{\xi}(r) = \exp\left[ - \frac{r^2}{R_{\xi}^2}\right]$ with an autocorrelation length $R_{\xi}$. $R_{\xi}$ is now linked to the length $R$ of the $\epsilon(r)$ and $\gamma(r)$ processes in a first-order sense. Considering Taylor approximations of $\mathcal{C}_{\xi}(r)$, $\mathcal{C}_{\gamma}(r)$, and $\mathcal{C}_{\epsilon}(r)$ with only the first term included, one obtains
\begin{equation}
    1 - \frac{r^2}{R_{\xi}^2} \approx \frac{\sum_{k=0}^N\left( \sigma_k^2 \left(1 - \frac{r^2}{R^2}\right) + \mu_k^2\right) \left( 1 - \frac{r^2}{R^2}\right)^k 2^k k! \sqrt{\pi}}{\sum_{k=0}^N \left[\sigma_k^2 + \mu_k^2 \right] 2^k k! \sqrt{\pi}} \approx 1 - \frac{\sum_{k=0}^N \left[ (1+k) \sigma_k^2 + k \mu_k^2\right] 2^k k! }{\sum_{k=0}^N \left[\sigma_k^2 + \mu_k^2 \right] 2^k k!} \left(\frac{r^2}{R^2}\right).
\end{equation}
The relationship between $R$ and $R_{\xi}$ then follows immediately as
\begin{equation} \label{eq:AutoLength}
    R \approx  R_{\xi} \sqrt{\frac{\sum_{k=0}^N \left[\sigma_k^2 + \mu_k^2 \right] 2^k k!}{\sum_{k=0}^N \left[ (1+k) \sigma_k^2 + k  \mu_k^2\right] 2^k k! }}.
\end{equation}
To now link the length $R_{\xi}$ to the recorded PSD of the surface sample in question, one employs the Wiener-Khinchin-Einstein theorem \citep{Wiener1930}. A procedure for this is further discussed in \cref{subsec:Wave_Solution}, and is based on the known fact in wave scattering theory, that in the close vicinity of a reference point where $r \rightarrow 0$, the autocorrelation function of any surface resembles a Gaussian autocorrelation function \citep{Beckmann1973}. Hence, it has been shown in this section that by extending the poly-Gaussian model of \citet{Litvak2012}, the features relevant to scattering off any arbitrary surface, i.e. the height and slope distributions, as well as two-point autocorrelation functions can be modelled by a small set of $2N+3$ parameters, namely $\sigma_k$ and $\mu_k$ with $k \in \{0, \dots, N\}$ and an autocorrelation length $R$. In practice, however, $R$ need not be selected based on the autocorrelation length of a real surface, and instead can be set to the value of 1. Then, a different set of parameters $\overline{\mu_k} = \frac{\mu_k}{R}$ and $\overline{\sigma_k} = \frac{\sigma_k}{R}$ should be sought, which are defined as the autocorrelation-normalised poly-Gaussian parameters and are adimensional. However, for notation simplicity, in the rest of this study, we assume $R=1$ and the parameters $\mu_k$ and $\sigma_k$ to already be normalised and adimensional.

\subsection{Scattering Solution through a Modified Kirchhoff Approximation} \label{subsec:Wave_Solution}

\begin{figure}[h]
    \centering
    \includegraphics[width=.6\linewidth]{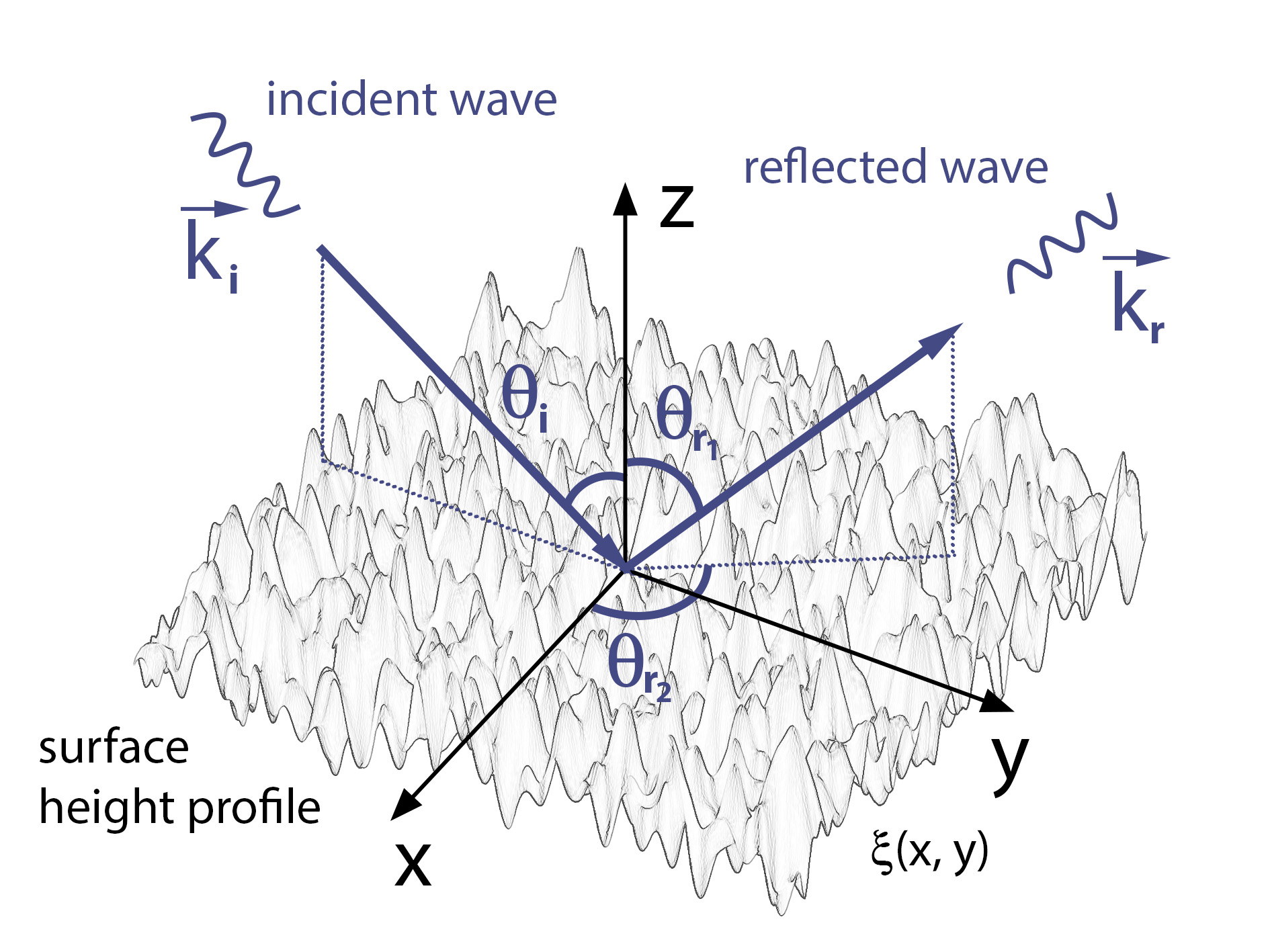}
    \caption{A sketch of the coordinate system employed by the Kirchhoff approximation \citep{Beckman1987-kr}, together with the angles that describe the incident and reflected wave directions. The incident wave is assumed to be in the zy plane.}
    \label{pp_1:fig:angle_definitions}
\end{figure}

In \cref{subsec:Wave_Approx}, it was shown under certain assumptions that gas-surface dynamics can be reduced to a classical electromagnetic wave scattering problem described by the Helmholtz Equation (\cref{eq:WaveEq}). Furthermore, the boundary of this problem, i.e. the height profile of the surface, has been characterised in a general form in \cref{subsec:Surface_Modelling}. The task at hand is to find an expression for the power of the scattered wave field, assuming a normalised incident plane wave of the form
\begin{equation}
    \Psi_i(\mathbf{x}) = e^{i\mathbf{k_i} \cdot \mathbf{x}}, \quad \text{where} \enspace \mathbf{k_i} = \frac{2\pi}{\lambda} \begin{bmatrix}
        \sin\left(\theta_i\right) & 0 & -\cos\left(\theta_i\right)
    \end{bmatrix}^T.
\end{equation}
The reflected wave will then be a function of the incidence angle $\theta_i$ and two reflected angles $\theta_{r_1}$ and $\theta_{r_2}$ defined as in \cref{pp_1:fig:angle_definitions}. The domain of \cref{eq:WaveEq} is henceforth defined as $\Omega$, and the surface profile is hence denoted with the open boundary $\partial \Omega$. To compute this wave, the Kirchhoff approximation introduced by \citet{Beckman1987-kr} is employed. We begin this section with the derivation of this method from \citep{Beckman1987-kr} for a general surface, and then we proceed to extend it for poly-Gaussian surfaces. According to it, the scattered field solution can be expressed using the Helmholtz integral as a function of the wave function values on the surface itself, i.e.
\begin{equation} \label{eq:Helmholtz}
    \Psi_r(P) = \frac{1}{4\pi}\iint_{\partial \Omega}\left(\Psi_r(\mathbf{x}) \frac{\partial \mathcal{G}(\mathbf{x})}{\partial \mathbf{n_L}} - \mathcal{G}(\mathbf{x}) \frac{\partial \Psi_r(\mathbf{x})}{\partial \mathbf{n_L}}\right)dS,
\end{equation}
where
\begin{equation}
    \mathcal{G}(\mathbf{x}) = \frac{e^{i \mathbf{k_r} \cdot \mathbf{x}}}{\lVert\mathbf{x}\rVert}
\end{equation}
is the Green function of the surface boundary, 
\begin{equation}
    \mathbf{k_r} = 2\pi \frac{m \mathbf{v_r}}{\hbar}
\end{equation}
is the wave vector of the reflected wave, and $\mathbf{v_r}$ is the velocity of the reflected particle. A detailed derivation for obtaining the Helmholtz integral mentioned above is given in Appendix \ref{ap:Helmholtz}. 

\begin{figure}[h]
    \centering
    \includegraphics[width=.6\linewidth]{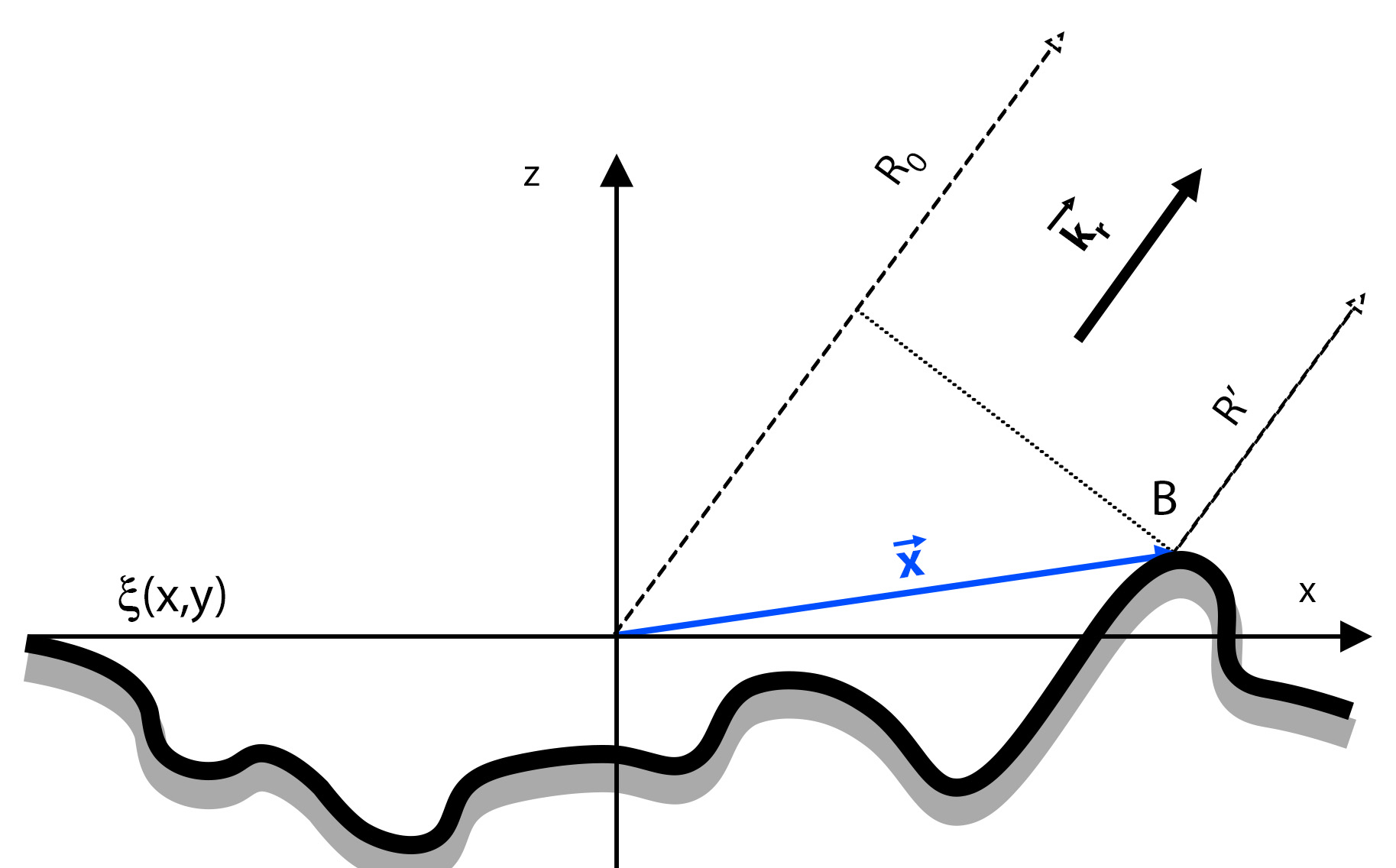}
    \caption{A sketch of a plane wave reflecting off a rough surface, observed from a point, P, situated infinitely far away from the surface in the direction of $\mathbf{k_r}$. Here, $R'$ is the distance between the observer and point $B$, $R_0$ is the distance between the same observer and the origin, $\xi(x, y)$ is the height profile of the surface and $\mathbf{k_r}$ is the reflected wave vector.}
    \label{pp_1:fig:Frauhonder_Zone}
\end{figure}

\noindent At the heart of the Kirchhoff approximation lies the assumption that the reflected wave function $\Psi_r(\mathbf{x})$, $\mathbf{x} \in \partial\Omega$, and its normal derivative on the surface boundary $\frac{\partial \Psi_r(\mathbf{x})}{\partial \mathbf{n_L}}$, $\mathbf{x} \in \partial \Omega$, are locally described by the solution of a wave reflected from a smooth, infinite plane \citep{Beckman1987-kr}.  Hence, $\Psi_r(\mathbf{x})$ and $\frac{\partial \Psi(\mathbf{x})}{\partial \mathbf{n_L}}$ are given by
\begin{equation}
    \Psi_r(\mathbf{x}) = 0 \quad \text{and} \quad \frac{\partial \Psi_r(\mathbf{x})}{\partial \mathbf{n_L}} = 2 i \Psi_i(\mathbf{x}) \frac{k_r}{k_i} \mathbf{k_i} \cdot \mathbf{n_L},
\end{equation}
respectively, where $\mathbf{n_L}$ is the local surface normal, $\mathbf{k_i}$ is the incident wave vector, and $k_r = \lVert \mathbf{k_r}\rVert$ and $k_i = \lVert\mathbf{k_i}\rVert$ are the wave numbers of the reflected and incident waves. This approximation is accurate when the incident wavelength $\lambda_i$ is much  smaller than the smallest surface curvature $\mathcal{R}(x, y)$ at point
$\begin{bmatrix}
    x & y & \xi(x, y)
\end{bmatrix}^T$, i.e. $\lambda_i \ll \mathcal{R}(x, y)$. Suppose the wave is reflecting off a surface at a point $B$, and suppose an observer is situated at an infinite distance away from this point, in the Fraunhofer zone of diffraction, where the reflected wave can safely be assumed a plane wave \citep{Beckman1987-kr}. Let $R'$ be the distance from point $B$ to the observer, and $R_0$ the distance from the origin to the same observer. In this situation, illustrated in \cref{pp_1:fig:Frauhonder_Zone}, the Green function of the surface is given by
\begin{equation}
    \mathcal{G}(\mathbf{x}) = \frac{e^{i k_r R'}}{R'} \approx \frac{e^{i k_r R_0 - i\mathbf{k_r} \cdot \mathbf{x}}}{R_0}, \quad \text{with} \quad R' = R_0 - \frac{\mathbf{k_r} \cdot \mathbf{x}}{k_r}
\end{equation}

\noindent Substituting the Green function above, together with the smooth plane solutions into Eq. (\ref{eq:Helmholtz}), results in
\begin{equation}
    \Psi_r(P) = \frac{1}{4 \pi}\iint_{\partial \Omega}\frac{e^{i k_r R_0 - i \mathbf{k_r} \cdot \mathbf{x}}}{R_0} \left( 2 i e^{i \frac{k_r}{k_i} \mathbf{k_i} \cdot \mathbf{x}} \frac{k_r}{k_i} \mathbf{k_i}\right) \cdot \mathbf{n_L}\, \dif S = -\frac{i e^{i k_r R_0}}{2\pi R_0} \iint_{\partial \Omega} e^{i \mathbf{v} \cdot \mathbf{x}} \mathbf{k_i} \cdot \mathbf{n_L}\, \dif S.
\end{equation}
In the equation above, the vectors $\mathbf{k_i}$ and $\mathbf{v} = \mathbf{k_i} - \mathbf{k_r}$ can be deduced from \cref{pp_1:fig:angle_definitions} as
\begin{equation}
    \mathbf{k_i} = k_i \begin{bmatrix}
        \sin(\theta_i) \\
        0\\
        - \cos(\theta_i) 
    \end{bmatrix}
    \quad \text{and} \quad
    \mathbf{v} = k_r \begin{bmatrix}
       \sin(\theta_i) -  \sin(\theta_{r_1}) \cos(\theta_{r_2}) \\
        - \sin(\theta_{r_1}) \sin(\theta_{r_2}) \\
        - \cos(\theta_i) - \cos(\theta_{r_1})
    \end{bmatrix},
\end{equation}
while the local surface normal vector is defined by
\begin{equation}
     \quad \mathbf{n_L} = \begin{bmatrix}
        -\frac{\partial \xi(x, y)}{\partial x} \\
        -\frac{\partial \xi(x, y)}{\partial y} \\
        1
    \end{bmatrix}.
\end{equation}
Here, the $\mathbf{v}$ vector should not be confused with the incident and reflected particle velocity vectors from \cref{subsec:Existing_Methods_GSI}, i.e. $\mathbf{v_i}$ and $\mathbf{v_r}$. Making these substitutions into the previous equation and assuming the surface $\partial \Omega$ is a rectangle with limits $(-X, X)$ and $(-Y, Y)$ centred around $(0, 0)$, one arrives at
\begin{equation}
    \Psi_r(\theta_i, \theta_{r_1}, \theta_{r_2}) = \frac{i e^{i k_r R_0}}{2\pi R_0} \int_{-X}^{X}\int_{-Y}^{Y} e^{i \mathbf{v} \cdot \mathbf{x}} \left( \frac{\partial \xi(x, y)}{\partial x} \sin(\theta_i) - \cos(\theta_i)\right) \, \dif y \dif x.
\end{equation}
Integrating the above equation by parts and ignoring the 'edge effects` for $X >> 1$ and $Y >> 1$ \citep{Beckman1987-kr} gives
\begin{equation}
    \Psi_r(\theta_i, \theta_{r_1}, \theta_{r_2}) = \frac{i e^{i k_r R_0}}{2\pi R_0} \left( - \frac{v_x}{v_z} \sin(\theta_i) - \cos(\theta_i) \right) \int_{-X}^X \int_{-Y}^{Y} e^{i \mathbf{v} \cdot \mathbf{x}}\, \dif y \dif x.
\end{equation}
At this point, it is convenient to normalise the reflected wave function with the wave function
\begin{equation}
    \Psi_{r0}(\theta_i) = \frac{i k_r e^{i k_r R_0} XY \cos(\theta_i)}{\pi R_0}
\end{equation}
equivalent to a specular reflection off a smooth, infinite plane, i.e. in the direction defined by $\theta_{r_1} = \theta_i$ and $\theta_{r_2} = 0$.

Defining the quantity $\rho(\theta_i, \theta_{r_1}, \theta_{r_2}) = \frac{\Psi_r(\theta_i, \theta_{r_1}, \theta_{r_2})}{\Psi_{r0}(\theta_i)}$ analogously to \cite{Beckman1987-kr} yields
\begin{equation}
    \rho(\theta_i, \theta_{r_1}, \theta_{r_2}) = \frac{1}{2 XY \cos(\theta_i)} \frac{1 + \cos(\theta_i) \cos(\theta_{r_1}) - \sin(\theta_i)\sin(\theta_{r_1}) \cos(\theta_{r_2})}{\cos(\theta_i) + \cos(\theta_{r_1})} \int_{-X}^{X}\int_{-Y}^{Y} e^{i \mathbf{v} \cdot \mathbf{x}}\, \dif y \dif x.
\end{equation}
Defining
\begin{equation}
    F_k = \frac{1}{2 \cos(\theta_i)} \frac{1 + \cos(\theta_i) \cos(\theta_{r_1}) - \sin(\theta_i)\sin(\theta_{r_1}) \cos(\theta_{r_2})}{\cos(\theta_i) + \cos(\theta_{r_1})}
\end{equation}
and
\begin{equation}
    A = XY
\end{equation}
leads to the final expression for the normalised scattered field:
\begin{equation}
    \rho(\theta_i, \theta_{r_1}, \theta_{r_2}) = \frac{F_k}{A} \int_{-X}^X\int_{-Y}^Y e^{i \mathbf{v} \cdot \mathbf{x}}\, \dif y \dif x
\end{equation}
Up to this point, we have closely followed the derivation from \citep{Beckman1987-kr}, and obtained an expression for the scattered wave function in the Schr\"odinger equation. Indeed, the quantity of interest for GSI problems is not function, but the probability of a particle scattering at a given angle. This is given by the mean scattered power of this wave function, defined by $\langle\rho \rho^*\rangle$, where $\langle \rangle$ defines space averaging. For a stochastic surface height profile, this is given by
\begin{equation}
    \langle \rho \rho^* \rangle(\theta_i, \theta_{r_1}, \theta_{r_2}) = \frac{F_k^2}{A^2} \left\langle \int_{-X}^{X}\int_{-X}^{X}\int_{-Y}^{Y}\int_{-Y}^{Y} \exp\left[i(v_x ( x_2 - x_1) + v_y (y_2 - y_1) + v_z (\xi_2 - \xi_1))\right] \, \dif y_1 \dif y_2 \dif x_1 \dif x_2\right\rangle,
\end{equation}
where the scattered power is integrated over two arbitrary points $P_1 = (x_1, y_1)$ and $P_2 = (x_2, y_2)$ on the surface. Now, we make the Kirchhoff model compatible with poly-Gaussian surfaces by assuming the statistics of $\xi(x, y)$ to be given by \cref{pp_1:eq:polyGaussian_height} and \cref{pp_1:eq:SlopePDF}. Then, we can separate the deterministic and stochastic components in the above expression. Keeping in mind that for a poly-Gaussian surface, the local mean surface height $\mu(x, y)$ need not be zero, the height profile can be split as $\xi(x, y) = \langle \xi \rangle + \xi' = \mu(x, y) + \xi'(x, y)$, where $\xi'(x, y)$ is the randomly fluctuating component of the surface with a mean of $\langle \xi'\rangle = 0$. Then, the mean scattered power becomes
\begin{equation}
    \langle \rho \rho^* \rangle(\theta_i, \theta_{r_1}, \theta_{r_2}) = \frac{F_k^2}{A^2} \int_{-X}^{X}\int_{-X}^{X}\int_{-Y}^{Y}\int_{-Y}^{Y} \exp\left[i(v_x ( x_2 - x_1) + v_y (y_2 - y_1) + v_z (\mu_2 - \mu_1))\right] \, \left\langle\exp\left[i v_z \left( \xi_2' - \xi_1'\right) \right] \right\rangle \, \dif y_1 \dif y_2 \dif x_1 \dif x_2,
\end{equation}
where $\mu_1$ and $\mu_2$ are the local means of the surface at points $P_1$ and $P_2$. Furthermore, the deterministic and stochastic terms of the exponential have been separated. The reader may now recognize the definition of a characteristic function in the exponential term on the right side. Hence, the equation can be rewritten as
\begin{align}
    \langle \rho \rho^* \rangle(\theta_i, \theta_{r_1}, \theta_{r_2}) & = \frac{F_k^2}{A^2} \int_{-X}^{X}\int_{-X}^{X}\int_{-Y}^{Y}\int_{-Y}^{Y} \exp\left[i (v_x ( x_2 - x_1) + v_y (y_2 - y_1) + v_z (\mu_2 - \mu_1))\right]  \chi'\left(v_z, -v_z, \sigma_1, \sigma_2\right)  \, \dif y_1 \dif y_2 \dif x_1 \dif x_2 \\
    \begin{split}
    & = \frac{F_k^2}{A^2} \int_{-X}^{X}\int_{-X}^{X}\int_{-Y}^{Y}\int_{-Y}^{Y} \exp\left[i (v_x ( x_2 - x_1) + v_y (y_2 - y_1) + v_z (\mu_2 - \mu_1))\right] \\ & \myQuad \times \exp \left[-\frac{v_z^2}{2}\left( \sigma_1^2 - 2 \mathcal{C}_{\epsilon} \sigma_1 \sigma_2 + \sigma_2^2\right)\right] \, \dif y_1 \dif y_2 \dif x_1 \dif x_2
    \end{split} \\
    \begin{split}
    & = \frac{F_k^2}{A^2} \int_{-X}^{X}\int_{-X}^{X}\int_{-Y}^{Y}\int_{-Y}^{Y} \exp\left[i(v_x ( x_2 - x_1) + v_y (y_2 - y_1) + v_z (\mu_2 - \mu_1))\right] \\ & \myQuad \times \exp \left[-\frac{v_z^2}{2}\left( \sigma_2 - \sigma_1\right)^2 + 2 \sigma_1 \sigma_2 \left( 1 - \mathcal{C}_{\epsilon}\right)\right] \, \dif y_1 \dif y_2 \dif x_1 \dif x_2,
    \end{split}
\end{align}
where $\sigma_1$ and $\sigma_2$ are the local variances of the surface at points $P_1$ and $P_2$. Continuing to a closed-form solution for the scattered field intensity now requires a few approximations. First, the mean process $\mu(x, y)$ at points $P_1$ and $P_2$ is approximated with a Taylor series expanded around an arbitrary reference point $P_0 = (x_0, y_0)$ and truncated after the linear term, i.e.
\begin{equation} \label{eq:mu_linear}
    \mu_1 = \mu_0 + \Delta \mu \approx \mu_0 + \frac{\dif \mu}{\dif x} (x_1 - x_0) + \frac{\dif \mu}{\dif y} (y_1 - y_0)
\end{equation}
and
\begin{equation*}
    \mu_2 = \mu_0 + \Delta \mu \approx \mu_0 + \frac{\dif \mu}{\dif x} (x_2 - x_0) + \frac{\dif \mu}{\dif y} (y_2 - y_0),
\end{equation*}
where $\mu_0$ is the value of the mean at point $P_0$. An equivalent linearlisation is performed for the $\sigma$ process, resulting in
\begin{equation} \label{eq:sigma_linear}
    \sigma_1 = \sigma_0 + \Delta \sigma \approx \sigma_0 + \frac{\dif \sigma}{\dif x} (x_1 - x_0) + \frac{\dif \sigma}{\dif y} (y_1 - y_0)
\end{equation}
and
\begin{equation*}
    \sigma_2 = \sigma_0 + \Delta \sigma \approx \sigma_0 + \frac{\dif \sigma}{\dif x} (x_2 - x_0) + \frac{\dif \sigma}{\dif y} (y_2 - y_0).
\end{equation*}
Substituting these approximations into the scattered intensity equation and taking $P_1 = P_0$ for convenience results in
\begin{multline}
    \langle \rho \rho^* \rangle(\theta_i, \theta_{r_1}, \theta_{r_2}) = \frac{F_k^2}{A^2} \int_{-X}^{X}\int_{-X}^{X}\int_{-Y}^{Y}\int_{-Y}^{Y} \exp\left[i \left( v_x  + \derivative{\mu}{x} v_z\right) ( x_2 - x_1) + i \left( v_y  + \derivative{\mu}{y} v_z\right) (y_2 - y_1)\right] \\ \times \exp \left[-\frac{v_z^2}{2}\left(\derivative{\sigma}{x} (x_2 - x_1) + \derivative{\sigma}{y} (y_2 - y_1)\right)^2 - 2 v_z^2 \sigma_1 \left(\sigma_1 + \derivative{\sigma}{x}(x_2 - x_1) + \derivative{\sigma}{y}(y_2 - y_1) \right) \left( 1 - \mathcal{C}_{\epsilon}\right)\right] \, \dif y_1 \dif y_2 \dif x_1 \dif x_2.
\end{multline}
The above equation can be simplified even further by making a change of variables to $\Delta x = x_2 - x_1$ and $\Delta y = y_2 - y_1$. Additionally, the limits of $X$ and $Y$ are taken to infinity, i.e. the surface sample is considered to have an infinite size, which gives
\begin{multline}
    \langle \rho \rho^* \rangle(\theta_i, \theta_{r_1}, \theta_{r_2}) = \frac{F_k^2}{A^2} \int_{-\infty}^{\infty}\int_{-\infty}^{\infty} \exp\left[i \left( v_x  + \derivative{\mu}{x} v_z\right) \Delta x + i \left( v_y  + \derivative{\mu}{y} v_z\right) \Delta y\right] \\ \times \exp \left[-\frac{v_z^2}{2}\left(\derivative{\sigma}{x} \Delta x + \derivative{\sigma}{y} \Delta y\right)^2 - 2 v_z^2 \sigma_1 \left(\sigma_1 + \derivative{\sigma}{x}\Delta x + \derivative{\sigma}{y} \Delta y \right) \left( 1 - \mathcal{C}_{\epsilon}\right)\right] \, \dif \Delta y \dif \Delta x.
\end{multline}
Furthermore, the scattered field intensity is expanded into a Gaussian mixture at point $P_0$, described by the control process $\gamma$ and its derivatives $\Dot{\gamma}_{x}$ and $\Dot{\gamma}_{y}$:
\begin{align}
     \begin{split}
        \langle \rho \rho^* \rangle(\theta_i, \theta_{r_1}, \theta_{r_2}) & = \frac{F_k^2}{A^2} \infint{\infint{\infint{P(\gamma)P(\Dot{\gamma}_x)P(\Dot{\gamma}_y)
        \left[
            \infint\infint\exp\left[ i \left( v_x  + \derivative{\mu}{x} v_z\right) \Delta x + i \left( v_y  + \derivative{\mu}{y} v_z\right) \Delta y\right]
        \right. \\
        & \left. \times \exp \left[-\frac{v_z^2}{2}\left(\derivative{\sigma}{x} \Delta x + \derivative{\sigma}{y} \Delta y\right)^2 - 2 v_z^2 \sigma \left(\sigma + \derivative{\sigma}{x}\Delta x + \derivative{\sigma}{y} \Delta y \right) \left( 1 - \mathcal{C}_{\epsilon}\right)\right] \, \dif \Delta y \dif \Delta x \right] \, \dif \gamma \dif \Dot{\gamma}_x \dif \Dot{\gamma}_y }}}
    \end{split} \\
    \begin{split}
        & = \frac{F_k^2}{A^2} \infint{\infint{\infint{P(\gamma)P(\Dot{\gamma}_x)P(\Dot{\gamma}_y)\left[\infint\infint\exp\left[i \left( v_x  + \mu_{\gamma} \Dot{\gamma}_x v_z\right) \Delta x + i \left( v_y  + \mu_{\gamma} \Dot{\gamma}_y v_z\right) \Delta y\right] \right. \\ & \left. \times \exp \left[-\frac{v_z^2}{2}\left(\sigma_{\gamma} \Dot{\gamma}_x \Delta x + \sigma_{\gamma} \Dot{\gamma}_y \Delta y\right)^2 - 2 v_z^2 \sigma \left(\sigma + \sigma_{\gamma} \Dot{\gamma}_x \Delta x + \sigma_{\gamma} \Dot{\gamma}_y  \Delta y \right) \left( 1 - \mathcal{C}_{\epsilon}\right)\right] \, \dif \Delta y \dif \Delta x \right] \, \dif \gamma \dif \Dot{\gamma}_x \dif \Dot{\gamma}_y }}}
     \end{split} \\
     \begin{split}
        & = \frac{F_k^2}{A^2} \infint{\infint{\infint{P(\gamma)P(\Dot{\gamma}_x)P(\Dot{\gamma}_y)\left[\infint\infint\exp\left[i \left( v_x  + \mu_{\gamma} \Dot{\gamma}_x v_z\right) \Delta x + i \left( v_y  + \mu_{\gamma} \Dot{\gamma}_y v_z\right) \Delta y\right] \right. \\ & \left. \times \exp \left[-\frac{v_z^2}{2}\left(\sigma_{\gamma} \Dot{\gamma}_x \Delta x + \sigma_{\gamma} \Dot{\gamma}_y \Delta y\right)^2 - 2 v_z^2 \sigma \left(\sigma + \sigma_{\gamma} \Dot{\gamma}_x \Delta x + \sigma_{\gamma} \Dot{\gamma}_y  \Delta y \right) \left( 1 - \exp\left[ -\frac{\Delta x^2 + \Delta y^2}{R^2}\right]\right)\right] \, \dif \Delta y \dif \Delta x \right] \, \dif \gamma \dif \Dot{\gamma}_x \dif \Dot{\gamma}_y }}}
    \end{split}
\end{align}
Next, $\overline{v_x} = v_x + \mu_{\gamma} \Dot{\gamma}_x v_z$ and $\overline{v_y} = v_y + \mu_{\gamma} \Dot{\gamma}_y v_z$ are introduced for brevity and the autocorrelation function $\mathcal{C}_{\epsilon}$ is approximated with a Taylor series truncated after the linear term, i.e. $\mathcal{C}_{\epsilon} \approx 1 - \frac{\Delta x^2 + \Delta y ^ 2}{R^2}$, leading to
\begin{align}
     \begin{split}
        \langle \rho \rho^* \rangle(\theta_i, \theta_{r_1}, \theta_{r_2}) & = \frac{F_k^2}{A^2} \infint{\infint{\infint{P(\gamma)P(\Dot{\gamma}_x)P(\Dot{\gamma}_y)\left[\infint\infint\exp\left[i \overline{v_x} \Delta x + i \overline{v_y} \Delta y\right] \right. \\ & \left. \times \exp \left[-\frac{v_z^2}{2}\left(\sigma_{\gamma} \Dot{\gamma}_x \Delta x + \sigma_{\gamma} \Dot{\gamma}_y \Delta y\right)^2 - 2 v_z^2 \sigma \left(\sigma + \sigma_{\gamma} \Dot{\gamma}_x \Delta x + \sigma_{\gamma} \Dot{\gamma}_y  \Delta y \right) \left(\frac{\Delta x^2 + \Delta y^2}{R^2}\right)\right] \, \dif \Delta y \dif \Delta x \right] \, \dif \gamma \dif \Dot{\gamma}_x \dif \Dot{\gamma}_y }}}
    \end{split} \\
    \begin{split}
         & = \frac{F_k^2}{A^2} \infint{\infint{\infint{P(\gamma)P(\Dot{\gamma}_x)P(\Dot{\gamma}_y)\left[\infint\infint\exp\left[i \overline{v_x} \Delta x + i \overline{v_y} \Delta y\right] \right. \\ & \left. \times \exp \left[-\frac{v_z^2 \sigma_{\gamma}^2}{2}\left( \Dot{\gamma}_x^2 \Delta x^2 + 2\Dot{\gamma}_x \Dot {\gamma}_y \Delta x \Delta y + \Dot{\gamma}_y^2 \Delta y^2\right) - 2  v_z^2 \sigma^2 \left(\frac{\Delta x^2 + \Delta y^2}{R^2}\right)\right] \, \dif \Delta y \dif \Delta x \right] \, \dif \gamma \dif \Dot{\gamma}_x \dif \Dot{\gamma}_y }}}.
    \end{split}
\end{align}
Neglecting the term correlating the x-axis statistics with the y-axis statistics, $2\Dot{\gamma}_x \Dot{\gamma}_y \Delta x \Delta y$, allows one to separate the x and y integrals:
\begin{align}
    \begin{split}
         \langle \rho \rho^* \rangle(\theta_i, \theta_{r_1}, \theta_{r_2}) & = \frac{F_k^2}{A^2} \infint{\infint{\infint{P(\gamma)P(\Dot{\gamma}_x)P(\Dot{\gamma}_y)\left[\infint\infint\exp\left[i \overline{v_x} \Delta x + i \overline{v_y} \Delta y\right] \right. \\ & \left. \times \exp \left[-\frac{v_z^2 \sigma_{\gamma}^2}{2}\left( \Dot{\gamma}_x^2 \Delta x^2 + \Dot{\gamma}_y^2 \Delta y^2\right) - 2  v_z^2 \sigma^2 \left(\frac{\Delta x^2 + \Delta y^2}{R^2}\right)\right] \, \dif \Delta y \dif \Delta x \right] \, \dif \gamma \dif \Dot{\gamma}_x \dif \Dot{\gamma}_y }}}
    \end{split} \\
    \begin{split}
          & = \frac{F_k^2}{A^2} \infint{\infint{\infint{P(\gamma)P(\Dot{\gamma}_x)P(\Dot{\gamma}_y)\left[\infint\infint\exp\left[i \overline{v_x} \Delta x - \frac{v_z^2}{2}\Delta x ^2\left(\sigma_{\gamma}^2 \Dot{\gamma}_x^2 + 2 \left( \frac{\sigma}{R}\right)^2 \right)\right] \right. \\ & \left. \times \exp\left[i \overline{v_y} \Delta y - \frac{v_z^2}{2} \Delta y ^2\left(\sigma_{\gamma}^2 \Dot{\gamma}_y^2 + 2 \left( \frac{\sigma}{R}\right)^2 \right)\right] \, \dif \Delta y \dif \Delta x \right] \, \dif \gamma \dif \Dot{\gamma}_x \dif \Dot{\gamma}_y }}}
    \end{split} \\
    \begin{split}
          & = \frac{F_k^2}{A^2} \infint{\infint{\infint{P(\gamma)P(\Dot{\gamma}_x)P(\Dot{\gamma}_y)\left[\infint{\exp\left[i \overline{v_x} \Delta x - \frac{v_z^2}{2}\Delta x ^2\left(\sigma_{\gamma}^2 \Dot{\gamma}_x^2 + 2 \left( \frac{\sigma}{R}\right)^2 \right)\right] \, \dif \Delta x } \right. \\ & \times \left. \infint{ \exp\left[i \overline{v_y} \Delta y - \frac{v_z^2}{2} \Delta y ^2\left(\sigma_{\gamma}^2 \Dot{\gamma}_y^2 + 2 \left( \frac{\sigma}{R}\right)^2 \right)\right] \, \dif \Delta y} \right] \, \dif \gamma \dif \Dot{\gamma}_x \dif \Dot{\gamma}_y }}}
    \end{split}
\end{align}
At this point, the reader may again recognise the form of a Gaussian characteristic function in both the x and y integrals. Thus, the integration is trivial and leads to 
\begin{equation} \label{eq:Kirchhoff_Solution_uncorr}
    \langle \rho \rho^* \rangle(\theta_i, \theta_{r_1}, \theta_{r_2}) = \frac{F_k^2}{4\pi v_z^2 A^2} \infint{\infint{\infint{ \frac{P(\gamma)P(\Dot{\gamma}_x)P(\Dot{\gamma}_y)}{\sqrt{\left( \left(\frac{\sigma}{R} \right)^2  + \frac{1}{2} \sigma_{\gamma}^2 \Dot{\gamma}_x^2\right) \left( \left(\frac{\sigma}{R} \right)^2  + \frac{1}{2} \sigma_{\gamma}^2 \Dot{\gamma}_y^2 \right)}} \exp \left[- \frac{(v_x + \mu_{\gamma} \Dot{\gamma}_x v_z)^2}{4 v_z^2 \left( \left(\frac{\sigma}{R} \right)^2  + \frac{1}{2} \sigma_{\gamma}^2 \Dot{\gamma}_x^2\right)} - \frac{(v_y + \mu_{\gamma} \Dot{\gamma}_y v_z)^2}{4 v_z^2 \left( \left(\frac{\sigma}{R} \right)^2  + \frac{1}{2} \sigma_{\gamma}^2 \Dot{\gamma}_y^2\right)}\right] \, \dif \gamma \dif \Dot{\gamma}_x \dif \Dot{\gamma}_y. }}}
\end{equation}
\Cref{eq:Kirchhoff_Solution_uncorr} is an important result, as it represents a closed-form, analytic expression of the scattered power of the gas particle wave function, depending only on the incidence and reflection angles, accompanied by the poly-Gaussian surface parameters $\sigma_k$ and $\mu_k$ and the autocorrelation length of the control process $\gamma$. After normalisation, this represents the angular scattering probability of a particle, which is the building block of the scattering kernel and iterative algorithm proposed in \cref{subsec:algorithm}. Another important aspect is that the equation resembles the product of two Gaussians, namely the PDF of the slopes in the x and y-axis, $P(\Dot{\xi}_x)$ and $P(\Dot{\xi}_y)$. This confirms that the Kirchhoff theory reduces to a classical mechanics analysis when the contributions of $\mu(x, y)$ and $\sigma(x, y)$ to the scattered field are linearised around point $P_0$. Indeed, in the domain of geometric optics, the scattering behaviour from rough surfaces is dictated by the slope PDF as stated by \citet{Beckmann1973}. 

On a different note, one underlying assumption of \cref{eq:Kirchhoff_Solution_uncorr} is that only one collision with the surface takes place. For very rough surfaces, however, this may not be true. If multi-reflections are to be taken into account, the Kirchhoff solution has to be correlated to the previous scattering angles, denoted by $\theta_{i_{old}}, \theta_{r_{1_{old}}}$, and $\theta_{2_{2_{old}}}$. Employing the properties of Gaussian PDFs, this can be done approximately by
\begin{multline} \label{eq:Kirchhoff_Solution}
    \langle \rho \rho^* \rangle(\theta_i, \theta_{r_1}, \theta_{r_2} \, | \, \theta_{i_{old}}, \theta_{r_{1_{old}}}, \theta_{r_{2_{old}}}) = \frac{F_k^2}{4\pi v_z^2 A^2} \infint{\infint{\infint{ \frac{P(\gamma \, | \, \gamma_{old})P(\Dot{\gamma}_x \, | \, \Dot{\gamma}_{x_{old}})P(\Dot{\gamma}_y \, | \, \Dot{\gamma}_{y_{old}})}{w_{x} w_{x_{old}} w_{y} w_{y_{old}} (1 - \mathcal{C}_{\epsilon}^2)} \\ \times \exp \left[-\frac{1}{2(1-\mathcal{C}_{\epsilon}^2)}\left(\left(\frac{\overline{v_x}^2}{v_z^2 w_x^2} + \frac{\overline{v_{x_{old}}}^2}{v_{z}^2 w_{x_{old}}^2} -\frac{2 \mathcal{C}_{\epsilon} \overline{v_x} \overline{v_{x_{old}}}}{ v_z w_{x} w_{x_{old}}} \right) + \left(\frac{\overline{v_y}^2}{v_z^2 w_y^2} + \frac{\overline{v_{y_{old}}}^2}{v_{z} w_{y_{old}}^2} -\frac{2 \mathcal{C}_{\epsilon} \overline{v_y} \overline{v_{y_{old}}}}{v_z v_{z} w_{y}  w_{y_{old}}} \right)\right)\right] \, \dif \gamma \dif \Dot{\gamma}_x \dif \Dot{\gamma}_y }}}, \\
    w_x = \sqrt{2\left( \frac{\sigma}{R}\right)^2 + \sigma_{\gamma}^2\Dot{\gamma}_x^2}, \quad w_{x_{old}} = \sqrt{2\left( \frac{\sigma_{old}}{R}\right)^2 + \sigma_{{\gamma}_{old}}^2\Dot{\gamma}_{x_{old}}^2}, \quad w_y = \sqrt{2\left( \frac{\sigma}{R}\right)^2 + \sigma_{\gamma}^2\Dot{\gamma}_y^2}, \quad w_{y_{old}} = \sqrt{2\left( \frac{\sigma_{old}}{R}\right)^2 + \sigma_{{\gamma}_{old}}^2\Dot{\gamma}_{y_{old}}^2},
\end{multline}
where subscript '$old$` denotes quantities related to the old collision point $P_{old}$ with control process values of $\gamma_{old}$, $\Dot{\gamma}_{x_{old}}$, and $\Dot{\gamma}_{y_{old}}$. A final aspect to address is the autocorrelation length parameter $R$. In \cref{subsec:Surface_Modelling}, a relationship was established through \cref{eq:AutoLength}, between this parameter and the physical autocorrelation length of the surface, $R_{\xi}$, under the assumption of a Gaussian autocorrelation function. If one considers a surface sample with radius $R_S$ and a discrete, arbitrary radially-averaged PSD, then through the Wiener-Khinchin-Einstein theorem \citep{Wiener1930}, its autocorrelation function is given by the inverse Fourier transform as the cosine Fourier series
\begin{equation}
    \mathcal{C}_{\xi}(r) = \sum_{n=0}^M \left(A_{n}^2 + B_{n}^2\right) \cos\left( n \frac{2\pi}{R_S} r \right) = \sum_{n=0}^M P_n \cos\left( n \frac{2\pi}{R_S} r \right), \quad \text{where} \quad  \xi(x, y) = \sum_{n=0}^M \left(A_{n} + i B_{n} \right) \exp\left[{i \left( \frac{2 n\pi}{R_S} r \right)}\right]
\end{equation}
with Fourier coefficients $A_n$ and $B_n$, and the squared amplitude $P_n = A_{n}^2 + B_{n}^2$.
For small values of $r$, this cosine series can be approximated as
\begin{equation}
    \mathcal{C}_{\xi}(r) \approx \sum_{n=0}^M P_n \left(1 - \frac{\pi^2 n^2 r^2}{R_S^2} \right) = 1 - \left[\sum_{n=0}^M P_n \frac{\pi^2 n^2}{R_S^2}\right] r^2 \approx \exp \left\{- \frac{r^2}{\left[ \sum_{n=0}^M P_n \frac{\pi^2 n^2}{R_S^2}\right]^{-1}} \right\} \quad \text{assuming} \quad \sum_{n=0}^M P_n = 1.
\end{equation}
As the scattering behaviour of very rough surfaces is influenced primarily by local features (i.e. small $r$), it has been shown that these can always be approximated in the second order through a Gaussian autocorrelation function with $R_{\xi} = \left[ \sum_{n=0}^M P_n \frac{\pi^2 n^2}{R_S^2}\right]^{-1}$. Hence, the adjustable parameter $R$ can be approximated from a real surface's PSD as
\begin{equation} \label{pp_1:eq:auto_length_approx}
    R \approx  \left[ \sum_{n=0}^M P_n \frac{\pi^2 n^2}{R_S^2}\right]^{-1} \sqrt{\frac{\sum_{k=0}^N \left[\sigma_k^2 + \mu_k^2 \right] 2^k k!}{\sum_{k=0}^N \left[ (1+k) \sigma_k^2 + k  \mu_k^2\right] 2^k k! }}.
\end{equation}
A different approach to obtaining all adjustable parameters $\sigma_k$, $\mu_k$, and $R$ simultaneously is employing multiparametric optimisation algorithms \citep{Litvak2012}.

\subsection{Self-Shadowing and Multi-reflections} \label{subsec:shadowing_multireflection}

\begin{figure}[h]
    \centering
    \includegraphics[width=.6\linewidth]{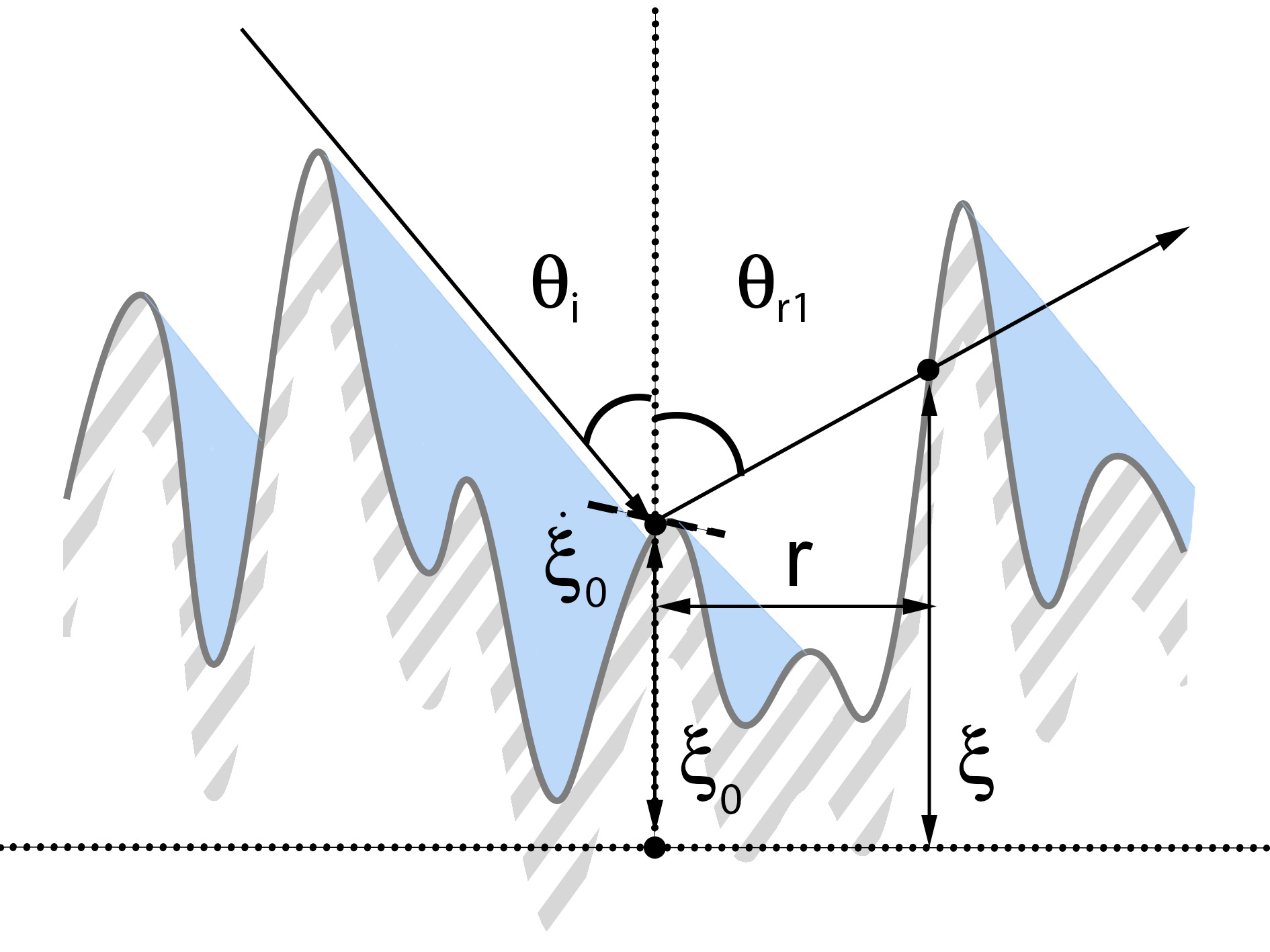}
    \caption{A sketch of a rough surface's self-shadowing effects on gas particles. The blue areas represent the shadowed parts for the first reflection.}
    \label{pp_1:fig:Shadowing_Sketch}
\end{figure}

The previous section presented in great detail a derivation of the scattered field of the gas particle wave functions following one surface reflection. However, for very rough surfaces, multiple wave reflections are expected to occur, which can substantially modify the features of the reflected particle's angular PDF. It is therefore of interest to quantify the probability of a scattered particle to reflect again off of the surface, given a reflection wave vector $\mathbf{k_r}$ and the height and slope component values. Under the assumption of isotropy, this analysis can be performed in one dimension. Then, the particle trajectory is fully described by the quantities $\xi_0$, $\Dot{\xi}_0$, $\Dot{\xi}_{0_p}$ and $\theta_{r_1}$, with $\Dot{\xi}_{0_p}$ being the surface slope perpendicular to the radial direction, as sketched in \cref{pp_1:fig:Shadowing_Sketch}. Several authors \citep{Beckmann1965, Smith1967, Brown1980} have tackled a similar problem in the context of electromagnetic wave scattering. For the purposes of this paper, the method developed by \citet{Smith1967} is adopted due to its accuracy and simplicity, and modified for poly-Gaussian surfaces. To this end, the shadowing function $\mathcal{S} : [0, \infty) \rightarrow [0, 1]$, where $\mathcal{S} = \mathcal{S}(\theta_{r_1}, \xi_0, \Dot{\xi}_0, \Dot{\xi}_{0_p}, r)$, is defined as the probability that a gas particle will not collide with any part of the surface for distances shorter than $r$, given its initial aforementioned parameters. Then, the sought-for multi-reflection probability is
\begin{equation}
    P\left(\text{particle reflects again} \, | \, \theta_{r_1}, \xi_0, \Dot{\xi}_0, \Dot{\xi}_{0_p}\right) = \lim_{r \rightarrow \infty} \mathcal{S}\left(\theta_{r_1}, \xi_0, \Dot{\xi}_0, \Dot{\xi}_{0_p}, r\right) = \mathcal{S}\left(\theta_{r_1}, \xi_0, \Dot{\xi}_0, \Dot{\xi}_{0_p}\right).
\end{equation}
Following \citet{Smith1967}, this function can be developed into a simple ordinary differential equation by expressing its value at $r + \Delta r$ in terms of its value at $r$:
\begin{equation}
    \mathcal{S}\left(\theta_{r_1}, G, r + \Delta r\right) = \mathcal{S}\left(\theta_{r_1}, G, r\right) \left(1 - g(r) \Delta r \right)
\end{equation}
Here, the parameters $\xi_0$, $\Dot{\xi}_0$ and $\Dot{\xi}_{0_p}$ were lumped together into the proxy parameter $G$ and $g(r) \Delta r$ represents the probability that the gas particle hits the surface in the interval $\left(r, r + \Delta r\right)$ provided that it has not at distance $r$.

\noindent Expanding the left-hand-side term into a Taylor series truncated after the linear term leads to a first order ordinary differential equation with the solution
\begin{equation} \label{eq:Shadowing_ODE_Solution}
    \mathcal{S}\left(\theta_{r_1}, G, r\right)
    = \mathcal{S}\left(\theta_{r_1}, G, 0\right)
    \exp\left[- \int_{0}^{r} g(s) \, \dif s \right]
    = h(\eta - \Dot{\xi}_0) \exp\left[- \int_{0}^{r} g(s) \, \dif s \right],
\end{equation}
where $\eta = \cot(\theta_{r_1})$ and $h(x)$ is the unit step function defined as
\begin{equation}
    h(x) = \begin{cases}
        0, & x < 0 \\
        1, & x \geq 0
    \end{cases}.
\end{equation}
The function $g(r) \Delta r$ is constructed as the conditional probability of two events. The first is that the gas particle is not shadowed and, therefore, is above the surface at $r$, i.e. $\xi_0 + \eta r > \xi(r)$. The other is that the gas particle is shadowed at $r + \Delta r$, i.e. $\xi_0 + \eta (r + \Delta r) \le \xi(r + \Delta r)$ and $\eta \leq \Dot{\xi}(r + \Delta r)$. Hence,
\begin{eqnarray}
    g(r) \Delta r & = & P\left(\xi_0 + \eta (r + \Delta r) < \xi(r + \Delta r), \eta \leq \Dot{\xi}(r + \Delta r) \, | \,  \xi_0 + \eta r > \xi(r)\right) \nonumber \\
    & = & \frac{P\left(\xi_0 + \eta (r + \Delta r) < \xi(r + \Delta r), \eta \leq \Dot{\xi}(r + \Delta r),  \xi_0 + \eta r > \xi(r) \right)}{P\left( \xi_0 + \eta r > \xi(r)\right)}.
\end{eqnarray}
Expanding the probabilities in the denominator and numerator leads to
\begin{equation}
    g(r) \Delta r = \frac{\Delta r \int_{\eta}^{\infty}{\left(\Dot{\xi} - \eta\right) \left[P(\xi, \Dot{\xi} \, | \, G, r)\right]_{\xi = \xi_0 + \eta r} \, \dif\Dot{\xi}}}{\infint{ \int_{-\infty}^{\xi_0 + \eta r} P(\xi, \Dot{\xi} \, | \, G, r) \, \dif \xi \dif \Dot{\xi}}}
\end{equation}
 It has already been shown in \cref{subsec:Surface_Modelling} that the PDFs of the height and slopes are independent variables for a Gaussian surface. The same is now assumed for a poly-Gaussian one. It is further assumed that new collisions occur at a far enough distance $r$ that $\xi$ and $\Dot{\xi}$ are independent of $\xi_0$ and $\Dot{\xi}_0$ \citep{Smith1967}. Then, the expression reduces to
\begin{equation} \label{pp_1:eq:shadow_independence}
    g(r) \Delta r = \frac{\int_{\eta}^{\infty}{\left(\Dot{\xi} - \eta \right) P(\Dot{\xi}) \, \dif \Dot{\xi}}}{\int_{-\infty}^{\xi_0 + \eta r} P(\xi) \, \dif \xi} P(\xi_0 + \eta r) \Delta r = \frac{\Delta r}{F(\xi_0 + \eta r)} \left. \frac{\dif F}{\dif \xi} \right|_{\xi = \xi_0 + \eta r} \int_{\eta}^{\infty}{\left(\Dot{\xi} - \eta \right) P(\Dot{\xi}) \, \dif \Dot{\xi}} ,
\end{equation}
where $F(x)$ is the cumulative distribution function of the PDF $P(x)$ \citep{Brown1980}. Substituting the above formula into \cref{eq:Shadowing_ODE_Solution} and changing integration variables from $r$ to $\xi$ leads to
\begin{align} \label{eq:Shadowing_Func_Expression}
    \mathcal{S}\left( \theta_{r_1}, \xi_0, \Dot{\xi}_0\right) & = h\left(\eta - \Dot{\xi}_0\right) \exp\left\{ - \frac{1}{\eta} \int_{\eta}^{\infty}{\left(\Dot{\xi} - \eta \right) P(\Dot{\xi}) \, \dif \Dot{\xi}} \int_{\xi_0}^{\infty} \frac{1}{F(\xi)} \frac{\dif F}{\dif \xi}\left( \xi\right) \, \dif \xi \right\} \\ & = h\left(\eta - \Dot{\xi}_0\right) \exp\left\{ - \frac{1}{\eta} \int_{\eta}^{\infty}{\left(\Dot{\xi} - \eta \right) P(\Dot{\xi}) \, \dif \Dot{\xi}} \int_{F(\xi_0)}^{1} \dif \ln\left( F(\xi) \right) \right\} \\
    & = h\left(\eta - \Dot{\xi}_0\right) F(\xi_0)^{ \frac{1}{\eta} \int_{\eta}^{\infty}{\left(\Dot{\xi} - \eta \right) P(\Dot{\xi}) \, \dif \Dot{\xi}}}.
\end{align}
Up to this point, we have followed the derivations of the shadowing function from \citep{Brown1980}. Now, we adapt the expression of this function for the Kirchhoff model. In the case of gas particle scattering, the condition that $\eta \geq \Dot{\xi}_0$ for the reflected particles is always satisfied, for if not, they would scatter through the surface, which violates impermeability, and is physically impossible. As a result, the $h\left(\eta - \Dot{\xi}_0\right)$ term is dropped, and $\mathcal{S}$ loses its dependence on $\Dot{\xi}_0$. By inserting Eqs. \ref{pp_1:eq:polyGaussian_height} and \ref{pp_1:eq:SlopePDF}, $\mathcal{S}(\theta_{r_1}, \xi_0)$ becomes
\begin{equation}
    S(\theta_{r_1}, \xi_0) = \left\{ \infint{ P(\gamma) \cdot \frac{1}{2}\left[1 + \erf\left( \frac{\xi_0 - \mu(\gamma)}{\sigma(\gamma)\sqrt{2}}\right) \right]\, \dif \gamma}\right\}^{\frac{1}{\eta} \infint{\infint{\int_{\eta}^{\infty}{P(\gamma) P(\Dot{\gamma})\left(\Dot{\xi} - \eta \right) \frac{1}{\sqrt{2\pi\left( \Dot{\gamma}^2\sigma_{\gamma}(\gamma)^2 + 2 \left(\frac{\sigma(\gamma)}{T}\right)^2\right)}}\exp\left[ \frac{\left( \Dot{\xi} - \eta_{\gamma} \Dot{\gamma}\right)^2}{2 \left( \Dot{\gamma}^2\sigma_{\gamma}(\gamma)^2 + 2\left(\frac{\sigma(\gamma)}{T}\right)^2\right)}\right] \, \dif \Dot{\xi} \dif \Dot{\gamma} \dif \gamma}}}}.
\end{equation}
Working out the inner integral in the exponent leads to
\begin{multline}
    S(\theta_{r_1}, \xi_0) = \left\{ \infint{ P(\gamma) \cdot  \frac{1}{2}\left[1 + \erf\left( \frac{\xi_0 - \mu(\gamma)}{\sigma(\gamma)\sqrt{2}}\right) \right]\, \dif \gamma}\right\}^{\frac{1}{\eta} \infint{\infint{P(\gamma) P(\Dot{\gamma}) \Delta(\gamma, \Dot{\gamma}) \, \dif \Dot{\gamma} \dif \gamma}}}, \\ \text{where} \quad \Delta(\gamma, \Dot{\gamma}) = \frac{w(\gamma, \Dot{\gamma})}{\sqrt{2\pi}}\exp\left[ - \frac{(\eta - \mu_{\gamma} \Dot{\gamma})^2}{2 w(\gamma, \Dot{\gamma})^2}\right] - \frac{1}{2}\left( \eta - \mu_{\gamma} \Dot{\gamma}\right)\erfc\left[\frac{\eta - \mu_{\gamma} \Dot{\gamma}}{w(\gamma, \Dot{\gamma})\sqrt{2}} \right] \quad \text{and} \quad w(\gamma, \Dot{\gamma})^2 = 2 \left(\frac{\sigma(\gamma)}{T}\right)^2 + \sigma_{\gamma}(\gamma)^2 \Dot{\gamma}^2.
\end{multline}
The above equation represents the desired analytical expression of the probability of a particle colliding with the surface, given an initial reflection angle $\theta_{r_1}$ and height $\xi_0$. We now derive, from this shadowing function, another important quantity to the development of a multi-reflection GSI algorithm, namely the probability of a gas particle hitting the surface at height $\xi$, given its initial conditions described by $G$. This probability is immediately found as
\begin{equation}
    P\left(\xi_{0_{new}} \, | \, \theta_r, \xi_0\right) = \mathcal{S}(\theta_r, \xi_0, r) P(\xi).
\end{equation}
Expanding the expression of $\mathcal{S}(\theta_r, \xi_0, \tau)$ using \cref{eq:Shadowing_Func_Expression} results in
\begin{multline}
    P\left(\xi_{0_{new}} \, | \, \theta_r, \xi_0\right) = P(\xi) \left(\frac{F(\xi_0)}{F(\xi(\tau))}\right)^{\frac{1}{\eta} \infint{\infint{P(\gamma) P(\Dot{\gamma}) \Delta(\gamma, \Dot{\gamma}) \, \dif \Dot{\gamma} \dif \gamma}}} \\ = \infint{ P(\gamma) \frac{1}{\sigma(\gamma)\sqrt{2\pi}} \exp\left[ -\frac{\left(\xi_{0_{new}} - \mu(\gamma) \right)^2}{2\sigma(\gamma)^2}\right] \, \dif \gamma} \left( \frac{\infint{ P(\gamma) \frac{1}{2}\left[1 + \erf\left( \frac{\xi_0 - \mu(\gamma)}{\sigma(\gamma)\sqrt{2}}\right) \right]\, \dif \gamma}}{\infint{ P(\gamma) \frac{1}{2}\left[1 + \erf\left( \frac{\xi_{0_{new}} - \mu(\gamma)}{\sigma(\gamma)\sqrt{2}}\right) \right]\, \dif \gamma}}\right)^{\frac{1}{\eta} \infint{\infint{P(\gamma) P(\Dot{\gamma}) \Delta(\gamma, \Dot{\gamma}) \, \dif \Dot{\gamma} \dif \gamma}}}
\end{multline}
which is the final form of that probability. A final expression needed for the Kirchhoff model is the probability of a particle hitting the surface at a given height $\xi_0$ on its initial reflection, given the incidence angle $\theta_i$, which is is given by
\begin{multline}
    P(\xi_0 \, | \, \theta_i) = \frac{\mathcal{S}(\theta_i, \xi_0)}{\mathcal{S}(\theta_i)} P(\xi_0) \\ = \frac{1}{\mathcal{S}(\theta_i)}\left\{ \infint{ P(\gamma)  \frac{1}{2}\left[1 + \erf\left( \frac{\xi_0 - \mu(\gamma)}{\sigma(\gamma)\sqrt{2}}\right) \right]\, \dif \gamma}\right\}^{\frac{1}{\eta} \infint{\infint{P(\gamma) P(\Dot{\gamma}) \Delta(\gamma, \Dot{\gamma}) \, \dif \Dot{\gamma} \dif \gamma}}} \infint{ P(\gamma) \frac{1}{\sigma(\gamma)\sqrt{2\pi}} \exp\left[ -\frac{\left(\xi - \mu(\gamma) \right)^2}{2\sigma(\gamma)^2}\right] \, \dif \gamma} \\ \text{with} \quad \mathcal{S}(\theta_i) = \infint{\mathcal{S}(\theta_i, \xi_0) P(\xi_0) \, \dif \xi_0}.
\end{multline}
\subsection{Poly-Gaussian Mixture Self-Shadowing} \label{subsec:polygauss_mixture_shadowing}
Following the theoretical framework developed in \cref{subsec:shadowing_multireflection}, we now return to \cref{eq:Kirchhoff_Solution} and re-examine the poly-Gaussian mixture defined by the processes $\gamma(x, y)$ and $\Dot{\gamma}(x, y)$ through the lens of shadowing and multi-reflection phenomena. Specifically, we define the new joint PDF $P\left(\gamma, \Dot{\gamma}_x, \Dot{\gamma}_y \, | \, \theta_i\right)$, which describes the "apparent" mixture observed by the incoming gas particles, given the angle of incidence $\theta_i$:
\begin{align} \label{eq:gamma_shadowing}
    P\left(\gamma, \Dot{\gamma}_x, \Dot{\gamma}_y \, | \, \theta_i \right) & = P\left( \text{particle hits surface at} \left\{\gamma, \Dot{\gamma}_x, \Dot{\gamma}_y \right\}\right) \nonumber \\
    & = \left[\infint{\infint{\infint{\mathcal{S}\left(\theta_i, \xi, \Dot{\xi}_y\right) \left\lvert\frac{\mathbf{n_i} \cdot \mathbf{n_L}}{\mathbf{n_L} \cdot \mathbf{n_G}}\right\rvert P\left(\xi, \Dot{\xi}_x, \Dot{\xi}_y \, | \, \gamma, \Dot{\gamma}_x, \Dot{\gamma}_y\right) \, \dif \Dot{\xi}_x \dif \Dot{\xi}_y \dif \xi}}} \right] P(\gamma) P(\Dot{\gamma}_x) P(\Dot{\gamma}_y)  \nonumber \\
    & = \left[\infint{\infint{\mathcal{S}\left(\theta_i, \xi, \Dot{\xi}_y\right) \left|\frac{\cos\left(\theta_i - \arctan(\Dot{\xi}_y) \right)}{\cos\left( \arctan(\Dot{\xi}_y)\right)}\right| P\left(\xi \, | \, \gamma\right) P\left(\Dot{\xi}_y \, | \, \gamma, \Dot{\gamma}_y \right) \, \dif \Dot{\xi}_y \dif \xi}} \right] P(\gamma) P(\Dot{\gamma}_x) P(\Dot{\gamma}_y)  \nonumber \\
    & = \left[\infint{\int_{-\infty}^{\eta}{\left|\frac{\cos\left(\theta_i - \arctan(\Dot{\xi}_y) \right)}{\cos\left( \arctan(\Dot{\xi}_y)\right)}\right| F(\xi)^{ \frac{1}{\eta} \int_{\eta}^{\infty}{\left(\Dot{s} - \eta \right) P(\Dot{s}) \, \dif \Dot{s}}} P\left(\xi \, | \, \gamma\right) P\left(\Dot{\xi}_x \, | \, \gamma, \Dot{\gamma}_y \right) \, \dif \Dot{\xi}_y \dif \xi}} \right] P(\gamma) P(\Dot{\gamma}_x) P(\Dot{\gamma}_y)
\end{align}
where $\mathbf{n_i}$ is the incident unit vector, $\mathbf{n_G} = [0, 0, 1]^T$ is the global surface-normal vector and $\mathbf{n_L}$ is the local surface-normal vector. Furthermore, in the above equation, the y-axis invariance of the shadow function $\mathcal{S}(\theta_i, \xi, \Dot{\xi}_y)$ was employed. To simplify it further, the height-dependent term $F(\xi)^{ \frac{1}{\eta} \int_{\eta}^{\infty}{\left(\Dot{s} - \eta \right) P(\Dot{s}) \, \dif \Dot{s}}}$ is expanded in a Taylor series around the local mean $\mu(\gamma)$ and truncated after the linear term, i.e.
\begin{align}
    F(\xi)^D & \approx F(\mu)^D + D F(\mu)^{D-1} P(\mu) \left( \xi - \mu\right), 
\end{align}
where $D = \frac{1}{\eta} \int_{\eta}^{\infty}{\left(\Dot{s} - \eta \right) P(\Dot{s}) \, \dif \Dot{s}}$. Substituting this approximation into \cref{eq:gamma_shadowing} gives
\begin{align}
    P\left(\gamma, \Dot{\gamma}_x, \Dot{\gamma}_y \, | \, \theta_i \right) & = \left\{\infint{\left[ F(\mu)^{D} + D  F(\mu)^{D-1} P(\mu) \left( \xi - \mu\right)\right] P\left(\xi \, | \, \gamma\right) \dif \xi \int_{-\infty}^{\eta}{\left|\frac{\cos\left(\theta_i - \arctan(\Dot{\xi}_y) \right)}{\cos\left( \arctan(\Dot{\xi}_y)\right)}\right| P\left(\Dot{\xi}_y \, | \, \gamma, \Dot{\gamma}_x \right) \, \dif \Dot{\xi}_y}} \right\} P(\gamma) P(\Dot{\gamma}_x) P(\Dot{\gamma}_y) \\
    & = \left\{F(\mu)^{D} \int_{-\infty}^{\eta}{\left|\frac{\cos\left(\theta_i - \arctan(\Dot{\xi}_y) \right)}{\cos\left( \arctan(\Dot{\xi}_y)\right)}\right|  P\left(\Dot{\xi}_y \, | \, \gamma, \Dot{\gamma}_y \right) \, \dif \Dot{\xi}_y} \right\} P(\gamma) P(\Dot{\gamma}_x) P(\Dot{\gamma}_y) \\
    & = \left\{F(\mu)^{D} \int_{-\infty}^{\eta}{\left[\cos(\theta_i) + \Dot{\xi}_y \sin(\theta_i) \right] \frac{1}{w(\gamma, \Dot{\gamma}_y)\sqrt{2\pi}}\exp\left[ - \frac{\left( \Dot{\xi}_y - \mu_{\gamma} \Dot{\gamma}\right)^2}{2 w^2(\gamma, \Dot{\gamma}_y)}\right] \, \dif \Dot{\xi}_y} \right\} P(\gamma) P(\Dot{\gamma}_x) P(\Dot{\gamma}_y) \\
    & = F(\mu)^D \left[\frac{1}{2}\left( 1 + \erf\left(\frac{\eta + \mu_{\gamma} \Dot{\gamma}_y}{w(\gamma, \Dot{\gamma}_y) \sqrt{2}} \right) \right) \left( \cos(\theta_i) + \mu_{\gamma} \Dot{\gamma}_y \sin(\theta_i)\right) + \frac{w(\gamma, \Dot{\gamma}_y)}{\sqrt{2\pi}}\sin(\theta_i) \exp\left(-\frac{\left(\eta + \mu_{\gamma} \Dot{\gamma}_y\right)^2}{2 w^2(\gamma, \Dot{\gamma}_y)} \right) \right] P(\gamma) P(\Dot{\gamma}_x) P(\Dot{\gamma}_y) \\
    & = \mathcal{S}(\theta_i, \mu(\gamma)) \Theta(\gamma, \Dot{\gamma}_y, \theta_i) P(\gamma) P(\Dot{\gamma}_x) P(\Dot{\gamma}_y) \\
    & \text{with} \quad \mathcal{S}(\theta_i, \mu(\gamma)) = F(\mu)^D \\ & \text{and} \quad \Theta(\gamma, \Dot{\gamma}_y, \theta_i) = \frac{1}{2}\left( 1 + \erf\left(\frac{\eta + \mu_{\gamma} \Dot{\gamma}_y}{w(\gamma, \Dot{\gamma}_y) \sqrt{2}} \right) \right) \left( \cos(\theta_i) + \mu_{\gamma} \Dot{\gamma}_y \sin(\theta_i)\right) + \frac{w(\gamma, \Dot{\gamma}_y)}{\sqrt{2\pi}}\sin(\theta_i) \exp\left(-\frac{\left(\eta + \mu_{\gamma} \Dot{\gamma}_y\right)^2}{2 w^2(\gamma, \Dot{\gamma}_y)} \right)
\end{align}
where $\mathcal{S}(\theta_i, \mu(\gamma)) \Theta(\gamma, \Dot{\gamma}_y, \theta_i)$  denotes the poly-Gaussian mixture shadowing term as a function of the processes $\gamma$ and $\Dot{\gamma}_x$ and the incidence angle $\theta_i$. It modifies the poly-Gaussian mixture probability $P(\gamma) P(\Dot{\gamma}_x) P(\Dot{\gamma}_y)$ such that shadowing effects and multiple reflections are accounted for. Hence, the general Kirchhoff solution in \cref{eq:Kirchhoff_Solution} is adapted to include this modification as follows:
\begin{multline}
    \langle \rho \rho^* \rangle(\theta_i, \theta_{r_1}, \theta_{r_2}) \approx \frac{F_k^2}{4\pi v_z^2A^2}\infint{\infint{\infint{\frac{P(\gamma)P(\Dot{\gamma}_x)P(\Dot{\gamma}_y)\mathcal{S}(\theta_i, \mu(\gamma)) \Theta(\gamma, \Dot{\gamma}_x, \theta_i)}{\sqrt{\left( \left(\frac{\sigma}{T} \right)^2  + \frac{1}{2} \sigma_{\gamma}^2 \Dot{\gamma}_x^2\right) \left( \left(\frac{\sigma}{T} \right)^2  + \frac{1}{2} \sigma_{\gamma}^2 \Dot{\gamma}_y^2 \right)}} \\ \times \exp \left[- \frac{(v_x + \mu_{\gamma} \Dot{\gamma}_x v_z)^2}{4 v_z^2 \left( \left(\frac{\sigma}{T} \right)^2  + \frac{1}{2} \sigma_{\gamma}^2 \Dot{\gamma}_x^2\right)} - \frac{(v_y + \mu_{\gamma} \Dot{\gamma}_y v_z)^2}{4 v_z^2 \left( \left(\frac{\sigma}{T} \right)^2  + \frac{1}{2} \sigma_{\gamma}^2 \Dot{\gamma}_y^2\right)}\right] \, \dif \gamma \dif \Dot{\gamma}_x \dif \Dot{\gamma}_y}}}
\end{multline}

\subsection{Iterative Algorithm and Gaussian Simplification} \label{subsec:algorithm}

\begin{figure}[H]
    \centering
    \includegraphics[width=.5\linewidth]{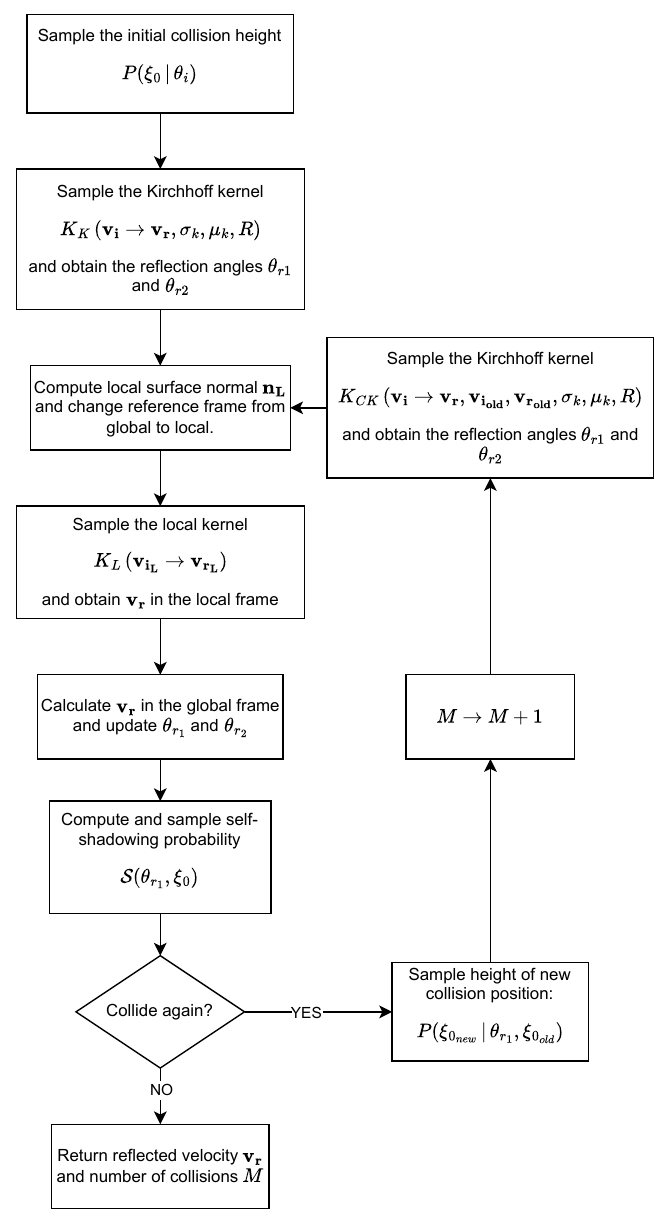}
    \caption{A flowchart of the proposed GSI model iterative process of accounting for multiple reflections.}
    \label{pp_1:fig:Flowchart}
\end{figure}

At a top level, the GSI model proposed in this paper should have the capacity to be written in the general scattering kernel form outlined in \cref{eq:KernelForm}, to represent a boundary condition to the Boltzmann transport equation describing free-molecular flow. The underlying probability of the reflected velocity $P(\mathbf{v_r})$ should capture multiple possible geometrical reflections and local interaction phenomena into one overarching expression between the incident and reflected velocities. The necessary components for such an expression have already been derived in \cref{subsec:Surface_Modelling,,subsec:Wave_Solution,,subsec:shadowing_multireflection}, where a modified poly-Gaussian surface model was used to derive the angular scattering probability $\langle \rho\rho^*\rangle\left(\theta_i, \theta_{r_1}, \theta_{r_2}\right)$ through the Kirchhoff approximation under the assumption of local specular reflection and, then, a probability of self-shadowing $\mathcal{S}(\theta_{r_1}, \xi_0)$ was included. Based on these components, the kernel for one reflection is
\begin{equation} \label{pp_1:eq:ngs_kirchhoff}
    K_{K}(\mathbf{v_i} \rightarrow \mathbf{v_r}) = \int_{\mathbf{v_i} \cdot \mathbf{n_L} < 0} K_L\left(\mathbf{v_{i_L}} \rightarrow \mathbf{v_{r_L}}\right) \frac{\partial \mathbf{v_{r_L}}}{\partial\mathbf{v_{r}}} P\left(\mathbf{n_L} \, | \, \mathbf{v_i}, \sigma_k, \mu_k, R \right)\frac{\partial \mathbf{n_L}}{\partial\mathbf{v_{r_L}}} \, \dif \mathbf{n_L},
\end{equation}
where the subscript $K$ stands for Kirchhoff, and
\begin{multline} 
    P\left(\mathbf{n_L} \, | \, \mathbf{v_i}, \sigma_k, \mu_k, R \right) = \langle \rho\rho^*\rangle(\theta_i, \mathbf{n_L}(\theta_{r_{1_s}}, \theta_{r_{2_s}}))  \frac{\partial \mathbf{n_L}}{\partial (\theta_i, \theta_{r_{1_s}}, \theta_{r_{2_s}})} = \frac{F_k^2}{4\pi v_z^2A^2}\infint{\infint{\infint{\frac{P(\gamma)P(\Dot{\gamma}_x)P(\Dot{\gamma}_y)\mathcal{S}(\theta_i, \mu(\gamma)) \Theta(\gamma, \Dot{\gamma}_x, \theta_i)}{\sqrt{\left( \left(\frac{\sigma}{R} \right)^2  + \frac{1}{2} \sigma_{\gamma}^2 \Dot{\gamma}_x^2\right) \left( \left(\frac{\sigma}{R} \right)^2  + \frac{1}{2} \sigma_{\gamma}^2 \Dot{\gamma}_y^2 \right)}} \\ \times \exp \left[- \frac{(v_x + \mu_{\gamma} \Dot{\gamma}_x v_z)^2}{4 v_z^2 \left( \left(\frac{\sigma}{R} \right)^2  + \frac{1}{2} \sigma_{\gamma}^2 \Dot{\gamma}_x^2\right)} - \frac{(v_y + \mu_{\gamma} \Dot{\gamma}_y v_z)^2}{4 v_z^2 \left( \left(\frac{\sigma}{R} \right)^2  + \frac{1}{2} \sigma_{\gamma}^2 \Dot{\gamma}_y^2\right)}\right] \, \dif \gamma \dif \Dot{\gamma}_x \dif \Dot{\gamma}_y}}} \sin(\theta_{r_{1_s}}).
\end{multline}
Here, a change of variable was performed in the Kirchhoff solution from \cref{eq:Kirchhoff_Solution}, from incidence and reflection angles to the local surface normal vector $\mathbf{n_L}$. However, the reflection angles $\theta_{r_{1_s}}$ and $\theta_{r_{2_s}}$ do not describe the final reflected particle direction, and instead represent the direction said particle would take assuming local specular reflections, i.e. $K_L\left(\mathbf{v_{i_L}} \rightarrow \mathbf{v_{r_L}}\right) \frac{\partial \mathbf{v_{r_L}}}{\partial\mathbf{v_{r}}} = 1$. Furthermore, $K_L(\mathbf{v_{i_L}} \rightarrow \mathbf{v_{r_L}})$ denotes the local scattering kernel with the velocities $\mathbf{v_{i_L}}$ and $\mathbf{v_{r_L}}$ in the local reference frame. Analogously, the kernel for any additional reflection is defined by
\begin{equation} \label{pp_1:eq:ngs_kirchhoff_cor}
    K_{CK}(\mathbf{v_i} \rightarrow \mathbf{v_r}) = \int_{\mathbf{v_i} \cdot \mathbf{n_L} < 0} K_L\left(\mathbf{v_{i_L}} \rightarrow \mathbf{v_{r_L}}\right) \frac{\partial \mathbf{v_{r_L}}}{\partial\mathbf{v_{r}}} P\left(\mathbf{n_L} \, | \, \mathbf{v_i}, \mathbf{n_{L_{old}}}, \mathbf{v_{i_{old}}}, \sigma_k, \mu_k, T \right)\frac{\partial \mathbf{n_L}}{\partial\mathbf{v_{r_L}}} \, \dif \mathbf{n_L}, 
\end{equation}
where the subscript $CK$ stands for correlated Kirchhoff and
\begin{multline}
    \quad P\left(\mathbf{n_L} \, | \, \mathbf{v_i}, \mathbf{n_{L_{old}}}, \mathbf{v_{i_{old}}}, \sigma_k, \mu_k, R \right) = \frac{F_k^2}{4\pi v_z^2A^2}\infint{\infint{\infint{\frac{P(\gamma \, | \, \gamma_{old})P(\Dot{\gamma}_x \, | \, \Dot{\gamma}_{x_{old}})P(\Dot{\gamma}_y \, | \, \Dot{\gamma}_{y_{old}}) \mathcal{S}(\theta_i, \mu(\gamma)) \Theta(\gamma, \Dot{\gamma}_x, \theta_i)}{w_{x} w_{x_{old}} w_{y} w_{y_{old}} (1 - \mathcal{C}_{\epsilon}^2)} \\ \times \exp \left[-\frac{1}{2(1-\mathcal{C}_{\epsilon}^2)}\left(\left(\frac{\overline{v_x}^2}{v_z^2 w_x^2} + \frac{\overline{v_{x_{old}}}^2}{v_{z}^2 w_{x_{old}}^2} -\frac{2 \mathcal{C}_{\epsilon} \overline{v_x} \overline{v_{x_{old}}}}{ v_z w_{x} w_{x_{old}}} \right) + \left(\frac{\overline{v_y}^2}{v_z^2 w_y^2} + \frac{\overline{v_{y_{old}}}^2}{v_{z} w_{y_{old}}^2} -\frac{2 \mathcal{C}_{\epsilon} \overline{v_y} \overline{v_{y_{old}}}}{v_z v_{z} w_{y}  w_{y_{old}}} \right)\right)\right]  \, \dif \gamma \dif \Dot{\gamma}_x \dif \Dot{\gamma}_y}}} \sin(\theta_{r_1}).
\end{multline}
Using the two expressions above for the uncorrelated and correlated Kirchhoff kernels together with $\mathcal{S}(\theta_{r_1}, \xi_0)$, one can define the multi-reflection
\begin{multline} \label{eq:GSI_Rough_Kernel}
    K_{MR}(\mathbf{v_i} \rightarrow \mathbf{v_r}) = \infint{\left\{\mathcal{S}(\mathbf{v_r}, \xi_0) K_{K}(\mathbf{v_i} \rightarrow \mathbf{v_r}) + \left(1 - \mathcal{S}(\mathbf{v_r}, \xi_0)\right)  \sum_{c=1}^{\infty} \left\{ \idotsint_{\mathbf{v_{r_{1\dots c}}} \mathbf{n_G} < 0} K_{CK}\left( \mathbf{v_{i}} \rightarrow \mathbf{v_{r_1}}\right) \left[\prod_{k=1}^{c-1}\left(1 - \mathcal{S}(\mathbf{v_{r_k}}, \xi_0)\right) K_{CK}\left( \mathbf{v_{r_k}} \rightarrow \mathbf{v_{r_{k+1}}}\right)\right] \right.\right. \\ \left.\left. \times \mathcal{S}\left(\mathbf{v_{r_c}}, \xi_0\right)K_{CK}\left( \mathbf{v_{r_c}} \rightarrow \mathbf{v_{r}}\right)  \, \dif \mathbf{v_{r_1}} \dots \dif \mathbf{v_{r_c}}\frac{\partial \mathbf{v_{r_c}}}{\partial \mathbf{v_r}} \right\}  \, \dif \xi_0 \right\}},
\end{multline}
where the subscript $MR$ stands for multi-reflection. The formula above describes the complete kernel proposed in this paper, and is prohibitively complex to express in a closed-form. It serves no purpose beyond being a starting point to a formal mathematical proof for the three required properties of scattering kernels: non-negativity, normalisation, and reciprocity. Thankfully, the first two can be inferred from the expressions for $K_K(\mathbf{v_i} \rightarrow \mathbf{v_r})$, $K_{CK}(\mathbf{v_i} \rightarrow \mathbf{v_r})$, and $K_{L}(\mathbf{v_{i_L}} \rightarrow \mathbf{v_{r_L}})$. Reciprocity, on the other hand, is less straightforward to infer based solely on the form of these expressions. Nevertheless, for the practical purpose of sampling velocities with this kernel, the iterative approach illustrated in \cref{pp_1:fig:Flowchart} is much simpler than using \cref{eq:GSI_Rough_Kernel} directly. In this approach, several non-standard PDFs need to be samples, i.e. $K_K(\mathbf{v_i} \rightarrow \mathbf{v_r}, \sigma_k, \mu_k, R)$, $K_{CK}(\mathbf{v_i} \rightarrow \mathbf{v_r}, \mathbf{v_{i_{old}}}, \mathbf{v_{r_{old}}}, \sigma_k, \mu_k, R)$, $K_L(\mathbf{v_{i_L}} \rightarrow \mathbf{v_{r_L}})$, $P(\xi_0 \, | \, \theta_i)$ and $P(\xi_{0_{new}} \, | \, \xi_{0_{old}}, \theta_{r_1})$. While sampling $K_L(\mathbf{v_{i_L}} \rightarrow \mathbf{v_{r_L}})$ can be decomposed into the sampling of a series of Gaussian and uniform PDFs through the graphical method outlined in \citep{Padilla2007}, the remaining PDFs require a more involved approach. For this, the Metropolis-Hastings algorithm as given in \citep{Metropolis1953} was employed, due to its computational efficiency. Another advantage of using this method is its laxity on the normalisation condition, as PDFs sampled with it need not be normalised. 
The iterative algorithm in \cref{pp_1:fig:Flowchart} begins by sampling the initial height where a gas particle approaching from incidence angle $\theta_i$ is likely to collide with the surface based on the PDF $P(\xi_0 \, | \, \theta_i)$. Next, the initial scattering angles $\theta_{r_1}$ and $\theta_{r_2}$, which arise from the surface roughness, are determined through sampling the Kirchhoff kernel $K_K(\mathbf{v_i} \rightarrow \mathbf{v_r})$. From these angles, a local surface normal is computed and, after transforming from global to local surface frame, the local scattering kernel $K_L(\mathbf{v_i} \rightarrow \mathbf{v_r})$ is also sampled to obtain the reflected particle's velocity relative to this normal. After transforming this velocity back to the global surface frame, where $\mathbf{n_G} = [0 \, 0 \, 1]^T$, the rescattering PDF is sampled through $\mathcal{S}(\theta_{r_1}, \xi_0)$. If a rescattering occurs, a new height is sampled through $P(\xi_{0_{new}} \, | \, \theta_{r_1}, \xi_{0_{old}})$ and the process repeats with another Kirchhoff sampling, this time using the kernel $K_{CK}$. Otherwise, the current reflected velocity $\mathbf{v_r}$ and number of collisions $N$ are returned. This simple algorithm samples the kernel $K_{MK}$ in \cref{eq:GSI_Rough_Kernel} while also simulating the 3D trajectory of a gas particle and the number of collisions with the surface. It can further be used to infer the reciprocity of the proposed Kirchhoff model. In a physical sense, reciprocity states that the number of particles incident to the surface must equal the number of reflected particles, i.e. no particles can be absorbed, on average. This condition is obviously fulfilled when the algorithm presented in \cref{pp_1:fig:Flowchart} converges. The analyses in \cref{subsec:Verification,subsec:Experimental,subsec:ModelApplication}will experimentally prove that the algorithm converges, confirming the reciprocity of the proposed Kirchhoff model. 

While the derivations behind the method may seem involved, the proposed GSI model still requires only between 1 and 4 coefficients $\sigma_k$ and $\mu_k$ and the autocorrelation length $R$ to describe the geometric scattering off of almost any surface. Special attention should be given to the case of a surface with a purely Gaussian height PDF, which has the same form as \cref{pp_1:eq:polyGaussian_height}, with the exception that $\sigma(\gamma)$ and $\mu(\gamma)$ become constants. In this situation, the model greatly simplifies, with the Kirchhoff solution and shadowing PDFs become
\begin{multline} \label{pp_1:eq:gs_kirchhoff}
    \langle \rho\rho^*\rangle(\theta_i, \theta_{r_1}, \theta_{r_2}) = \frac{F_k^2 R}{4\pi A^2 v_z^2 \sigma^2} \exp\left[-\frac{v_{xy}^2 R^2}{4 v_z^2 \sigma^2}\right] \, , \quad \mathcal{S}(\theta_{r_1}, \xi_0) = \left[ \frac{1}{2}\left(1 + \erf\left(\frac{\xi_0}{\sigma \sqrt{2}} \right) \right)\right]^{\frac{\Delta}{\eta}} \\ P(\xi_{0_{new}} \, | \, \theta_{r_1}, \xi_0) = \frac{1}{\sigma\sqrt{2\pi}}\exp\left[ -\frac{\xi_{0_{new}}}{2\sigma^2}\right] \times \left(\frac{1 + \erf\left( \frac{\xi_0}{\sigma\sqrt{2}}\right)}{1 + \erf\left( \frac{\xi_{0_{new}}}{\sigma\sqrt{2}}\right)}\right)^{\frac{\Delta}{\eta}} \, \quad P(\xi_0 \, |\, \theta_i) = \frac{1}{\sigma\sqrt{2\pi}}\exp\left[ -\frac{\xi_{0}}{2\sigma^2}\right] \times \left( \frac{1 + \erf\left( \frac{\xi_0}{\sigma\sqrt{2}}\right)}{2}\right)^{\frac{\Delta}{\eta}} \\ \text{with} \quad \Delta = \frac{w}{\sqrt{2\pi}}\exp\left[ \frac{\eta}{2w^2}\right] - \frac{\eta}{2}\erfc\left[ \frac{\eta}{w\sqrt{2}}\right] \, , \quad w = \frac{\sigma}{R}\sqrt{2},
\end{multline}
where only one model parameter is required to describe the scattering kernel $K_{MR}$, namely the roughness parameter $\frac{\sigma}{R}$. Furthermore, it can be easily shown that $R = R_{\xi}$.  Such a simplification is particularly attractive if one seeks to solve the inverse GSI problem, i.e. estimate this parameter based on in-orbit acceleration data through a procedure similar to that of \citet{March2021}. Therefore, \cref{sec:Results} investigates the discrepancy in the scattering indicators caused by the Gaussian surface assumption and its effect on the drag coefficient $C_D$ of a given satellite geometry.

\subsection{Model Limitations and Range of Validity} \label{subsec:ValidityRange}

Several assumptions and simplifications were made in the development of the GSI kernel in \cref{eq:GSI_Rough_Kernel}, some of which have important implications that could limit the range of applicability. By far, the most important one is the analogy between gas particles and waves, which holds if the trajectories of the former remain undisturbed until contact with the surface is achieved. This imposes a lower limit on the roughness scales where the model is applicable. These scales, defined by their minimum radius of curvature $\mathcal{R}$, must be much larger than the depth of the potential well $\mathcal{W}$ that characterises the surface, i.e. $\mathcal{R} \gg \mathcal{W}$, as described in \cref{subsec:Existing_Methods_GSI}. Interactions occurring due to roughness at smaller scales, e.g. atomic corrugations, should be considered "local" and accounted for through a model such as the Washboard model of \citet{Liang2018}.

Another limiting assumption of the model comes from the derivation of the Kirchhoff approximation, which assumes the scattered wave field at the surface boundary to resemble that of an infinitely-flat and smooth plane (cf. \cref{subsec:Roughness_GSI}). \citet{Beckman1987-kr} claim this hypothesis is valid if the wavelength of the incident particle is much lower than the smallest radius of curvature on the surface, $\mathcal{R}$, i.e
\begin{equation}
    4\pi \cdot \mathcal{R} \cdot \cos(\nu) \gg \lambda_i \quad \text{with} \quad \nu = \measuredangle \left(- \mathbf{k_i}, \mathbf{n_L} \right).
\end{equation}
If one now considers the largest wavelength encountered abundently in the atmosphere, given by the Helium atom with molar mass $\mathcal{M}_{He}=$ \SI{4}{\gram\per\mol}, and assumes a representative velocity of \SI{7000}{\meter\per\second}, then this wavelength is approximately $\lambda_i=$ \SI{0.14}{\angstrom}. Even for a grazing incidence angle of \SI{85}{\degree}, this wavelength is approximately 20 times smaller than the lattice parameter of an iron crystal, which is $a  =$ \SI{2.85}{\angstrom}. As such, this assumption is valid for almost all angles, gasses and materials encountered in GSI problems, i.e. it does not limit the model's applicability.

A third, very important shortcoming of the Kirchhoff model comes from the assumption of inverse-squared exponential autocorrelation functions for $\mathcal{C}_{\gamma}(r)$ and $\mathcal{C}_{\epsilon}(r)$. This limits the PSD of a surface generated with this model to the form
\begin{equation}
    \PSD(\nu) = \mathcal{F}\left\{ C(r) \right\}(\nu) = \infint{\exp\left[ -\frac{r^2}{R^2} - 2 \pi i \nu r\right] \, \dif r} = R\sqrt{\pi} \exp\left[ - \pi^2 R^2 \nu^2\right],
\end{equation}
where $C(r)$ stands for either $\mathcal{C}_{\gamma}(r)$ or $\mathcal{C}_{\epsilon}(r)$. This equation exhibits a different behaviour compared to the typical fractal surface model used in many studies (\citet{Karan2008, Cutler2021}), which is characterised by the Hurst coefficient $H(q)$. Consequently, the poly-Gaussian surface approach may misrepresent the larger length scales of the surface, as illustrated in \cref{pp_1:fig:psd_rough_surfaces}. Therefore, careful selection of the length-scale range is necessary for experimental fitting to 3D surface data. However, this is usually not a concern, as demonstrated in \cref{pp_1:eq:auto_length_approx} by the $\frac{\pi^2n^2}{R_S^2}$ factor, which shows that the highest frequency components of the PSD (and thus the smallest length scales) predominantly influence the autocorrelation length $R$. In other words, surface roughness is largely determined by smaller length scales.

A final limitation to the kernel stems from the poly-Gaussian surface model itself. It was proven in \cref{subsec:Surface_Modelling} that the slope PDF $P(\Dot{\xi})$ of the surface depends on the control process $\gamma(x, y)$, which is a free parameter of the model. Since this PDF is directly linked to the scattering behaviour, the choice of $\gamma$ may significantly limit the range of slope statistics that can be captured. In particular, if $\gamma$ is the standard Gaussian process, and its derivative $\Dot{\gamma}$ is also a Gaussian process, then this imposes a symmetry constraint around zero on the quantities $\sigma_{\gamma}(\gamma)\Dot{\gamma}$ and $\mu_{\gamma}(\gamma)\Dot{\gamma}$, which, in turn, imposes a symmetry constraint on the entire slope PDF $P(\Dot{\xi})$. In fact, any symmetrical process $\Dot{\gamma}$ results in the same limitation. This, in turn, excludes any anisotropic surfaces from being modelled, such as those analysed by \citet{Shoda2022}, exhibiting a "saw-tooth pattern". To circumvent this, the Gaussian processes of the model would have to be generalized in terms of a "global orientation angle" parameter, and a different choice of $\gamma$ would have to be made as well, at the expense of analytical simplicity. 

\section{Results and Discussion} \label{sec:Results}

\subsection{Model Verification with Ray Tracing Simulations} \label{subsec:Verification}

To verify the mathematical accuracy of the Kirchhoff model described in \cref{sec:Model}, an extensive series of tests was performed using an alternative benchmark method across a portion of the parameter space likely to be encountered in the thermosphere environment. For benchmarking, we employed a Test Particle Monte Carlo (TPMC) approach, where incident particles with a given velocity vector $\mathbf{v_i}$ scatter from a rectangular geometry sample of a surface defined by $2N + 2$ normalised coefficients $\mu_k$ and $\sigma_k$, $k \in \overline{0, N}$, using a local kernel $K_L(\mathbf{v_i} \rightarrow \mathbf{v_r})$. The CLL kernel, as described by \citet{Lord1995}, was chosen for modelling the local interactions due to its ability to empirically replicate the quasi-specular behaviour observed in many experimental scattering results from atomically smooth or "clean" surfaces, where impulsive scattering is the dominant interaction mechanism \citep{Xu2023, Murray2015, Murray2017}. The software implementing the TPMC algorithm used in this study, named "GSI\_ToolBox," has been published under the Apache 2.0 license on the 4TU.ResearchData repository \citep{TPMC}. {\sethlcolor{green}\hl{Note to editor and reviewers: We will make the software in form of source code public upon acceptance of the manuscript. This is to ensure that the published software version agrees with the final  version of the manuscript. For now, a question mark will serve as a placeholder.}}

\begin{figure}[H]
    \centering
    \includegraphics[width=0.99\linewidth]{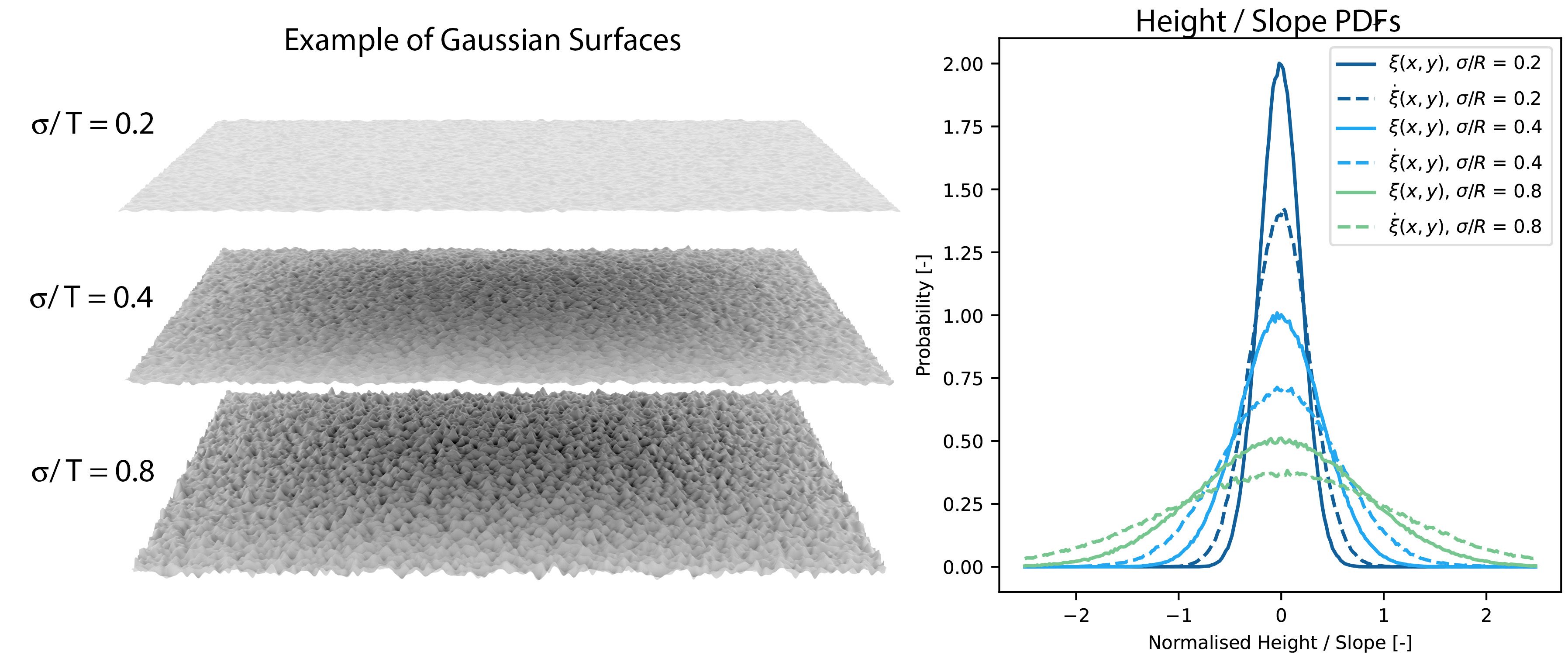}
    \caption{On the left: illustrations of three Gaussian surface samples with roughness values of 0.2, 0.4 and 0.8. On the right: the height ($\xi$) and slope ($\Dot{\xi}$) PDFs of these surfaces.}
    \label{pp_1:fig:gs_surf_pdf}
\end{figure}

The first set of verification tests focused on the Gaussian version of the Kirchhoff model as given in \cref{pp_1:eq:gs_kirchhoff}. For this analysis, the varied parameters included the surface roughness parameter $\sigma/R$, the normal energy accommodation coefficient $\alpha_N$, the tangential momentum accommodation coefficient $\sigma_T$ of the local CLL kernel, and the angle of incidence $\theta_i$. Their tested ranges are $[0.0, 1.0]$, $[0.0, 1.0]$, $[0.0, 1.0]$, and $[0^{\circ}, 75^{\circ}]$, respectively. Several representative cases from this analysis are presented in this section, with the model input parameters provided in \cref{tab:gaussian_analysis}.
\begin{table}[H]
\centering
\caption{Input parameters for the Gaussian model analysis.}
\label{tab:gaussian_analysis}
\begin{tabular}{lrrlrr}
\hline\hline
\textbf{Parameter Name} & \textbf{Value}    & \textbf{Unit}        & \textbf{Parameter Name} & \textbf{Value} & \textbf{Unit} \\ \hline 
$\lVert \mathbf{v_i}\rVert$               & 7000                             & [m/s]                               & $\sigma/R$             & 0.2, 0.4, 0.8                      & [-]                               \\
$\mathcal{M}_{He}$                   & 4                                  & [g/mol]                           & $\theta_i$              & 15, 45, 75                         & [$^{\circ}$]                      \\
$N_{particles}$         & 100000                             & [-]                               & CLL, $\alpha_N$              & 0.0, 1.0                           & [-]                               \\
$T_S$                   & 300                                & [K]                                 & CLL, $\sigma_T$              & 0.0, 1.0                           & [-]                               \\ \hline\hline
\end{tabular}
\end{table}
 Since it is impossible to present all analysed cases, only the extreme cases of the CLL kernel parameter space are presented, as they show the most diverging behaviour between the Kirchhoff model and the TPMC simulations. Furthermore, the incidence angles $\theta_i = 15^{\circ}$, $\theta_i = 45^{\circ}$ and $\theta_i = 75^{\circ}$ are chosen to represent most flow-facing surfaces of satellites in the thermosphere. In a similar manner, three surfaces with roughness parameters of $\sigma/R = 0.2$, $\sigma / R = 0.4$ and $\sigma / R = 0.8$ were selected, as they succinctly capture the effect of geometric roughness on the scattering behaviour. Renders of the rectangular geometry samples for these roughness parameters are shown in \cref{pp_1:fig:gs_surf_pdf} on the left\footnote{The renders were generated using the Blender 2.0 software from \citet{Blender2}.}. On the right side of the same figure, the height and slope PDFs of these surfaces are plotted. These were normalised on the horizontal axis with the autocorrelation lengths of the surfaces.

\begin{figure}[H]
    \centering
    \includegraphics[width=0.92\linewidth]{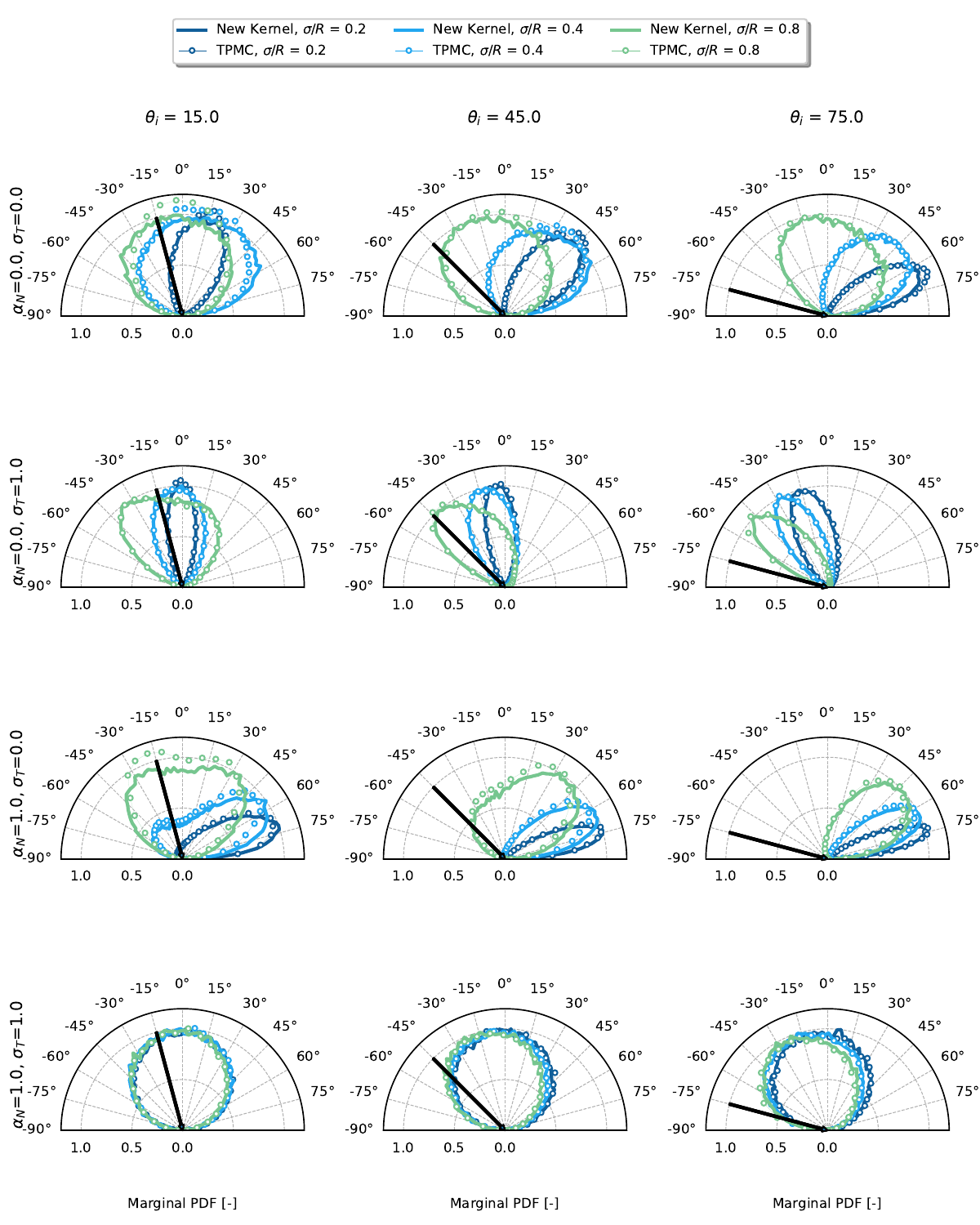}
    \caption{The YZ-plane marginal angular PDFs of Helium gas particles with different incidence angles, scattering from Gaussian surfaces with different levels of roughness. The model-generated results are shown in solid lines, while the TPMC simulation results are given in dotted lines. The incident direction is depicted with a black arrow.}
    \label{pp_1:fig:gs_angular_pdfs}
\end{figure}

Angular scattering plots were generated for each of the test cases outlined above using a number of particles of $N_{particles} = 100000$ in the TPMC simulations. \Cref{pp_1:fig:gs_angular_pdfs} shows the marginals of the 3D angular scattering PDF onto the YZ plane (cf. \cref{pp_1:fig:angle_definitions} for the definition of the YZ plane). Overall, good agreement is observed between the Kirchhoff model and TPMC angular PDFs across all considered test cases. An increase in accuracy can be noticed for increasing incidence angles across all CLL parameter combinations. This is expected because of the assumption of independence between the pairs $\xi_0$, $\Dot{\xi}_0$ and $\xi$, $\Dot{\xi}$ in \cref{pp_1:eq:shadow_independence}. If a particle has a higher initial incidence angle, it is likely to travel further in the horizontal direction after its first reflection, making that assumption more valid for higher incidence angles. A particle under normal incidence, however, is more likely to remain in close vicinity to its original scattering point as it "penetrates" the surface deeper, hence resulting in an underestimation of the shadowing probability $\mathcal{S}(\theta_{r_1}, \xi_0)$. A second important feature visible in \cref{pp_1:fig:gs_angular_pdfs} is the larger discrepancy of test cases with zero tangential momentum accommodation, i.e. $\sigma_T = 0.0$ and high normal energy accommodation, i.e. $\alpha_N \approx 1$. This may also be explained through the assumption of independence in \cref{pp_1:eq:shadow_independence}. In this scenario, a particle incident to a surface at a high speed loses most of its normal momentum, but preserves its tangential component, resulting in a reflection direction that is almost tangent to the local surface plane. As a result, if the particle originally hits a "valley" of the surface, it is bound to experience a second collision close to its original one, which goes against the assumption of independence. Such a discrepancy is most evident for the case where $\theta_i = 15^{\circ}$, $\alpha_N = 1.0$, $\sigma_T = 0.0$ in \cref{pp_1:fig:gs_angular_pdfs}. The reader should note that the YZ marginals shown in the aforementioned figure represent only part of the picture. Similar scatter plots for the XY marginals are presented and discussed in Appendix \ref{ap:Ad_results}.

\begin{figure}[H]
    \centering
    \includegraphics[width=0.99\linewidth]{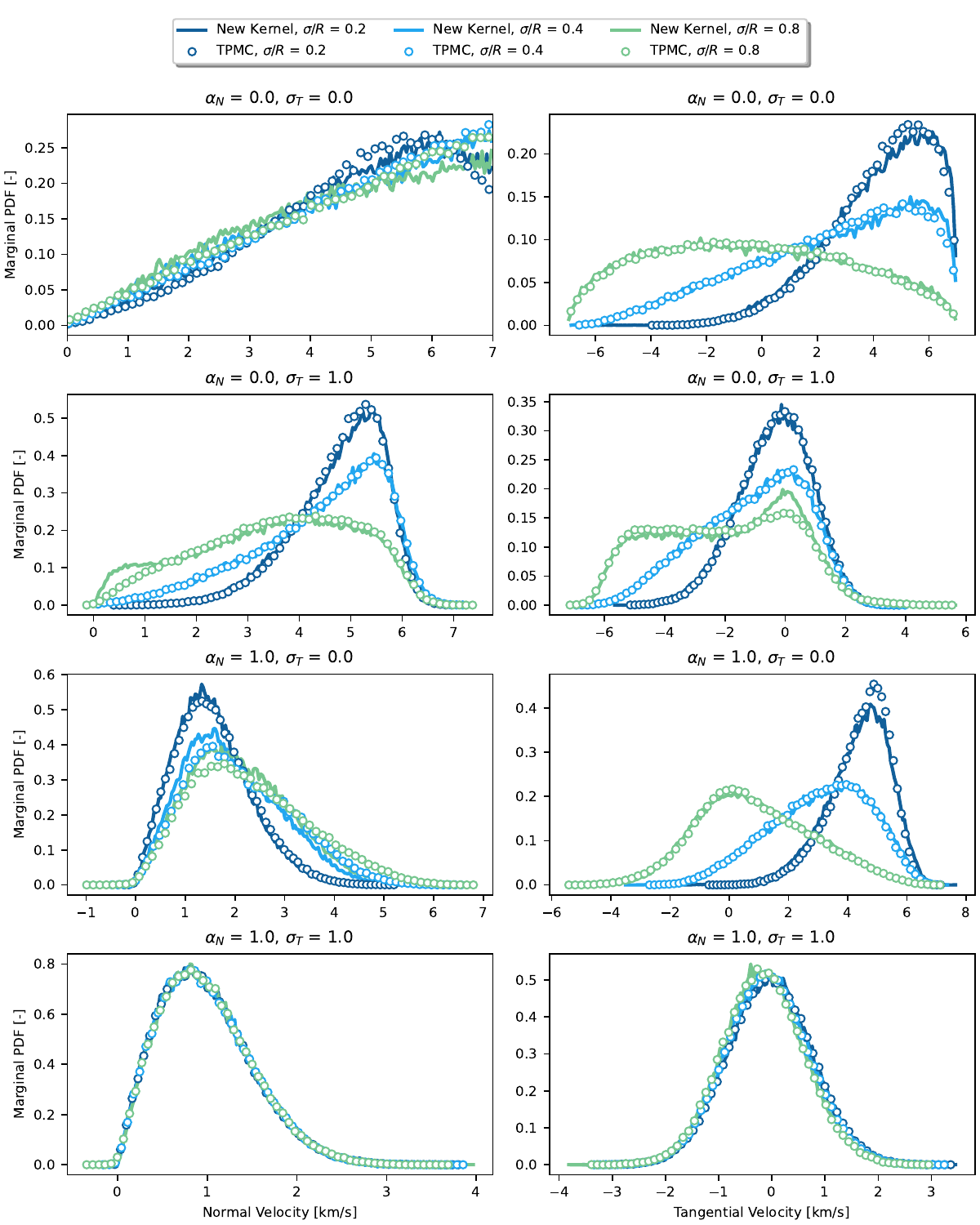}
    \caption{The YZ-plane marginal normal and tangential velocity PDFs of helium gas particles with an incidence angle of $\theta_i = 45^{\circ}$, scattering from Gaussian surfaces with different levels of roughness. The model-generated results are shown in solid lines, while the TPMC simulation results are given in dotted lines.}
    \label{pp_1:fig:gs_velocity_pdfs}
\end{figure}

From a phenomenological perspective, increasing the roughness parameter introduces considerable backscattering across all CLL parameter combinations, which aligns with experimental observations of noble gas scattering from various rough surfaces \citep{Erofeev2012, Erofeev2014-kk, Shoda2022, Ozhgibesov2013, Liao2018, Liu1979}. It is important not to confuse this effect with the "rainbow scattering" discussed by \citet{Livadiotti2020} and observed in experiments such as those by \citet{Pollak2009}, which results from the sinusoidal nature of the "apparent" atomic surface due to the lattice structure of most metals. Another noticeable effect of geometric roughness is the broadening of the quasi-specular scattering lobes across all varied parameters, leading to a more diffuse scattering for rougher surfaces. This effect has been observed in numerous ground-based experiments, such as those by \citet{Comsa1980} and \citet{Sazhin2001}, which recorded global tangential momentum accommodation coefficients close to unity for many metal surfaces, including titanium and silver, under noble gas scattering. In the absence of adsorbate contamination, this behaviour can be attributed to geometric imperfections in the test samples. To better understand both of the aforementioned roughness-induced mechanisms, one should study the statistics of the reflected gas particle velocity vectors per normal/tangential component. For the sake of brevity, such PDFs are provided in \cref{pp_1:fig:gs_velocity_pdfs} only for an incidence angle of $\theta_i = 45^{\circ}$. Analysing the effect of roughness on these PDFs per component reveals a significantly stronger diffusive effect in the tangential direction compared to the normal direction. This is particularly evident in the cases of $ (\alpha_N, \sigma_T) = (0.0, 0.0) $ and $ (\alpha_N, \sigma_T) = (1.0, 0.0) $, where the mean of the tangential velocity PDF shifts sharply to the left and develops a negative velocity tail. Conversely, a much smaller, opposite effect is observed in the normal direction, where for $(\alpha_N, \sigma_T) = (1.0, 0.0)$ the mean of the PDFs shifts toward higher velocity values when the surface is rougher. This is likely due to surface shadowing effects, which focus the reflected particle stream upward, resulting in less "global" normal momentum accommodation. The $ (\alpha_N, \sigma_T) = (0.0, 0.0) $ case is particularly interesting. In this scenario, with no local accommodation, the scattered gas particles retain their incident velocity magnitude. However, as roughness increases, the tangential velocity PDF shifts its mean towards 0, whereas the normal velocity is hardly affected. This means that the reflection changes from specular to diffuse as the roughness increases. This behaviour closely resembles the DRIA model with an accommodation coefficient of \( \alpha = 0.0 \). In the case of complete local accommodation, i.e. $ (\alpha_N, \sigma_T) = (1.0, 1.0) $, the effects of roughness are negligible for both velocity components. Since the local accommodation already results in a diffuse reflection, it cannot get more diffuse due to roughness. Instead, only a minor backscattering effect is observed. If full local accommodation is understood as an already atomically corrugated surface, this invariance to geometric roughness supports \cref{pp_1:eq:auto_length_approx}, which predicts that the larger wavelength components of a surface profile have a much smaller impact on scattering behaviour compared to the smaller wavelength components. If, however, this accommodation is the result of the chemisoprtion processes typically encountered in atomic oxygen scattering, this may also explain why the DRIA model has seen such wide success in modelling satellite aerodynamics at lower altitudes, as portrayed in \citet{Walker2014, Mehta2014, Mehta2014_2}, where atomic oxygen is the prevalent atomic species. 

\begin{figure}[H]
\centering
\hfill
\subfigure[Normal momentum accommodation coefficients.]{\includegraphics[width=.45\linewidth]{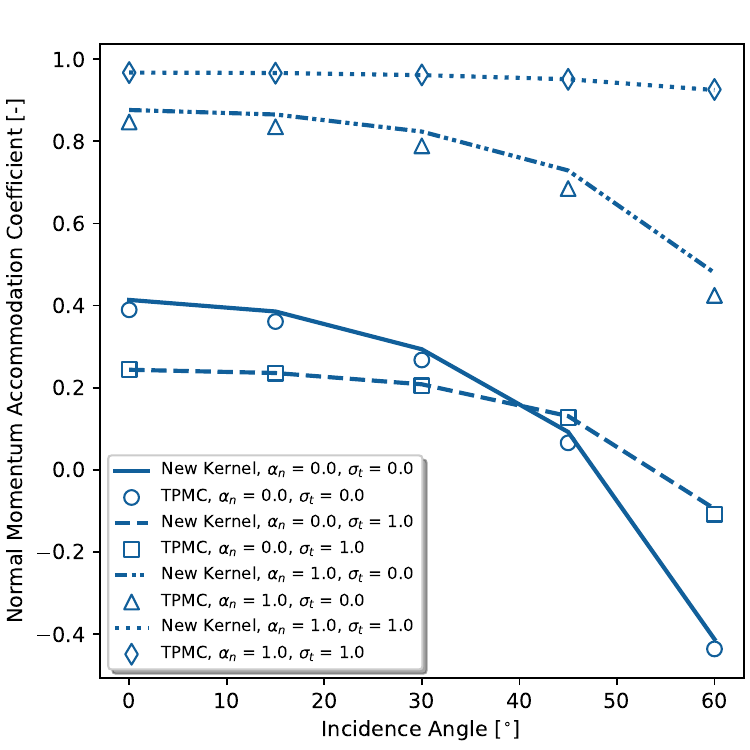}}
\hfill
\subfigure[Tangential momentum accommodation coefficients.]{\includegraphics[width=.45\linewidth]{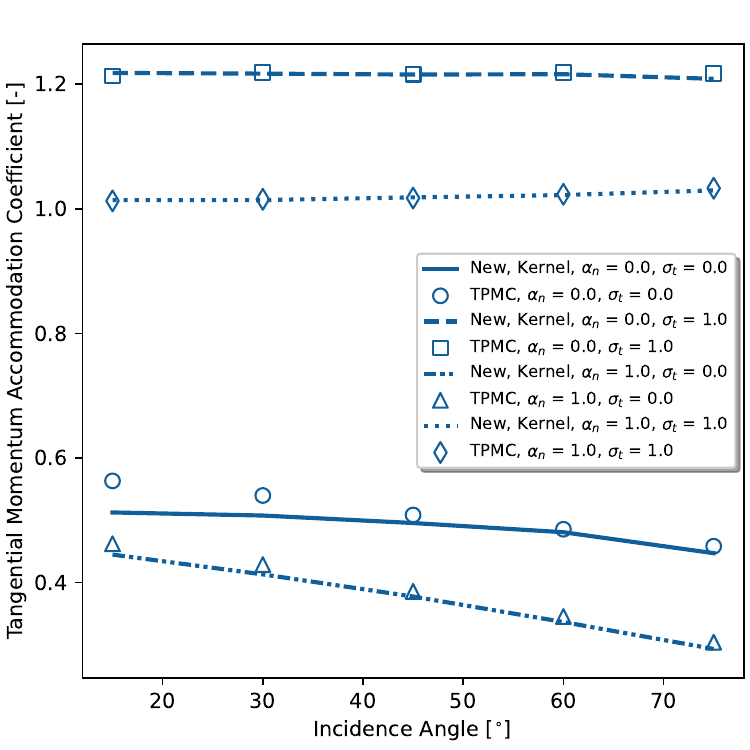}}
\hfill
\caption{The variation of the global normal and tangential momentum accommodation coefficients with respect to the incidence angle for a Gaussian surface with a roughness level of $\sigma / R = 0.4$. The extremes of the local CLL kernel parameter space ($\alpha_N$,$\sigma_T$) are plotted. }
\label{pp_1:fig:gs_acc_variation}
\end{figure}

A more detailed analysis of the global momentum accommodation coefficient behaviour for the Gaussian Kirchhoff model is shown in \cref{pp_1:fig:gs_acc_variation}, where the variation of these coefficients with respect to the incidence angle $\theta_i$ is presented for the extreme values of the local CLL parameters $\alpha_N$ and $\sigma_T$ (cf. \cref{tab:gaussian_analysis}) and a roughness parameter of $\sigma / R = 0.4$. One key observation from these plots is that the global normal momentum accommodation coefficient decreases with increasing incidence angle, regardless of the local CLL parameters. This effect is most pronounced for cases with zero local accommodation $\alpha_N = 0.0$ and $\sigma_T = 0.0$) and is minimized with complete local accommodation $\alpha_N = 1.0$ and $\sigma_T = 1.0$), as expected. Similar trends have been reported in various experimental studies, such as \citet{knechtel1973normal}, who investigated the accommodation behaviour of nitrogen ions on technical-quality satellite surfaces, as well as \citet{cook1994aiaa}, \citet{Cook1997_2}, and \citet{Cook1998}, who conducted similar experiments for Kapton. Most notably, \citet{cook1994aiaa} observed negative momentum accommodation coefficients for a $\text{SiO}_2$ surface, typical for a solar panel, which aligns with the behaviour of the Kirchhoff model shown in \cref{pp_1:fig:gs_acc_variation}. On the other hand, the tangential momentum accommodation coefficient appears to remain relatively constant with varying incidence angle, with slight decreases seen in the $ (\alpha_N, \sigma_T) = (0.0, 0.0) $ and $ (\alpha_N, \sigma_T) = (1.0, 0.0) $ cases. This trend somewhat agrees with the experiments by \citet{knechtel1973normal}, although not enough angles were analysed in that study. Similarly, the results from \citet{cook1994aiaa} match the Kirchhoff model in magnitude, though not in trend, as their study shows a slight increase in the tangential momentum accommodation coefficient with incidence angle. This discrepancy could potentially be explained by differences in roughness values between the experimental surface samples and the model.

\begin{figure}[H]
    \centering
    \includegraphics[width=0.99\linewidth]{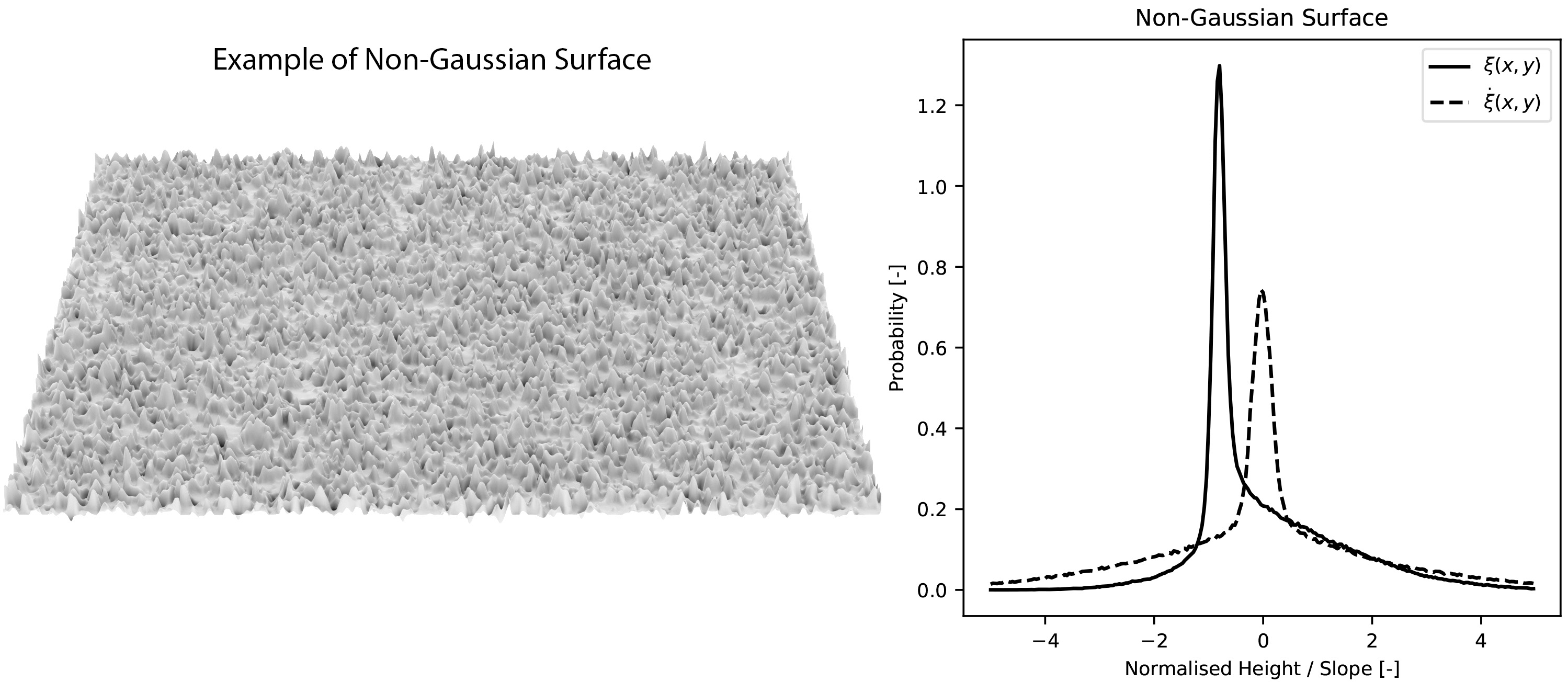}
    \caption{On the left: illustrations of a Non-Gaussian surface sample. On the right: the height ($\xi$) and slope ($\Dot{\xi}$) PDFs of the same surface.}
    \label{pp_1:fig:ngs_surf_pdf}
\end{figure}

The second and final set of verification tests focused on the poly-Gaussian version of the Kirchhoff model, as outlined in \cref{pp_1:eq:ngs_kirchhoff}, \cref{eq:GSI_Rough_Kernel}, and \cref{pp_1:fig:Flowchart}. This set of tests is very similar to those conducted for the Gaussian version, where the local CLL parameters, $\alpha_N$ and $\sigma_T$, and the incidence angle $\theta_i$ are varied within the same ranges. The key difference is that the surface roughness of a non-Gaussian surface is described by the functions $\mu(\gamma)$ and $\sigma(\gamma)$ instead of the roughness parameter $\sigma/R$. These functions have the same meaning as in \cref{sec:Model}, i.e. they define the local mean and variance as a function of the control process $\gamma(x, y)$. They are provided in \cref{tab:polygaussian_analysis} along with the other analysis inputs and chosen display cases.
\begin{table}[H]
\centering
\caption{Input parameters for the poly-Gaussian model analysis.}
\label{tab:polygaussian_analysis}
\begin{tabular}{lrrlrr}
\hline\hline
\textbf{Parameter Name} & \textbf{Value}    & \textbf{Unit}        & \textbf{Parameter Name} & \textbf{Value} & \textbf{Unit} \\ \hline
$\lVert\mathbf{v_i}\rVert$                & 7000                             & [m/s]                               & $\mu(\gamma)$           & $0.8 \cdot \erf(2 \gamma)$                        & [-]                               \\
$\mathcal{M}_{He}$                 & 4                                  & [g/mol]                           & $\sigma(\gamma)$        & $0.1 + 0.8 \cdot \left( 1 + \erf(2\gamma)\right)$ & [-]                               \\
$N_{particles}$         & 100000                             & [-]                               & $\theta_i$              & 15, 45, 75                                        & [$^{\circ}$]                      \\
$N$                     & 40                                 & [-]                               & CLL, $\alpha_N$              & 0.0, 1.0                                          & [-]                               \\
$T_S$                   & 300                                & [K]                                 & CLL, $\sigma_T$              & 0.0, 1.0                                          & [-]                               \\ \hline\hline
\end{tabular}
\end{table}
The $\mu(\gamma)$ and $\sigma(\gamma)$ functions defined in \cref{tab:polygaussian_analysis} were chosen to replicate a typical metallic surface, like polished aluminium. Such a surface exhibits regions of imperfections, as well as smooth areas, in a distinctly non-Gaussian way, as seen in \citet{Mwema2018}. To merge these two regions in a controllable way, the error function $\erf(c \cdot \gamma)$ is employed in both transformations, where $c$ dictates the rate of transition. A render of the generated sample geometry for this surface is shown \cref{pp_1:fig:ngs_surf_pdf} on the left. On the right of the same figure, the height and slope PDFs of this surface are given. Indeed, both of these resemble mixtures of two different Gaussians corresponding to the smooth and rough areas.

\begin{figure}[H]
    \centering
    \includegraphics[width=0.99\linewidth]{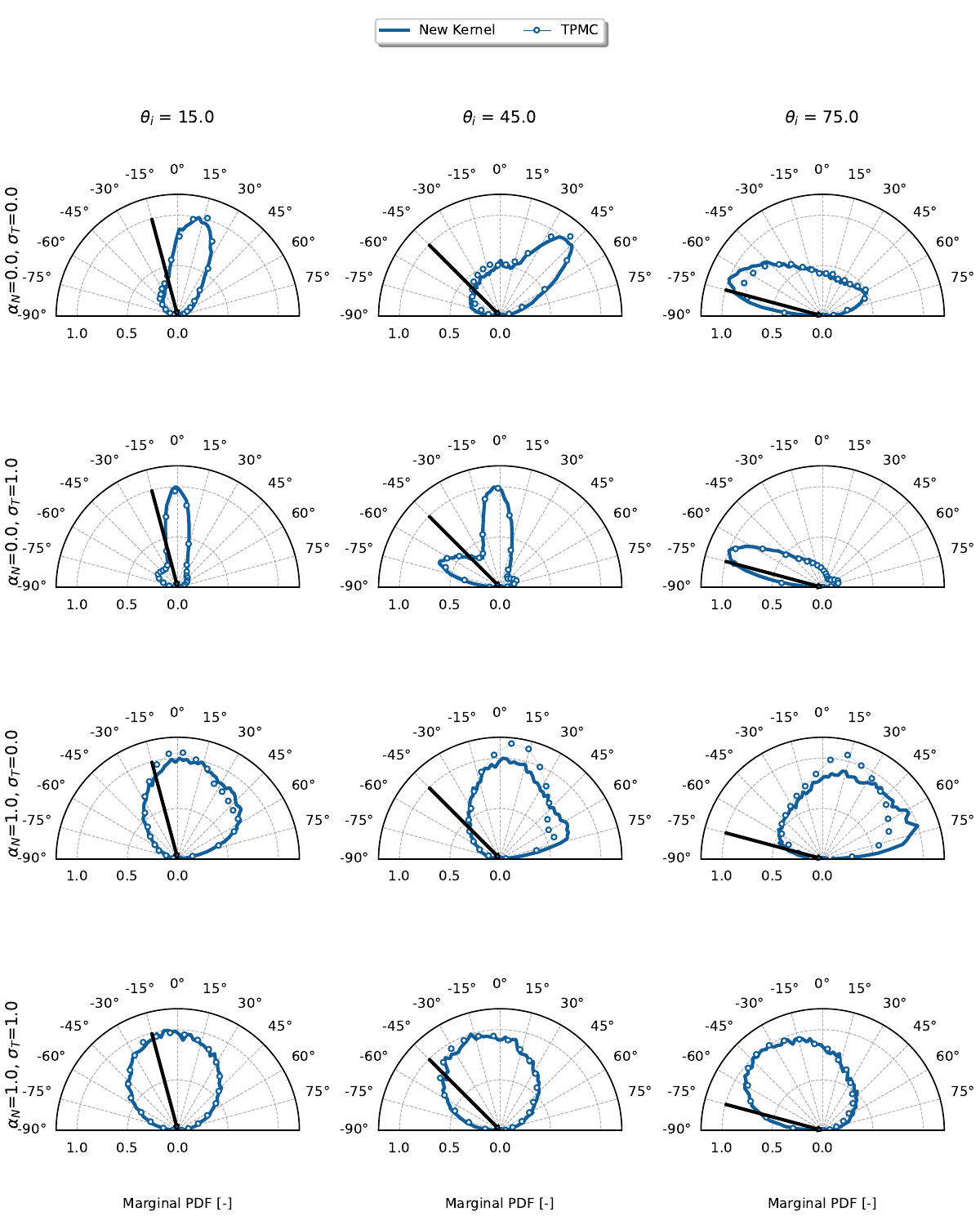}
    \caption{The YZ-plane marginal angular PDFs of helium gas particles with different incidence angles, scattering from a Non-Gaussian surface. The model-generated results are shown as solid lines while the TPMC simulation results are given as dotted lines. The incident direction is depicted with a black arrow.}
    \label{pp_1:fig:ngs_angular_pdfs}
\end{figure}

\begin{figure}[H]
    \centering
    \includegraphics[width=0.99\linewidth]{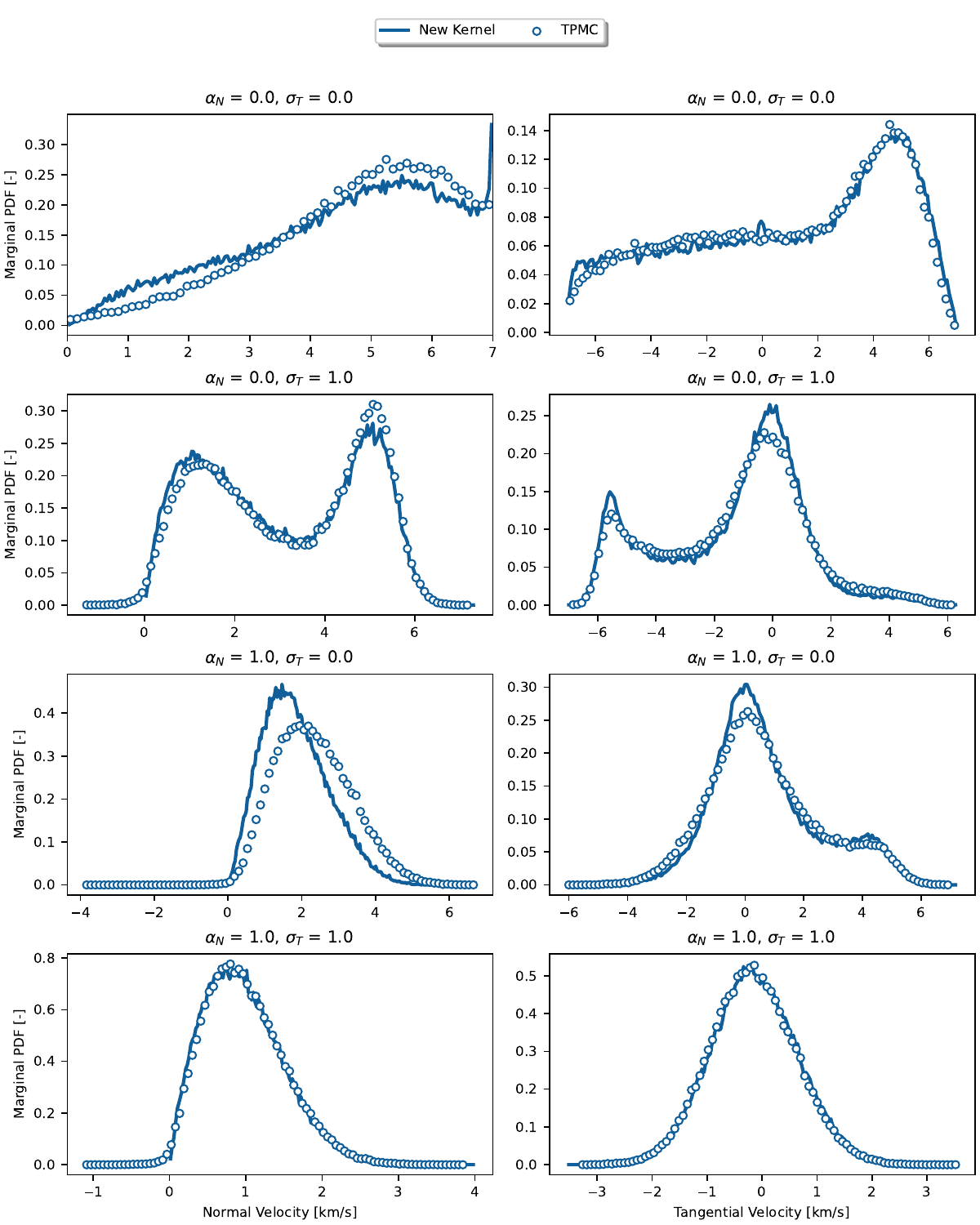}
    \caption{The YZ-plane marginal normal and tangential velocity PDFs of helium gas particles with an incidence angle of $\theta_i = 45^{\circ}$, scattering from a Non-Gaussian surface. The model-generated results are shown in solid lines, while the TPMC simulation results are given in dotted lines.}
    \label{pp_1:fig:ngs_velocity_pdfs}
\end{figure}

\Cref{pp_1:fig:ngs_angular_pdfs} depicts the YZ-plane marginal angular PDFs of the Kirchhoff model and TPMC code. Much like the Gaussian version of the model, the poly-Gaussian version is overall in good agreement with the TPMC simulations. The highest discrepancies are again observed when $\alpha_N = 1.0$ and $\sigma_T = 0.0$, and they appear to increase with incidence angle. This is also visible in the velocity PDFs shown in \cref{pp_1:fig:ngs_velocity_pdfs}, particularly for the global normal momentum accommodation. In this instance, however, the root cause is two-fold. On top of the independence assumption between $(\xi_0, \Dot{\xi}_0)$ and $(\xi, \Dot{\xi})$ in the derivation of the shadowing function $\mathcal{S}(\theta_{r_1}, \xi_0)$, the poly-Gaussian model employs the additional assumption that the height and slope PDFs are globally independent processes (see \cref{subsec:shadowing_multireflection}). Although this was proven mathematically in \cref{subsec:Surface_Modelling} for the case of Gaussian surfaces, it may be a poor approximation for non-Gaussian ones. Indeed, if the variance process $\sigma(\gamma)$ has a low global magnitude compared to the mean process $\mu(\gamma)$, i.e. $\left\langle \sigma(\gamma)^2 \right\rangle \ll \left\langle \mu(\gamma)^2 \right\rangle$ for $\gamma \in \mathbb{R}$, then the statistics of both $\xi$ and $\Dot{\xi}$ are correlated to the value of $\gamma$. 

\begin{figure}[H]
\centering
\hfill
\subfigure[Normal momentum accommodation coefficients.]{\includegraphics[width=.45\linewidth]{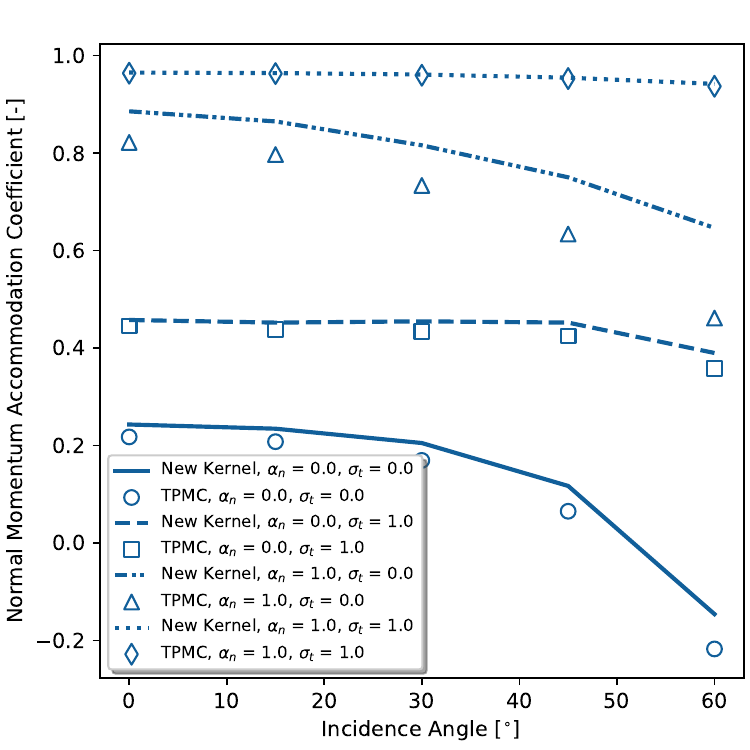}}
\hfill
\subfigure[Tangential momentum accommodation coefficients.]{\includegraphics[width=.45\linewidth]{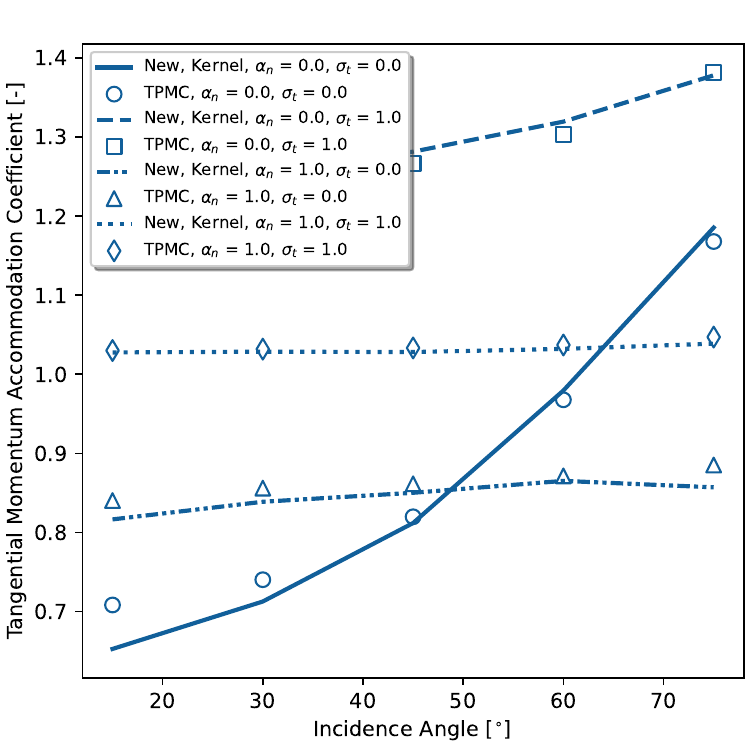}}
\hfill
\caption{The variation of the global normal and tangential momentum accommodation coefficients with respect to the incidence angle, for a Non-Gaussian surface. The extremes of the CLL kernel parameter space $(\alpha_N,\sigma_T)$ are plotted.}
\label{pp_1:fig:ngs_acc_variation}
\end{figure}

The best agreement with the TPMC results is again observed for the $(\alpha_N, \sigma_T) = (0.0, 1.0)$ case. This is because full tangential momentum accommodation for particles with a high incident kinetic energy results in reflections that are almost perpendicular to the local surface plane, hence increasing the distance, on average, with the next collision point. Similar error behaviours between the Kirchhoff model and the TPMC simulations are observed in Figs. \ref{pp_1:fig:ngs_velocity_pdfs} and \ref{pp_1:fig:ngs_acc_variation}. In \cref{pp_1:fig:ngs_velocity_pdfs}, the normal and tangential velocity PDFs for all combinations of local parameters and an incidence angle of $\theta_i = 45^{\circ}$ are shown. Much like for the Gaussian tests, the Kirchhoff model in the case of $(\alpha_N, \sigma_T) = (1.0, 0.0)$ displays noticeably more normal accommodation than the TPMC simulation, while the $(\alpha_N, \sigma_T) = (0.0, 1.0)$ case shows the best agreement between the two methods. This is portrayed even more clearly through the momentum accommodation variation plots in \cref{pp_1:fig:ngs_acc_variation}.

\begin{figure}[H]
    \centering
    \includegraphics[width=0.99\linewidth]{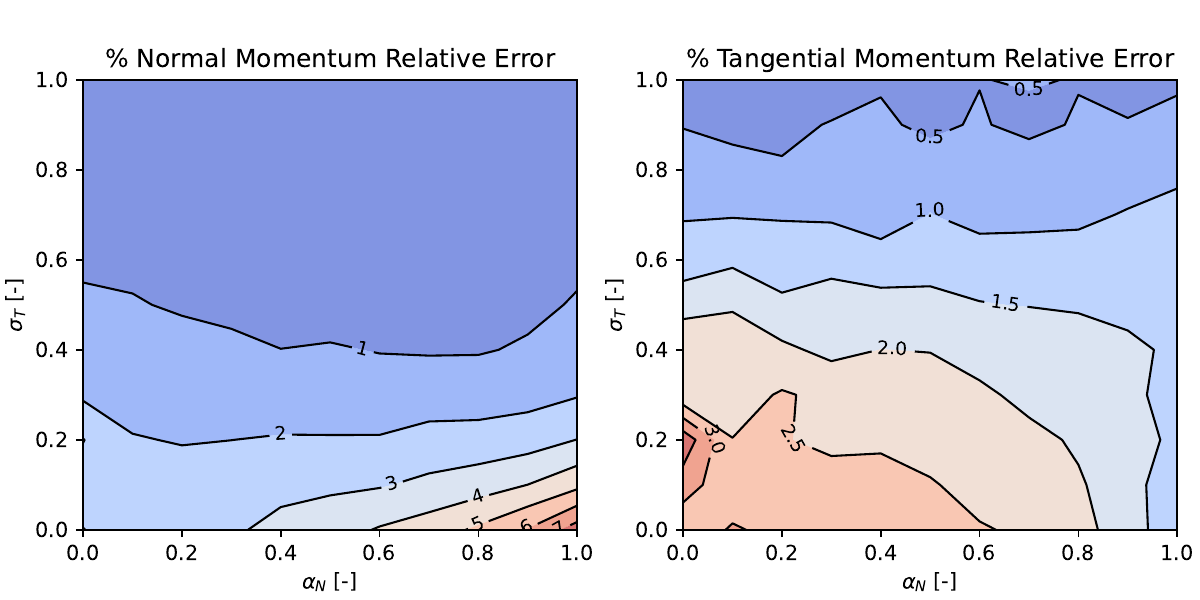}
    \caption{The global momentum relative error between the poly-Gaussian Kirchhoff model and the TPMC simulation results as a function of the local CLL kernel parameters $\alpha_N$ and $\sigma_T$, for a given non-Gaussian surface and an incidence angle of $\theta_i = 45^{\circ}$.}
    \label{pp_1:fig:ngs_error}
\end{figure}

To study this error behaviour in more detail, a relative error map between Kirchhoff model and the TPMC simulations has been generated and is illustrated in \cref{pp_1:fig:ngs_error} for all possible CLL parameters and a fixed incidence angle of $\theta_i = 45^{\circ}$. The poly-Gaussian Kirchhoff formulation with the example non-Gaussian surface given in \cref{tab:polygaussian_analysis} has been chosen for this analysis, as it appears to be a "worst case", exhibiting the highest disagreement out of all the rough surfaces analysed in this study. The global normal momentum plot on the left exhibits the expected rise in relative error for lower $\sigma_T$ values, with an error maximum of 7\% at the corner of the parameter space $(\alpha_N, \sigma_T) = (1.0, 0.0)$. The tangential momentum error, on the other hand, exhibits its highest error for $(\alpha_N, \sigma_T) = (0.0, 0.0)$, of around 3\%. Assuming the TPMC simulations are "the ground truth", the error plots suggest that the Kirchhoff model is very accurate for very rough surfaces under the condition that $\sigma_T > 0.5$, as a relative error below 1\% is expected in the aerodynamic coefficients in this region. Thus, provided that the assumptions in \cref{subsec:ValidityRange} are valid and that CLL accurately describes local interactions, if the in-orbit roughness of a surface is known and $\sigma_T > 0.5$, the Kirchhoff model is expected to provide far more accurate aerodynamic coefficients than empirically tuned DRIA and CLL models, which have an accuracy ranging from several percent to a few tens of percent according to \citet{Bernstein2022}, \citet{Mehta2023}, and \citet{ SIEMES2024}. If, however, the in-orbit roughness is unknown or $\sigma_T < 0.5$, the Kirchhoff model's parameters could still be empirically tuned based on in-orbit data to increase the model's accuracy. It should be noted that the region of model validity within the CLL parameter space $(\alpha_N,\sigma_T)$ is expected to increase for less rough surfaces. This is analysed in Appendix~\ref{ap:Ad_results}.

Phenomenologically, the poly-Gaussian surface's scattering lobes in \cref{pp_1:fig:ngs_angular_pdfs} display much more intricate behaviour compared to their Gaussian counterparts. For most combinations of local parameters, such as $(\alpha_N, \sigma_T) = (0.0, 0.0)$ and $(\alpha_N, \sigma_T) = (0.0, 1.0)$, the scattering lobe appears as a superposition of two quasi-specular lobes with different peaks. This can be attributed to the nature of the example poly-Gaussian surface, which, as previously mentioned, features smooth regions at a positive local mean height that abruptly transition to smooth regions at a different, negative local mean height. This relief creates distinct areas with either very small or very large slopes — in effect, a superposition of smooth and rough surfaces. 

In the specific case of $(\alpha_N, \sigma_T) = (0.0, 0.0)$ at incidence angles of $\theta_i = 15^{\circ}$ and  $\theta_i = 45^{\circ}$, the angular scattering PDFs show a diffuse component superimposed on a quasi-specular component. This behaviour aligns closely with nanoscale roughness theory from \citet{Chen2024} and \citet{Barker1984} and mirrors the scattering lobes experimentally observed by \citet{Erofeev2012} and \citet{Shoda2022}. The implication is that the attempts at modelling roughness mentioned in \cref{subsec:Roughness_GSI} indirectly assume the specific kind of surface described above, but do not take into account the effects of shadowing and multiple reflections, i.e. the interference between the smooth and rough patches of surface. This latter effect is already obvious from the case with local specular reflection, i.e. $\alpha_N = 0.0$ and $\sigma_T = 0.0$, where the smooth patches of the surface are occluded for $\theta_i = 75^{\circ}$, and most particles only "see" the high-slope regions, resulting in a high degree of backscattering.

Where the current model diverges from the others available in literature is for the $(\alpha_N, \sigma_T) = (1.0, 0.0)$ case. Indeed, the increased likelihood of multiple reflections induced by this local parameter combination results in a diffuse peak with a preferential direction towards $\theta = 0^{\circ}$. Looking in \cref{pp_1:fig:ngs_velocity_pdfs} at the normal and tangential velocity PDFs for $\theta_i = 45^{\circ}$, one can observe a more pronounced roughness effect compared to the same PDFs for a Gaussian surface in \cref{pp_1:fig:gs_velocity_pdfs}. A clear distinction between the forward scattering and backscattering peaks in the tangential velocity component is particularly noticeable in the $(\alpha_N, \sigma_T) = (0.0, 1.0)$ case. This separation is less evident in the normal direction, but even here, the accommodation is higher compared to the roughest surface in the Gaussian verification tests. 
When examining the momentum accommodation behaviour in both the normal and tangential directions for the poly-Gaussian case in \cref{pp_1:fig:ngs_acc_variation}, the tangential behaviour is especially distinct compared to the Gaussian model in \cref{pp_1:fig:gs_acc_variation}. First, the poly-Gaussian case shows a consistently higher level of tangential accommodation across all tested local parameter combinations. This is consistent with the experimental findings of \citet{Liu1979}, who observed $ \sigma_T \approx 1$ for all incidence angles. Second, unlike the Gaussian case, the global tangential accommodation coefficient increases with the incidence angle \( \theta_i \). This trend is particularly noticeable in the $(\alpha_N, \sigma_T) = (0.0, 0.0)$ and $(\alpha_N, \sigma_T) = (1.0, 0.0)$ cases and aligns closely with the experimental results of \citet{knechtel1973normal} and \citet{cook1994aiaa}. This suggests that the surfaces tested in those experiments likely resembled a poly-Gaussian surface similar to the one analysed here, i.e. a combination of smooth regions and defect-filled areas, rather than a purely Gaussian surface.

\subsection{Consistency Analysis with Experimental Results} \label{subsec:Experimental}

\begin{figure}[H]
    \centering
    \includegraphics[width=0.99\linewidth]{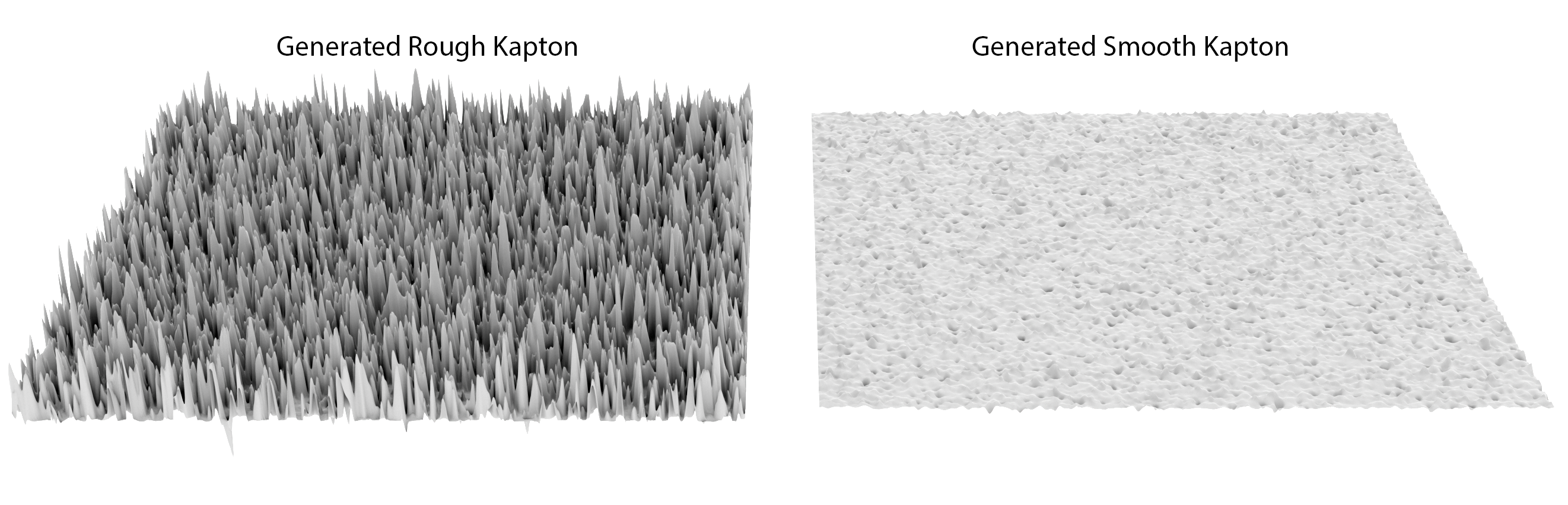}
    \caption{Illustrations of non-Gaussian surfaces resembling atomic oxygen-eroded Kapton (left) and pristine Kapton (right), both generated with the Kirchhoff model.}
    \label{pp_1:fig:exp_kapton_render}
\end{figure}

\begin{figure}[H]
    \centering
    \includegraphics[width=0.99\linewidth]{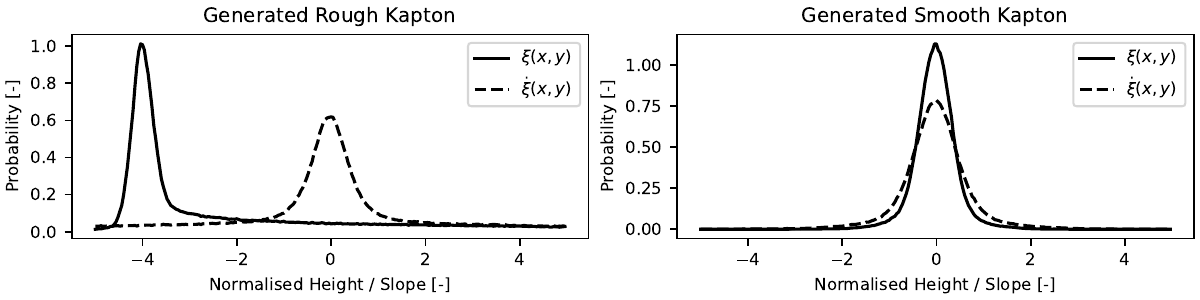}
    \caption{The height ($\xi$) and slope ($\Dot{\xi}$) PDFs of non-Gaussian surfaces resembling atomic oxygen-eroded Kapton (left) and pristine Kapton (right), both generated with the Kirchhoff model.}
    \label{pp_1:fig:exp_surface_pdfs}
\end{figure}

\begin{figure}[H]
    \centering
    \includegraphics[width=0.99\linewidth]{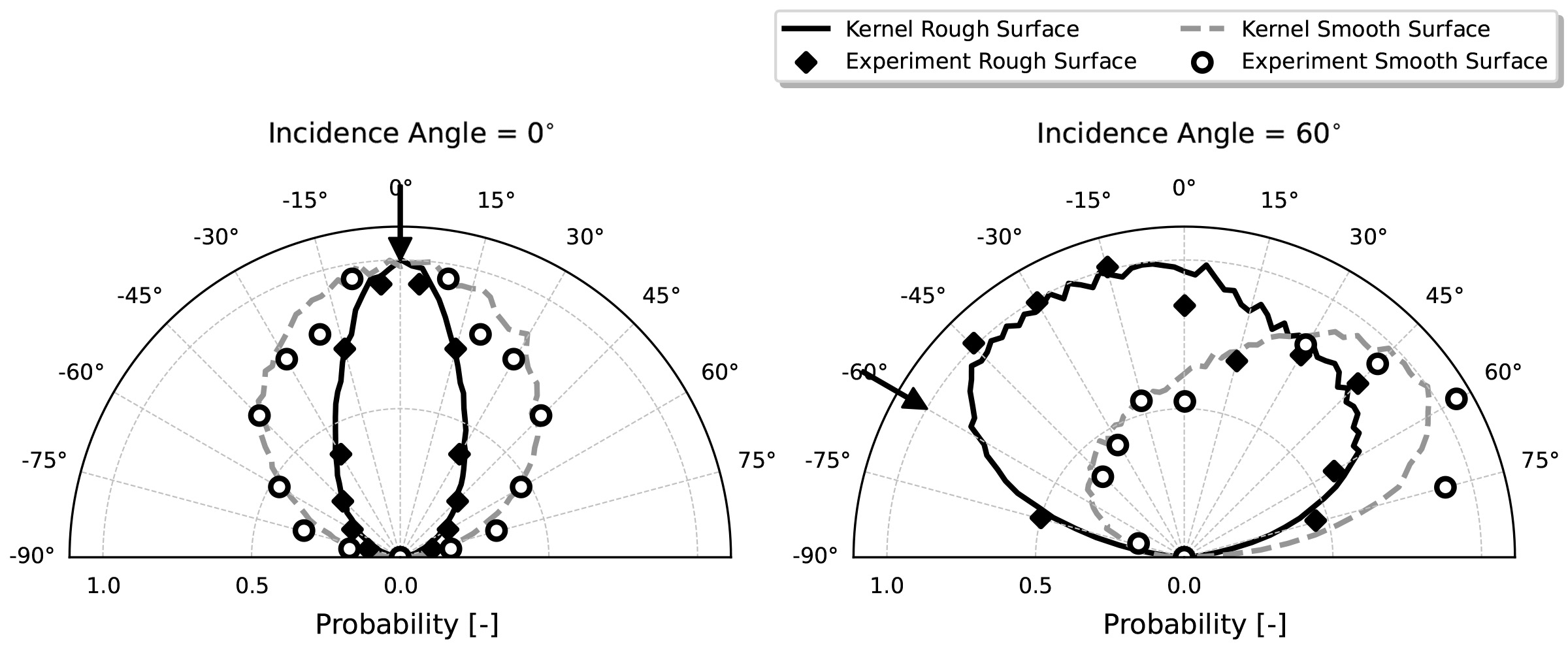}
    \caption{A comparison between the argon-Kapton scattering results from \citet{Erofeev2012} and the poly-Gaussian Kirchhoff model applied with the two previously-generated surfaces, for incidence angles of $\theta_i = 0^{\circ}$ and $\theta_i = 60^{\circ}$. The experimental results are shown as scatter plots while the model results are shown as continuous lines. The incident direction is denoted with a black arrow.}
    \label{pp_1:fig:exp_angular_plots}
\end{figure}

To further validate the proposed Kirchhoff model's ability to capture roughness effects, it was used to replicate a series of experimental results presented by \citet{Erofeev2012}, who examined gas scattering from rough surfaces. More specifically, they present two sets experiments: one analysing the angular scattering PDFs of argon atoms from a smooth Kapton surface and another analysing the same PDFs from a Kapton surface subjected to 2.5 hours of atomic oxygen erosion, both performed at incidence angles of $\theta_i = 0^{\circ}$ and $\theta_i = 60^{\circ}$. The detector was placed in the YZ plane as described by \cref{pp_1:fig:angle_definitions} at angle steps of $\Delta \theta_{r_1} = 10^{\circ}$ while the out-of-plane angle was kept at $\theta_{r_2} = 0^{\circ}$. Kapton is commonly found on the outer surfaces of satellites and can develop significant roughness after extended exposure to the atomic oxygen-rich environment of the thermosphere \citep{Banks2004}, which makes it a prime candidate for a validity analysis of the model proposed in this paper. The underlying procedure of this analysis was to fit the GSI parameters of the Kirchhoff model to the experimentally obtained smooth surfaces' angular PDFs from \citet{Erofeev2012} through a trial and error process, and then recover the oxidised Kapton PDFs by only altering the $\mu_k$ and $\sigma_k$ parameters, which control the roughness of the surface. The CLL kernel as given in \citet{Lord1995} was again used to describe the local scattering dynamics in an empirical way, due to its versatility in replicating many different quasi-specular behaviours. On top of this, a certain fraction of gas particles $f_{physisorption}$ is assumed to undergo physisorption locally, based on the findings of \citet{Chen2013, Chen2023}. This was modelled at a local level as in \citep{Bernstein2022}, by linearly superimposing two CLL kernels: one with quasi-specular parameters, and one with fully diffuse parameters. A full list of the input parameters for the conducted simulations is given in \cref{tab:Kapton_analysis}.
\begin{table}[H]
\centering
\caption{Poly-Gaussian input parameters for the Kapton experiment comparison.}
\label{tab:Kapton_analysis}
\begin{tabular}{lrrlrr}
\hline\hline
\textbf{Parameter Name} & \textbf{Value}    & \textbf{Unit}        & \textbf{Parameter Name} & \textbf{Value} & \textbf{Unit} \\ \hline
$\lVert\mathbf{v_i}\rVert$                   & 2500                               & [m/s]                             & Smooth $\mu(\gamma)$    & $0.0$                                                  & [-]                               \\
$\mathcal{M}_{Ar}$                 & 39.84                              & [g/mol]                           & Smooth $\sigma(\gamma)$ & $0.3 + 0.3 ( 1 + \erf(2\gamma - 2))$ & [-]                               \\
$N_{particles}$         & 1000000                            & [-]                               & Rough $\mu(\gamma)$     & $4 \erf(2 \gamma - 1)$                           & [-]                               \\
$N$                     & 40                                 & [-]                               & Rough $\sigma(\gamma)$  & $0.2 + 0.3 (1 + \erf(2 \gamma - 1))$             & [-]                               \\
$T_S$                   & 300                                & [K]                                 & $\theta_i$              & 0, 60                                                  & [$^{\circ}$]                      \\
$f_{physisorption}$     & 0.45                               & [-]                               & CLL, $\alpha_N$, $\sigma_T$  & 0.6, 0.2                                               & [-]                               \\ \hline\hline
\end{tabular}
\end{table}
In this table, the incident velocity magnitude $\vert\mathbf{v_i}\vert$, the gas molar mass $\mathcal{M}_G$, the surface temperature $T_S$, and the incidence angles $\theta_i$ are taken from \citet{Erofeev2012}. The local CLL parameters $\alpha_N$, $\sigma_T$ and the physisorption fraction  $f_{physisorption}$ are found through the aforementioned trial and error process. The poly-Gaussian transformations $\mu(\gamma)$ and $\sigma(\gamma)$ for the smooth and rough Kapton surfaces were chosen to visually replicate electron microscope pictures from \citet{Erofeev2012} and \citet{Banks2004}. These functions were parametrised using Hermite polynomial expansions with $N=40$ coefficients $\mu_k$ and $\sigma_k$. \Cref{pp_1:fig:exp_kapton_render} shows renders of generated surface geometries representing eroded (left) and smooth (right) Kapton, using these coefficients. Furthermore, in \cref{pp_1:fig:exp_surface_pdfs}, the corresponding height and slope PDFs for both of these surfaces are plotted. Finally, \cref{pp_1:fig:exp_angular_plots} shows recorded and modelled YZ plane cross-sections of the angular scattering PDFs for all experiments. The reader should not confuse the model-generated curves with the YZ plane marginal PDFs in \cref{subsec:Verification}. Instead of being integrated over the full $\theta_{r_2}$ range of $(0^{\circ}, 360^{\circ})$, these plots are generated by integrating only over an assumed detector angular width of $\Delta \theta_{r_2} = \pm 1^{\circ}$. 

It is clear from the aforementioned height and slope PDFs on the left of \cref{pp_1:fig:exp_surface_pdfs} that the rough, carpet-like Kapton surface exhibits a distinctly non-Gaussian probability distribution for both the height and slope profiles. Largely driven by the $\mu(\gamma)$ transformation, the surface consists of a smooth plane component centred around $\xi/R = -4$, along with a region of "spikes" visible in the height PDF through the large tail extending from $\xi/R = -4$ to $\xi/R = 4$. As a result, the expected scattering behaviour of this surface at an incidence angle of $\theta_i = 0^{\circ}$ features a quasi-specular lobe directed upwards, caused by the smooth regions, superimposed on a diffuse lobe biased toward $\theta_{r_1} = 0^{\circ}$, due to the pronounced shadowing effects from the spikes. This is precisely what is observed in \cref{pp_1:fig:exp_angular_plots} (left), both in the experimental data and in the model’s predictions. For the smooth Kapton surface, where the spikes are absent, the diffuse lobe is primarily caused by minor rough imperfections, the high CLL normal energy accommodation coefficient of $\alpha_N = 0.6$, and gas particle physisorption.

When the incidence angle is increased to $\theta_i = 60^{\circ}$, the shadowing effects dominate the scattering behaviour, as they fully occlude the smooth areas in the rough Kapton surface. Hence, a large degree of backscattering is expected as a result of the sides of the spikes being exposed to the incident gas flow. This is, again, exactly what is observed in \cref{pp_1:fig:exp_angular_plots} (right) in both the model PDFs and the experimental scatter plot. In contrast, the smooth Kapton surface exhibits very little shadowing and, due to the low level of tangential accommodation of $\sigma_T = 0.2$, develops a significant quasi-specular lobe in the specular direction. The small imperfections in the surface, together with the rate of physisorption, contribute to the small diffuse lobe, superimposed over the quasi-specular one.

\subsection{Model Application to Simple Shapes} \label{subsec:ModelApplication}

The final analysis conducted in this study revolves around the applicability and accuracy of the proposed Kirchhoff model in computing aerodynamic coefficients. For this purpose, two simple shapes were considered: a flat plate and a sphere. The flat plate was chosen as a study case because it constitutes the building block in calculating the aerodynamic behaviour of any convex satellite shape in free-molecular flow conditions, due to the linearity of the problem at hand. The sphere, on the other hand, was chosen due to the abundance of fitted aerodynamic data available in literature for spherical satellites at high altitudes \citep[e.g.,][]{Pardini2006, Pardini2010,Pilinski2013}. This allows for further validation of the model in the main atmospheric region of interest for this study, i.e. altitudes above 400 km. Besides this, a sphere is invariant to its attitude, assuming uniform surface properties, which greatly simplifies calculations. All investigations in this section have been conducted using the CLL kernel as given by \cite{Lord1995} to describe the local GSI. 

The investigation of the flat plate's aerodynamic behaviour consists of two analyses where different parameter sets are varied: the angle of attack (AOA), which coincides with the incidence angle $\theta_i$ for this shape, and the local kernel parameters $\alpha_N$ and $\sigma_T$. Both analyses are restricted to the Gaussian formulation of the Kirchhoff model for the sake of simplicity. \cref{tab:flat_plate_aoa} gives an overview of all the simulation inputs required for the first analysis. Local CLL parameters $(\alpha_N, \sigma_T) = (0.6, 0.2)$ were chosen because they are realistic for a typical satellite surface according to \citet{Goodman1965} as well as the analysis conducted in \cref{subsec:Experimental}. Roughness parameters up to $\sigma/R = 2.0$ were considered, as we expect most satellites to have a roughness lower than that. Finally, an incident velocity magnitude $v_i = \lVert\mathbf{v_i}\rVert$ = 7000 m/s and a surface temperature of 400 K were chosen to replicate the thermosphere environment, while the atmospheric temperature $T_{\infty}$ was set to a lower value of 200 K to obtain a high-altitude speed ratio value of $s = 15$, calculated by
\begin{equation}
    s = \frac{v_i}{\sqrt{\frac{2 \mathcal{R} T_{\infty}}{\mathcal{M}_{AO}}}},
\end{equation}
where $\mathcal{R}$ is the gas constant. In this analysis, the molar mass of atomic oxygen, $\mathcal{M}_{AO}$ is considered since this species still represents an important part of the atmospheric composition at altitudes where the Swarm, GRACE and GRACE-FO satellites are found. 
\begin{table}[H]
\centering
\caption{Input parameters for the flat plate angle of attack (AOA) analysis.}
\label{tab:flat_plate_aoa}
\begin{tabular}{lrrlrr}
\hline\hline
\textbf{Parameter Name} & \textbf{Value}    & \textbf{Unit}        & \textbf{Parameter Name} & \textbf{Value} & \textbf{Unit} \\ \hline
$\lVert\mathbf{v_i}\rVert$            & 7000                               & [m/s]                             & $\sigma/R$              & \begin{tabular}[c]{@{}r@{}}0.0, 0.2, 0.4, 0.6, \\ 0.8, 1.0, 2.0\end{tabular} & [-]                               \\
$\mathcal{M}_{AO}$                 & 15.999                             & [g/mol]                           & AOA range            & [-90, 90]                                                                    & [$^{\circ}$]                      \\
$N_{particles}$         & 100000                             & [-]                               & AOA step             & 5                                                                            & [$^{\circ}$]                      \\
$T_S$                   & 400                                & K                                 & DRIA $\alpha$           & 0.85                                                                         & [-]                               \\
$T_{\infty}$            & 200                                & K                                 & CLL, $\alpha_N$              & 0.6                                                                          & [-]                               \\
$S_{ref}$               & 1.0                                  & \SI{}{\meter\squared}          & CLL, $\sigma_T$              & 0.2                                                                          & [-]                               \\ \hline\hline
\end{tabular}
\end{table}

\begin{figure}[H]
    \centering
    \includegraphics[width=0.97\linewidth]{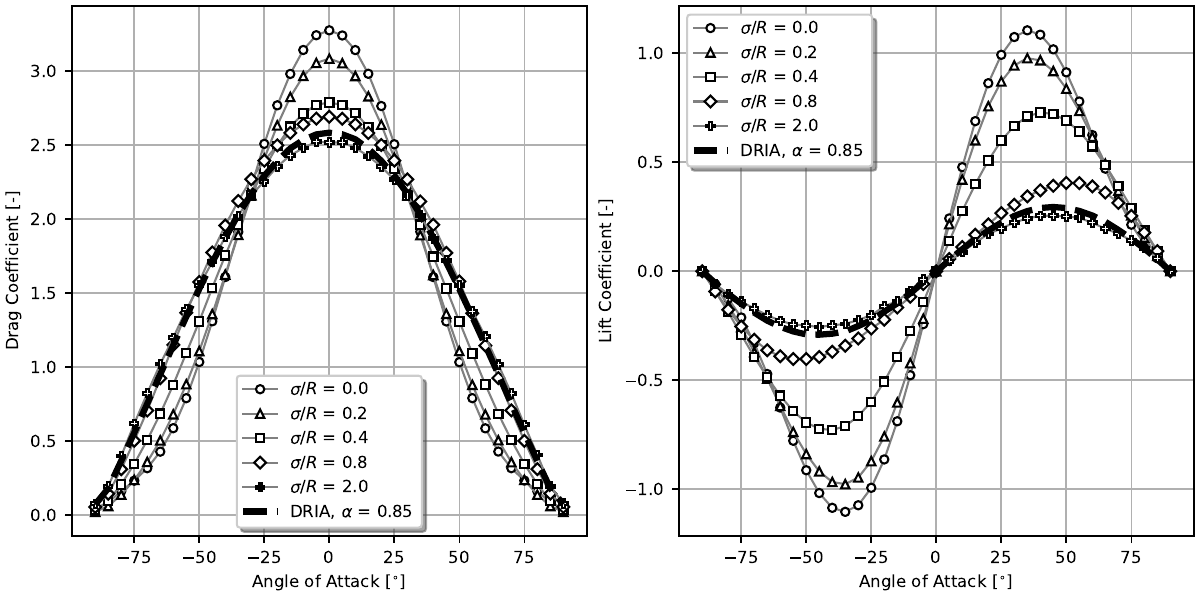}
    \caption{The variation of the drag (left) and lift (right) coefficients of a flat plate with respect to the angle of attack for different levels of Gaussian roughness levels. For comparison, the same variation produced with the DRIA model and an accommodation coefficient of $\alpha = 0.85$ (thick dashed line).}
    \label{pp_1:fig:cd_aoa_fp}
\end{figure}

The resulting variation in the lift and drag coefficient with respect to the AOA for the flat plate are given in \cref{pp_1:fig:cd_aoa_fp}. Similar curves are plotted for the DRIA model of \citet{Sentman1961FREEMF}, with an energy accommodation coefficient of $\alpha = 0.85$, for the sake of comparison. The value of this coefficient was chosen based on the conclusions of \citet{March2021}, who found this as the best-fit value for the aerodynamic modelling of the Swarm satellites, with the self-consistency of neutral density datasets as a metric. Studying the plots, it is clear that the Kirchhoff model is approaching DRIA as the roughness parameter $\sigma/R$ is increased. This indicates that the kernel proposed by \citet{Sentman1961FREEMF} empirically captures the effects of geometric roughness and may provide a physical explanation for the wide success it has achieved in satellite aerodynamic modelling. Of course, these conclusions may only be drawn under the assumption that most satellite surfaces in LEO exhibit a noticeable level of geometric roughness. This is further studied in the next analysis focusing on spherical satellites. Looking at the drag coefficient plot in particular (on the left), roughness appears to produce a "flattening" of the CLL kernel features, i.e. a decrease in drag for low incidence angles, and also a noticeable increase for the high incidence ones. It it likely that for even higher levels of roughness than those plotted, these features will get stronger than what is observed in the DRIA dashed curve. A similar effect is observed for the lift coefficient (on the right) , where an overall decrease of the lift coefficient is observed. These changes in aerodynamics directly correlate to the GSI phenomena described in \cref{subsec:Verification} and \cref{subsec:Experimental}, i.e. a widening of the quasi-specular lobes and backscattering at near-parallel incidence angles. 

The second flat plate analysis investigates the behaviour of the drag coefficient as the local kernel parameters $\alpha_N$ and $\sigma_T$ are varied at different levels of roughness. The simulation inputs for this are given in \cref{tab:flat_plate_acc}. Most simulation inputs are the same as in the previous analysis with a few exceptions. The angles of attack $0^{\circ}$, $30^{\circ}$, and $60^{\circ}$ were chosen to provide a comprehensive picture of $C_D$ changes for most satellite surface orientations. It was also chosen to fix one CLL parameter while the other one is varied, for the sake of clarity and because no deviating behaviours were observed in a similar analysis where both were varied. It should be noted that the tangential coefficient $\sigma_T$ is fixed to a value of 0, whereas the normal coefficient $\alpha_N$ is fixed to a value of 1. This was done in the interest of physical accuracy, as it is unphysical for a gas particle to accommodate its tangential momentum more than its normal momentum through thermal mechanisms alone \citep{Goodman1965, Logan1966, Logan1968, Rettner1991}. 

\begin{table}[H]
\centering
\caption{Input parameters for the flat plate local accommodation coefficient analysis.}
\label{tab:flat_plate_acc}
\begin{tabular}{lrrlrr}
\hline\hline
\textbf{Parameter Name} & \textbf{Value}    & \textbf{Unit}        & \textbf{Parameter Name} & \textbf{Value} & \textbf{Unit} \\ \hline
$\lVert\mathbf{v_i}\rVert$            & 7000                               & [m/s]                             & $\sigma/R$       & 0.0, 1.0, 2.0                      & [-]                               \\
$\mathcal{M}_{AO}$                 & 15.999                             & [g/mol]                           & AOA           & 0, 30, 60                          & [$^{\circ}$]                      \\
$N_{particles}$         & 100000                             & [-]                               & $\alpha_N$ range        & [0.0, 1.0]                         & [-]                               \\
$T_S$                   & 400                                & [K]                                 & $\sigma_T$ range        & [0.0, 1.0]                         & [-]                               \\
$T_{\infty}$            & 200                                & [K]                                 & $\alpha_N$ step         & 0.05                               & [-]                               \\
$S_{ref}$               & 1.0                                & [\SI{}{\meter\squared}]             & $\sigma_T$ step         & 0.05                               & [-]                               \\ \hline\hline
\end{tabular}
\end{table}

\begin{figure}[H]
    \centering
    \includegraphics[width=0.8\linewidth]{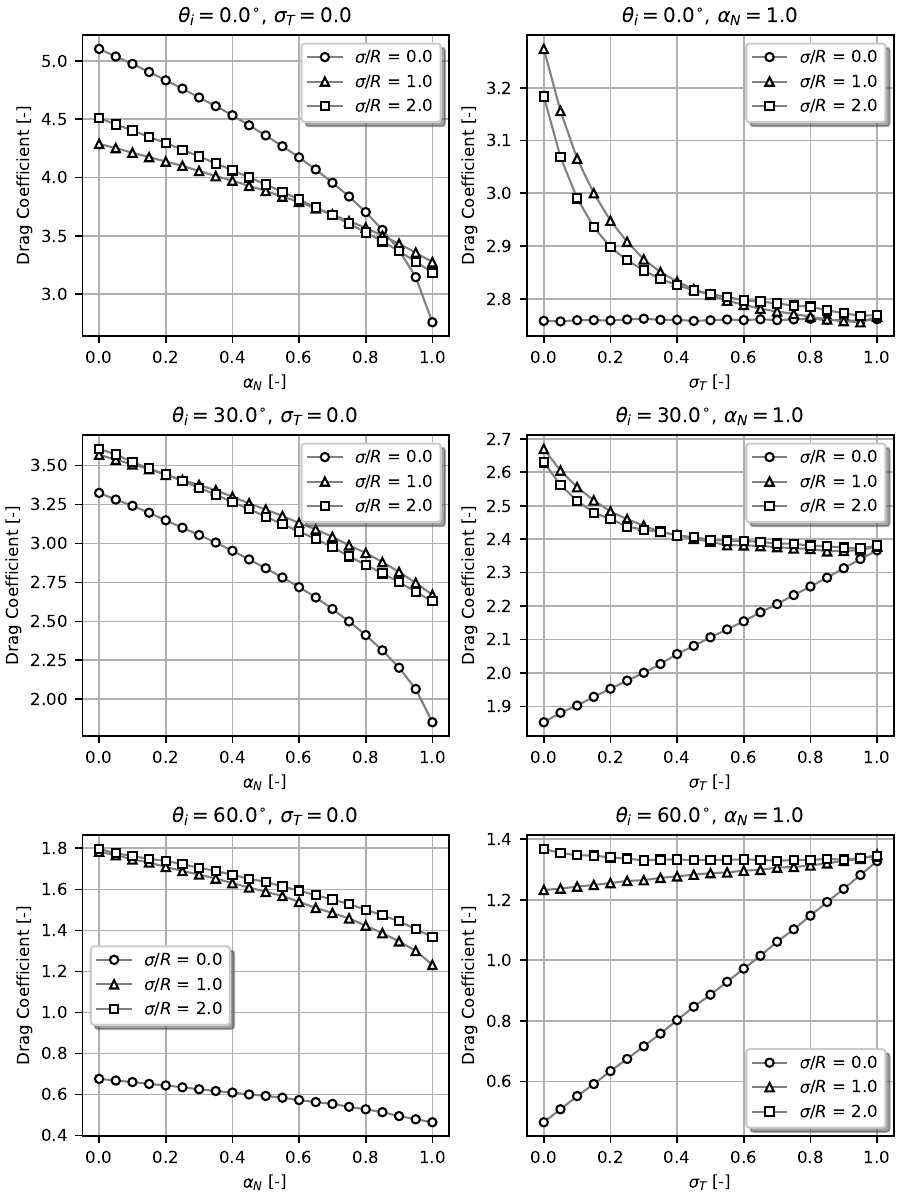}
    \caption{The variation of the drag coefficient of a flat plate with respect to the CLL parameters $\alpha_N$ and $\sigma_T$, for different levels of Gaussian roughness and incidence angles.}
    \label{pp_1:fig:cd_cll_fp}
\end{figure}

The resulting variations of $C_D$ with respect to $\alpha_N$ and $\sigma_T$ are shown in \cref{pp_1:fig:cd_cll_fp}. As a first observation about the variation with respect to $\alpha_N$ shown on the left, the drag coefficient appears to "lose" sensitivity to the local normal energy accommodation coefficient as the roughness parameter is increased, for all incidence angles. This is expected based on the multiple collisions a gas particle is likely to undertake as this parameter is raised. As each collision accommodates the momentum of the particle more, this has an additive effect that becomes more dependent on the surface roughness than the local scattering laws. Another important observation is that the drag coefficient appears to increase with roughness for near-parallel angles of attack. This is, of course, due to the backscattering phenomenon that was observed in previous analyses for high incidence angles.

In the variations of $C_D$ with respect to $\sigma_T$ shown on the right, an opposite trend emerges between the smooth surfaces and those with a non-zero roughness parameter. The smooth surfaces show a linear increase in $C_D$ as $\sigma_T$ increases, while the rougher surfaces start with a high $C_D$ and quickly converge to the same value as the smooth cases at $\sigma_T = 1.0$. This can, again, be attributed to backscattering, which affects the reflected gas particle directions in a similar way to tangential accommodation but without altering their velocity magnitudes. This results in a substantial increase in the drag coefficient, as more momentum is transferred to the surface, especially when $\sigma_T = 0.0$. As $\sigma_T$ increases, particles thermally accommodate to the surface more rapidly and reflect with an overall lower velocity magnitude, leading to a reduction in drag. Furthermore, as the incidence angle increases, the difference in $C_D$ between a smooth and rough surface at $\sigma_T = 0.0$ also grows, as more backscattering occurs.

For a final investigation, the Kirchhoff model was used to study the effects of surface roughness on $C_D$ in the altitude range of 200--1000 km for a spherical satellite, and compare the resulting $C_D$ with DRIA and CLL.  It is known from previous studies such as \citet{Walker2014}, \citet{Mehta2014}, and \citet{Mehta2014_2} that most GSI models not only diverge from each other, but also (partially) disagree with observations above an altitude of 400 km. It was observed that satellites with the same spherical shape exhibit vastly different $C_D$ values while orbiting at similar altitudes during comparable solar conditions, for example Stella and Gridsphere. It was concluded by \citet{Mehta2014_2} that this is likely due to different material properties of these two satellites, such as surface roughness. With that in mind, the Gaussian Kirchhoff model is incorporated into the Semi-Empirical Surface Accommodation Model (SESAM) methodology developed by \citet{Pilinski2013} to hopefully explain these discrepancies. SESAM empirically links the aerodynamic coefficients of an RSO to atmospheric properties derived from the NRLMSISE-00 atmospheric model for a given orbital position and time. It does so by assuming that a fraction $\Theta$ of satellite surfaces in the thermosphere is covered by atomic oxygen (AO). As stated in \citet{Pilinski2013}, gas particles reflect diffusely off an AO-covered surface, leading to an updated value of
\begin{equation}
    C_D = (1 - \Theta) C_{D_{s}} + \Theta C_{D_{ads}}
\end{equation}
where $C_{D_{s}}$ is the drag coefficient as a result of a clean surface, while $C_{D_{ads}}$ is that from the AO-covered parts, which is computed according to the assumption of full thermal accommodation using DRIA. To adapt this equation in the context of the Kirchhoff model, the AO coverage mechanism is assumed to happen locally, resulting in
\begin{equation}
    C_D = (1 - \Theta) C_{D_{L_s}} + \Theta C_{D_{L_{ads}}}
\end{equation}
where $C_{D_{L_s}}$ is the drag coefficient of a rough, locally clean surface, while $C_{D_{L_{ads}}}$ is the drag coefficient of a rough surface with complete local thermal accommodation as a result of AO coverage. SESAM further links the fraction $\Theta$ to the partial pressure of AO at a given point in orbit through an absorption equilibrium isotherm. \citet{Pilinski2013} specifically proposes the Langmuir isotherm for this purpose for its simplicity, while \citet{Mehta2014_2} additionally employ the Temkin and Freundlich isotherms. These isotherms are defined by
\begin{equation}
    \Theta_{Langmuir} = \frac{K P_O}{1 + K P_O}, \quad \Theta_{Temkin} = \frac{1}{B} \ln\left(\Xi P_O \right), \quad \text{and} \quad \Theta_{Freundlich} = A P_O^{\zeta},
\end{equation}
where $P_O$ is given by \citet{Pilinski2013} as $P_O = n_{AO} T_{\infty} \frac{\mathcal{R}}{\mathcal{M}_{AO}}$ and the constants $K$, $B$, $\Xi$, $A$, and $\zeta$ are empirical fitting parameters. The Freundlich isotherm is excluded from this analysis because its purpose in \citet{Mehta2014_2} was to loosely model roughness in an empirical way. All the parameters necessary for the final investigation are given in \cref{tab:sphere_alt}.
\begin{table}[H]
\centering
\caption{Input parameters for the sphere - altitude aerodynamic analysis.}
\label{tab:sphere_alt}
\begin{tabular}{lrrlrr}
\hline\hline
\textbf{Parameter Name} & \textbf{Value}    & \textbf{Unit}        & \textbf{Parameter Name} & \textbf{Value} & \textbf{Unit} \\ \hline
$K_{DRIA}$              & $3 \times 10^6$    & [\SI{}{\per\pascal}] & $\beta_{CLL}$           & 5.45           & [-]           \\
$K_{CLL}$               & $5 \times 10^6$    & [\SI{}{\per\pascal}] & $\gamma_{CLL}$          & 0.52           & [-]           \\
$K_K$                   & $3 \times 10^6$    & [\SI{}{\per\pascal}] & $\delta_{CLL}$          & 3.4            & [-]           \\
$B_{DRIA}$              & 13.8              & [\SI{}{\pascal}]     & $\mathcal{M}_S$                 & 26.982         & [g/mol]       \\
$B_{CLL}$               & 22.0              & [\SI{}{\pascal}]     & $T_S$                   & 300            & [K]           \\
$B_K$                   & 13.8              & [\SI{}{\pascal}]     & F10.7 Min               & 60             & [-]           \\
$\Xi_{DRIA}$           & $3 \times 10^{10}$ & [-]                  & F10.7 Max               & 145            & [-]           \\
$\Xi_{CLL}$            & $10^{14}$         & [-]                  & Kirchhoff Model         & Gaussian       & [-]           \\
$\Xi_K$                & $3 \times 10^{10}$ & [-]                  & Speed ratio range       & [2.0, 10.0]    & [-]           \\
$\xi_{CLL}$             & 30                & [-]                  & $\sigma/R$ range        & [0.2, 1.0]     & [-]           \\ \hline\hline
\end{tabular}
\end{table}
The Langmuir and Temkin parameters for the Kirchhoff model were not fitted to aerodynamic sphere data as done by \citet{Mehta2014_2}, for the sake of simplicity. Instead, the DRIA parameters from the same study are applied to the Kirchhoff model for both the Langmuir and Temkin isotherms, as an approximation, based on the similarities between DRIA and Kirchhoff observed in \cref{pp_1:fig:cd_aoa_fp}. The isotherm parameters for the dependence of the DRIA and CLL $C_D$ on altitude are also taken from \citet{Mehta2014_2}. The solar maximum and minimum F10.7 values were chosen based on the operational periods of both Stella and Gridsphere. Additionally, all atmospheric properties were averaged over all latitudes, i.e. from $-90^{\circ}$ to $90^{\circ}$, and over time from January 1$^{st}$ 1995 to January 1$^{st}$ 2001, while the longitude was kept constant at $0^{\circ}$. To compute the drag coefficient for the DRIA kernel, the closed-form expression
\begin{equation}
    C_{D_{DRIA}} = \frac{4s^4 + 4s^2 - 1}{2s^4} \, \text{erf}(s) + \frac{2s^2 + 1}{\sqrt{\pi s^3}} e^{-s^2} + \frac{2\sqrt{\pi}}{3s} \sqrt{\frac{T_{k,r}}{T_{\infty}}}, \qquad T_{k, r} = T_{k,i} (1 - \alpha) + T_{S} \alpha, \qquad T_{k,i} = \frac{\mathcal{M}_{\infty} v_{\infty}^2}{3 \mathcal{R}}
\end{equation}
 from \citet{Mehta2014_2} was used, where $s$ is the speed ratio, $\mathcal{M}_{\infty}$ is the molar mass of the atmospheric gas, $\alpha$ is the thermal energy accommodation coefficient,  and $v_{\infty}$ is approximated as the satellite velocity, i.e. thermospheric winds are neglected. The accommodation coefficient $\alpha$ is given by \cite{Sentman1961FREEMF} as $\alpha \approx \frac{2.4 M_R}{(1 + M_R)^2}$, where $M_R = \frac{\mathcal{M}_{G}}{\mathcal{M}_S}$. Here, $\mathcal{M}_G$ is the molar mass of the incoming gas species while $\mathcal{M}_S$ is the molar mass of the surface material. To compute the drag coefficient of the CLL kernel, the expression 
\begin{equation}
    C_{D_{CLL}} = \frac{1 - \sqrt{1 - \alpha_N} + \sigma_T}{2} \left( \frac{4s^4 + 4s^2 - 1}{2s^4} \, \erf(s) + \frac{2s^2 + 1}{\sqrt{\pi s^3}} e^{-s^2} \right) + \frac{2\sqrt{\pi}}{3s^2} e^{-\beta (1 - \alpha_N)/s^2} \left( \frac{T_S}{T_\infty} \right)^{0.5 + \delta}
\end{equation}
from \citet{Mehta2014_2} was used.
To generate the $C_D$ as a function of altitude using this kernel, the approach of \citet{Walker2014} is employed, i.e. the tangential momentum accommodation coefficient $\sigma_T$ is set to 1.0 and the normal energy accommodation coefficient is given as $\alpha_N = \max(0.5, 2 \alpha - 1)$. The Kirchhoff model aerodynamics are calculated using a different approach because it does not have a closed-form solution for any geometry. To compute its drag coefficient $C_{D_K}$, look-up tables were generated for the full parameter space of local accommodation coefficients $\alpha_N$ and $\sigma_T$, as well as speed ratios between 2.0 and 10.0. These were then linearly interpolated using a regular grid interpolator to approximate the coefficient for any set of input parameters. Furthermore, for physical accuracy, the local tangential accommodation coefficient $\sigma_T$ was assumed to be zero, based on \citet{Goodman1965}, \citet{Logan1966}, and \citet{Logan1968}. As a result, the normal energy accommodation coefficient is set to $\alpha_N = \alpha = \frac{2.4 M_R}{(1 + M_R)^2}$ following \citet{Sentman1961FREEMF}. 

\begin{figure}[H]
    \centering
    \includegraphics[width=0.5\linewidth]{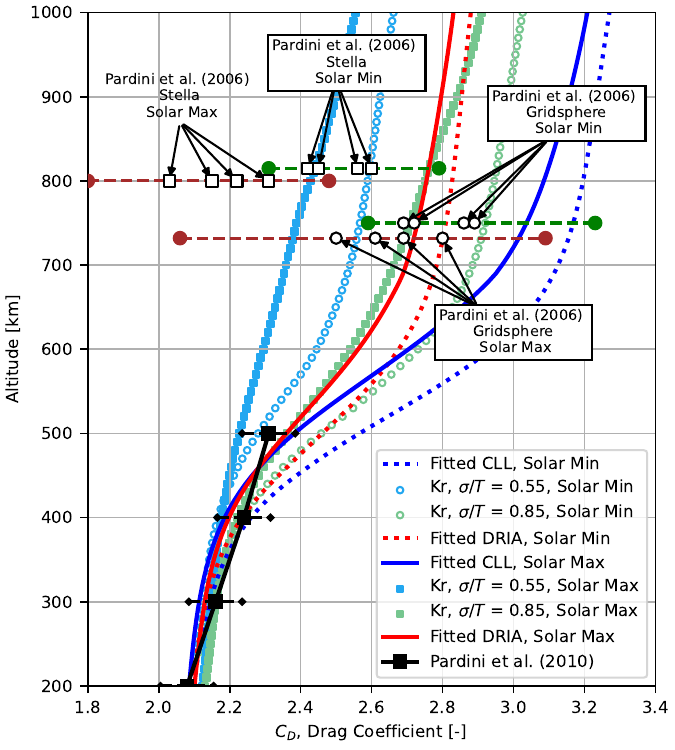}
    \caption{A comparison between the drag coefficient of a sphere computed with the closed forms of the DRIA and CLL kernels as given by \citet{Walker2014}, those generated with the Gaussian Kirchhoff model for different levels of roughness, and fitted drag coefficients of the Stella and Gridsphere spherical satellites reported by \citet{Pardini2006, Pardini2010}. A \textbf{Langmuir} isotherm was used to compute the $C_D$ variations with altitude.}
    \label{pp_1:fig:cd_sphere_langmuir}
\end{figure}

\begin{figure}[H]
    \centering
    \includegraphics[width=0.5\linewidth]{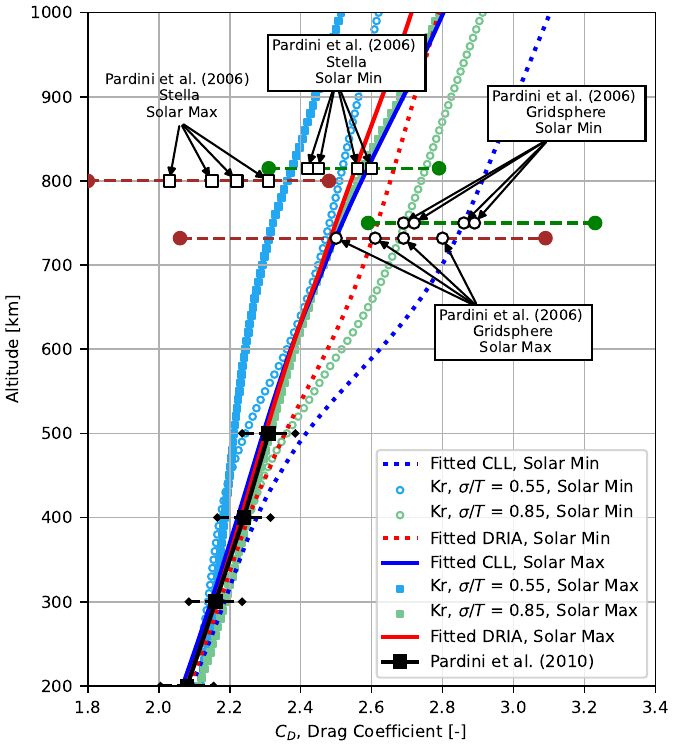}
    \caption{A comparison between the drag coefficient of a sphere computed with the closed forms of the DRIA and CLL kernels as given by \citet{Walker2014}, those generated with the Gaussian Kirchhoff model for different levels of roughness, and fitted drag coefficients of the Stella and Gridsphere spherical satellites reported by \citet{Pardini2006, Pardini2010}. A \textbf{Temkin} isotherm was used to compute the $C_D$ variations with altitude.}
    \label{pp_1:fig:cd_sphere_temkin}
\end{figure}

The drag coefficient $C_D$ was calculated as a function of the altitude for all there GSI models and the Langmuir and Temkin isotherms, as presented in \cref{pp_1:fig:cd_sphere_langmuir}  and \cref{pp_1:fig:cd_sphere_temkin}. Additionally, the figures show observed sphere aerodynamic coefficients for altitudes up to 500 km, sourced from \citet{Pardini2010}, as well as the $C_D$ estimations for the Stella and Gridsphere satellites, sourced from \citet{Pardini2006}. For the Kirchhoff model, two curves were generated in each figure, corresponding to roughness parameters of $\sigma/R = 0.55$ and $\sigma / R = 0.85$, which were optimised to fit the observations of the two high-altitude spherical satellites alongside the low-altitude data. Differences between the current DRIA and CLL curves and those presented by \cite{Mehta2014_2} may be attributed to the simplifications made to their methodology, i.e. different solar maximum and minimum F10.7 values, the use of latitude and time averaging instead of orbit propagation from two-line elements (TLEs), and the investigation of slightly different time frames. Both figures suggest that geometric roughness has little effect on the value of the drag coefficient at altitudes below 400 km, whereas it plays a significant role at higher altitudes, where local quasi-specular behaviour is expected in the helium-rich parts of the thermosphere. In these regions, it appears that a higher surface roughness results in substantially  larger aerodynamic drag for spherical satellites. This, however, may not hold true for other satellite shapes, as different drag effects were observed for low and high incidence angles in \cref{pp_1:fig:cd_cll_fp}. Comparing the two drag coefficients of the Kirchhoff model with the Stella and Gridsphere data points, it appears that Stella is best modelled by a roughness parameter of $\sigma / R = 0.55$ while Gridsphere corresponds to $\sigma / R = 0.85$. This makes physical sense because Stella's aluminium surface is equipped with 60 laser retroreflectors, which can safely be assumed to have a roughness parameter close to zero, covering about a third of its surface. Gridsphere, on the other hand, has a plain aluminium surface \citep{Pardini2006},  suggesting that its surface-averaged roughness parameter is likely higher than Stella's. Comparing  the figures between themselves, it appears that the choice of isotherm does not significantly change the effect of geometric roughness on the drag coefficient. Nevertheless, the $C_D$ based on the Temkin isotherm shows far less variation, which could be a result of either better modelling of the underlying absorption physics, or simply a result of fitting a more complex model to the available data, as discussed by \cite{Mehta2014_2}. A final observation may now be made about the validity of the assumption that $\sigma_T \approx 1$, which is employed in many studies such as \citet{Walker2014}, \cite{Mehta2014}, and \citet{Mehta2014_2}. Given the results in \cref{subsec:Verification} and the apparent convergence of the Kirchhoff model's $C_D$ to that of the CLL model for increasing levels of roughness, as portrayed in \cref{pp_1:fig:cd_sphere_langmuir} and \cref{pp_1:fig:cd_sphere_temkin}, we propose geometric roughness in the material surface samples as the root cause behind the high $\sigma_T$ values recorded by \citet{Liu1979}. This suggests that the methodology proposed by \citet{Walker2014} for implementing the CLL kernel can be regarded as an early empirical attempt at capturing the effects of roughness. Consequently, it supports the new Kirchhoff model presented in this study, which captures these effects with greater accuracy and more control.









\section{Conclusions} \label{sec:Conclusions}

This study brought attention to a crucial yet under-explored aspect of GSI in free molecular flow conditions, namely the scattering of gas particles due to the geometric surface roughness, which may arise from the macroscopic imperfections of real surfaces. It then introduced a new generalized model based on the Kirchhoff wave scattering theory, which accurately captures these scattering effects and outlined a method for integrating them with other GSI occurring at smaller scales. Consequently, the study presented an extensive verification of this new model against a series of TPMC simulations and, then, applied it to reproduce experimental scattering results as well as compute aerodynamic coefficients for simple shapes, focusing on the variation with altitude in the thermosphere environment.

The key innovation of the proposed model is its physics-based utilization of the wave-particle duality of rarefied gas particles, allowing for the application of the Kirchhoff approximation. This resulted in an analytic, closed-form expression for the reflected particle velocity PDF after a single reflection. Another novel aspect is the use of poly-Gaussian processes to statistically describe rough surfaces, offering a reduced-order, generalized framework for modelling any surface morphology under the assumption of isotropy. To robustly model high-roughness phenomena such as multiple particle reflections and surface shadowing, a simple, iterative algorithm was developed, making use of the one-reflection Kirchhoff scattering kernel alongside an analytic expression for the probability of rescattering. 

With an overall excellent agreement between the proposed model and the TPMC simulations for both Gaussian and poly-Gaussian surfaces, significant relative errors of up to 7\% were observed when gas particles locally accommodate their normal velocity component only. This is particularly important given that, in the impulsive scattering regime, most gas particles do not undergo tangential momentum accommodation. By far, the most likely culprit for these differences is the shadowing function $\mathcal{S}(\theta_{r_1}, \xi_0)$, which makes two assumptions of statistical independence: one between consecutive collision points and another between the height and slope PDFs of the surface. Thankfully, it is straightforward to mitigate this issue and avoid both assumptions either through Rice series expansion methods \citep{Aksenova2022} or a numerical treatment of $\mathcal{S}(\theta_{r_1}, \xi_0)$ at the cost of minor rise in computational complexity. Nevertheless, this mitigation is likely not needed for most satellite surfaces because their roughness is expected to be well below the level that resulted in the aforementioned error figure. To our knowledge, the only process by which such high roughness levels are attainable is an erosion of the exposed satellite surface, e.g. the oxidation of Kapton film. In such cases, the model should not be directly applied based on ground measurements of surface roughness. Instead, it should be "corrected" through an optimisation procedure of the surface roughness parameters using in-situ acceleration data. As a matter of fact, this method is the default used for empirical kernels such as DRIA and CLL. 

Consequently, the success of the model in replicating both experimental gas scattering results as well as in-orbit aerodynamic coefficient trends reinforces our belief that surface roughness as a crucial element of GSI has so far been mostly disregarded in existing scientific literature. It further strengthens our suspicion that most objects in LEO exhibit a level of roughness high enough to alter their aerodynamic behaviour from what is expected based on ground experiments. These beliefs are first confirmed through the model's ability to capture particle backscattering and its dominance at near-parallel incidence angles, a phenomenon which has been independently observed in many experimental studies but was never explained in a comprehensive manner. An immediate conclusion based on this finding is that most RSOs with angled, flow-exposed surfaces should exhibit a larger-than-expected drag coefficient in the helium-rich region of the thermosphere, where quasi-specular reflections are expected. This is exactly what was observed after using the model to investigate the altitude-dependence of the drag coefficient for spherical satellites with different roughness parameters. On top of this, the similarity in behaviour observed between this model and the DRIA kernel for high levels of roughness finally provides a physical explanation for the success of the latter in the same helium-rich regions where CLL was expected to be more accurate based on knowledge about the local interactions from many ground experiments. 

Future work will focus on employing both the Gaussian and poly-Gaussian implementations of the Kirchhoff model in modelling the aerodynamics of the Swarm, GRACE and GRACE-FO satellites, with the hope that this will result in more consistent neutral density and cross-wind datasets. To reduce the number of parameters to be fitted, the local interaction kernel parameters shall be fitted using a physics-based approach similar to SESAM. Besides this, additional research will be conducted into improving the accuracy of the model in its critical test cases, by implementing more accurate shadowing functions. The possibility of extending it for anisotropic surfaces shall also be investigated. 

\section*{Acknowledgements}
Funding for this work was provided by the Dutch Research Council (NWO), through grant number ENW.GO.001.008. 

\begin{appendices}

\section{Helmholtz Integral Derivation} \label{ap:Helmholtz}

\begin{figure}[H]
    \centering
    \includegraphics[width=0.4\linewidth]{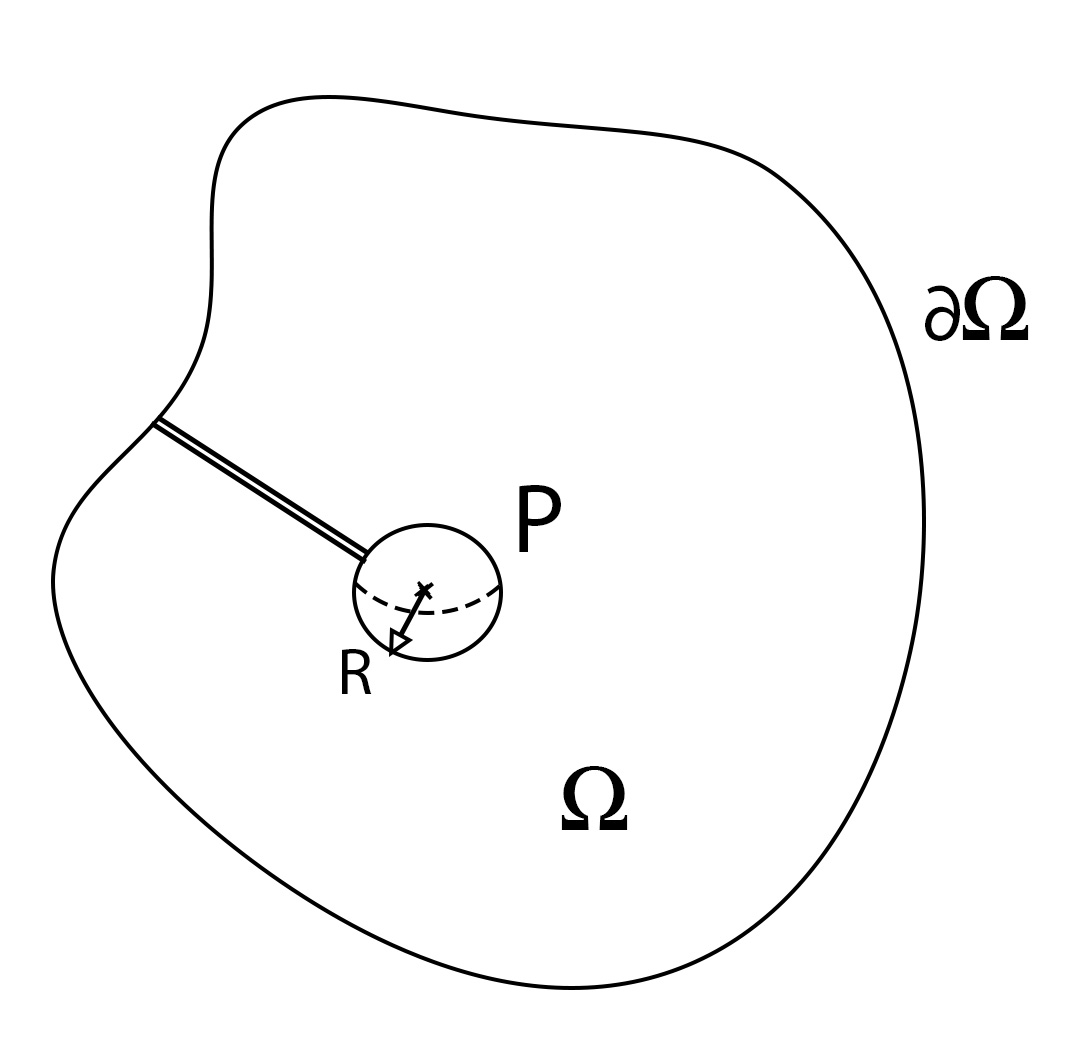}
    \caption{A sketch of a volume $\Omega$, bounded by a closed surface $\partial \Omega$. A point $P$ is shown inside this volume, at the centre of a sphere with radius $R$. The sphere surface and $\partial \Omega$ are connected through a cylindrical channel.}
    \label{pp_1:fig:Helmholtz_domain}
\end{figure}

Consider the classic wave equation defined through \cref{eq:WaveEq} on a closed domain $\Omega \subset \mathbb{R}^3$, bounded by a surface $\partial \Omega$. The purpose of this section is to find a relationship between the scattered wave field $\Psi_r(\mathbf{x})$ at an arbitrary point $P \in \Omega$, as a function of the values of the same scattered field on the boundary $\partial \Omega$. A similar derivation to the one presented here is given in \citet{Beckman1987-kr}. This situation is sketched in \cref{pp_1:fig:Helmholtz_domain}. For this, the Green function of the closed surface $\partial \Omega$ is introduced as
\begin{equation}
    \mathcal{G}(\mathbf{x}) = \frac{e^{i \mathbf{k_r} \cdot \mathbf{x}}}{\lVert\mathbf{x}\rVert},
\end{equation}
where $\mathbf{k_r} = 2\pi \frac{m \mathbf{v_r}}{h}$ is the wave vector of the reflected wave. It can be easily shown that $\mathcal{G}(\mathbf{x})$ satisfies the Helmholtz equation
\begin{equation}
    \nabla^2 \mathcal{G}(\mathbf{x}) + k_2^2 \mathcal{G}(\mathbf{x}) = 0
\end{equation}
for $\lVert\mathbf{x}\rVert > R$, where $R$ is an arbitrary positive number defining the radius of a sphere enclosed in $\Omega$. This is, of course, to avoid the singularity in $\mathcal{G}(\mathbf{x})$ at $\lVert\mathbf{x}\rVert = 0$. Applying the Divergence Theorem on the quantity $\Psi(\mathbf{x}) \cdot \nabla \mathcal{G}(\mathbf{x})$, one obtains
\begin{equation}
    \iiint_{\Omega} \nabla \cdot \left( \Psi(\mathbf{x}) \cdot \nabla \mathcal{G}(\mathbf{x}) \right) \, \dif V = \oiint_{\partial \Omega} \left(\Psi(\mathbf{x}) \cdot \nabla \mathcal{G}(\mathbf{x})\right) \cdot \mathbf{n} \, \dif S,
\end{equation}
where $\mathbf{n}$ here is the normal to the volume's surface $\partial \Omega$. Working out the equation above yields
\begin{equation}
    \iiint_{\Omega} \nabla \Psi(\mathbf{x}) \nabla \mathcal{G}(\mathbf{x}) \, \dif V = \oiint_{\partial \Omega} \Psi(\mathbf{x}) \frac{\partial \mathcal{G}(\mathbf{x})}{\partial \mathbf{n}} \, \dif S - \iiint_{\Omega} \Psi(\mathbf{x}) \nabla^2 \mathcal{G}(\mathbf{x}) \, \dif V.
\end{equation}
One can immediately notice the symmetry in the left-hand side of the expression above, and infer the same equality for $\Psi$ and $\mathcal{G}$ switching places:
\begin{equation}
    \iiint_{\Omega} \nabla \Psi(\mathbf{x}) \nabla \mathcal{G}(\mathbf{x}) \, \dif V = \oiint_{\partial \Omega} \mathcal{G}(\mathbf{x}) \frac{\partial \Psi(\mathbf{x})}{\partial \mathbf{n}} \, \dif S - \iiint_{\Omega} \mathcal{G}(\mathbf{x}) \nabla^2 \Psi(\mathbf{x})\, \dif V
\end{equation}
Combining the two equation above leads to the final form
\begin{equation}
    \iiint_{\Omega}\left( \Psi(\mathbf{x}) \nabla^2 \mathcal{G}(\mathbf{x}) - \mathcal{G}(\mathbf{X}) \nabla^2 \Psi(\mathbf{X}) \right)dV = \oiint_{\partial \Omega} \left( \Psi(\mathbf{x}) \frac{\partial \mathcal{G}(\mathbf{x})}{\partial \mathbf{n}} - \mathcal{G}(\mathbf{x}) \frac{\partial \Psi(\mathbf{x})}{\partial \mathbf{n}} \right) \, \dif S.
\end{equation}
As both $\Psi(\mathbf{x})$ and $\mathcal{G}(\mathbf{x})$ satisfy the Helmholtz equation in the volume between $\Omega$ and the sphere with radius $R$ and volume $\Omega_S$, one can make the following substitution after connecting the two enclosing surfaces  $\partial \Omega$ and $\partial \Omega_S$ with an infinitesimally thin tube, i.e.
\begin{equation} \label{eq:Helmholtz_1}
    \iiint_{\Omega - \Omega_S}\left( \Psi(\mathbf{x}) \cdot (- k_r^2 \mathcal{G}(\mathbf{x})) - \mathcal{G}(\mathbf{x}) \cdot (- k_r^2 \Psi(\mathbf{x}))\right) dV = \oiint_{\partial \Omega + \partial \Omega_S} \left(\Psi(\mathbf{x}) \frac{\partial \mathcal{G}(\mathbf{x})}{\partial \mathbf{n}} - \mathcal{G} \frac{\partial \Psi(\mathbf{x})}{\partial \mathbf{n}} \right) \, \dif S.
\end{equation}
Substituting the expression for $\mathcal{G}(\mathbf{x})$ into the surface integral of the spherical surface yields
\begin{equation}
    \oiint_{\partial \Omega_S} \left(\Psi(\mathbf{x}) \frac{\partial \mathcal{G}(\mathbf{x})}{\partial \mathbf{n}} - \mathcal{G} \frac{\partial \Psi(\mathbf{x})}{\partial \mathbf{n}} \right) dS = \oiint_{\partial \Omega_S} \left( \Psi(\mathbf{x}) \left(\frac{i k_r}{R} - \frac{1}{R^2} \right)e^{i \mathbf{k_r} \cdot \mathbf{x}} \cdot \mathbf{n_S} - \frac{e^{i \mathbf{k_r} \cdot x}}{R} \frac{\partial \Psi(\mathbf{x}}{\partial \mathbf{n_S}}\right) \, \dif S,
\end{equation}
where $\lVert \mathbf{x} \rVert = R$. Making $R \rightarrow 0$ enables the use of the Mean Value theorem, which results in
\begin{equation}
    \oiint_{\partial \Omega_S} \left(\Psi(\mathbf{x}) \frac{\partial \mathcal{G}(\mathbf{x})}{\partial \mathbf{n}} - \mathcal{G} \frac{\partial \Psi(\mathbf{x})}{\partial \mathbf{n}} \right) \, \dif S  = 4 \pi R^2 \lim_{R \rightarrow 0} \left( \left( \frac{i k_r}{R^2} - \frac{1}{R^2}\right)e^{i \mathbf{k_r} \cdot \mathbf{x}} \left\langle \Psi(\mathbf{x}) \right\rangle - \frac{e^{i \mathbf{k_r} \cdot \mathbf{x}}}{R} \left\langle \frac{\partial \Psi(\mathbf{x})}{\partial \mathbf{n_S}}\right\rangle \right)  = - 4 \pi \Psi(P).
\end{equation}
Finally, substituting this result back into Eq. (\ref{eq:Helmholtz_1}) and acknowledging that the left-hand-side is zero results in the final form of the Helmholtz integral for volume $\Omega$:
\begin{equation}
    \Psi(P) = \frac{1}{4\pi}\oiint_{\partial \Omega}\left(\Psi(\mathbf{x}) \frac{\partial \mathcal{G}(\mathbf{x})}{\partial \mathbf{n}} - \mathcal{G}(\mathbf{x}) \frac{\partial \Psi(\mathbf{x})}{\partial \mathbf{n}}\right)\, \dif S
\end{equation}

\section{Additional Results} \label{ap:Ad_results}

\begin{figure}[H]
    \centering
    \includegraphics[width=0.99\linewidth]{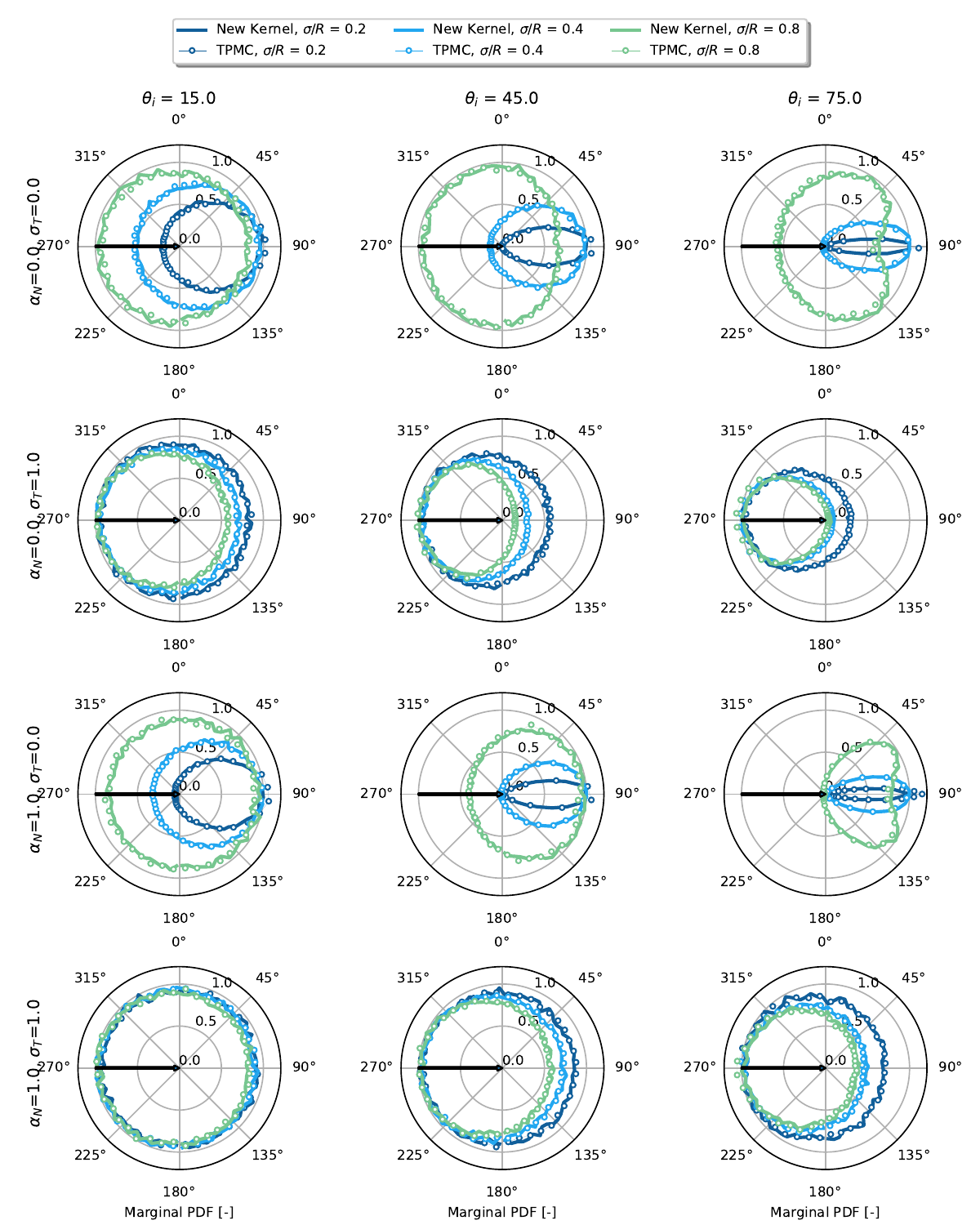}
    \caption{The XY-plane marginal angular PDFs of Helium gas particles with different incidence angles, scattering from Gaussian surfaces with different levels of roughness. The model-generated results are shown in solid lines, while the TPMC simulation results are given in dotted lines. The incident direction is depicted with a black arrow.}
    \label{pp_1:fig:gs_XY_angular_pdfs}
\end{figure}

\begin{figure}[H]
    \centering
    \includegraphics[width=0.99\linewidth]{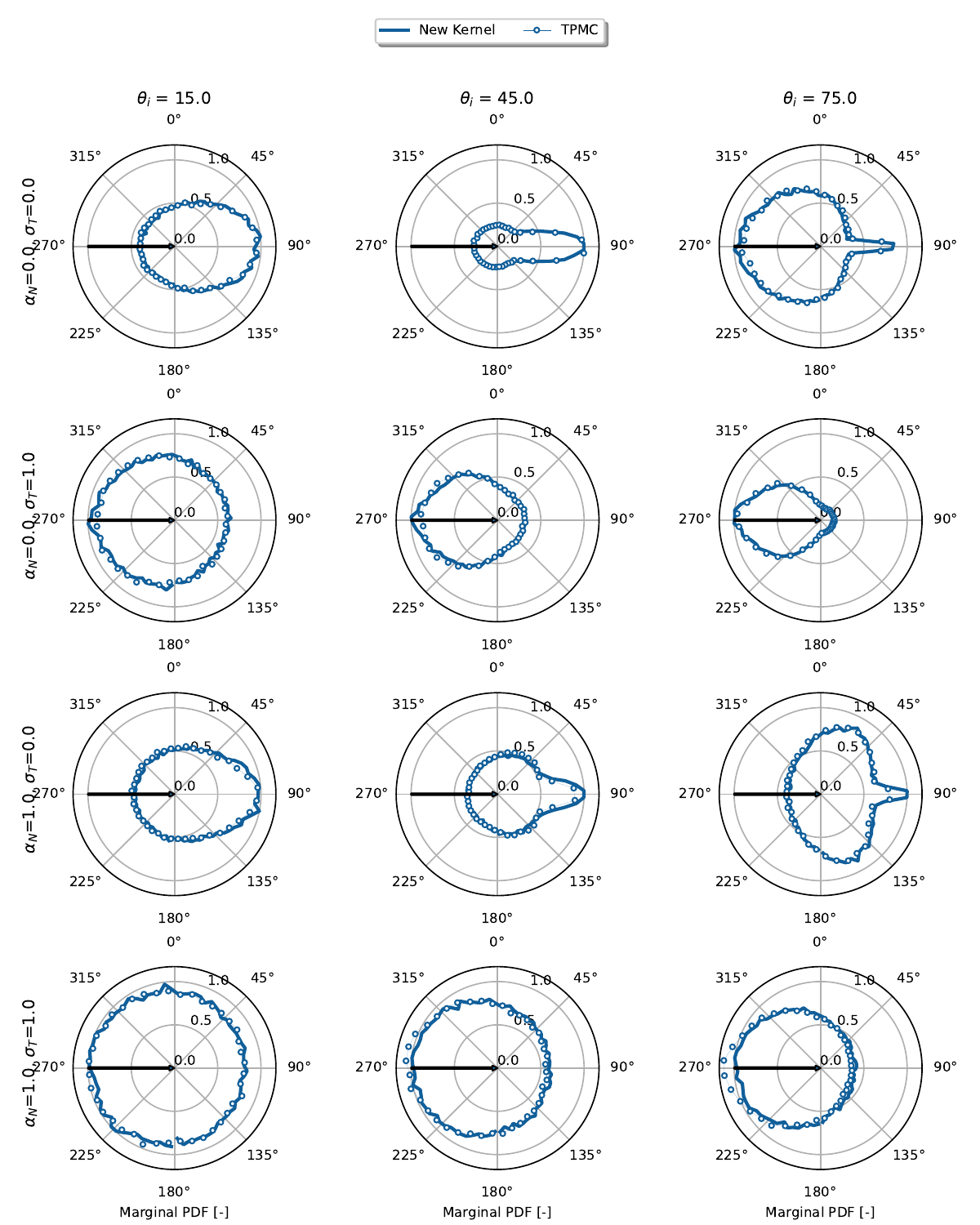}
    \caption{The XY-plane marginal angular PDFs of Helium gas particles with different incidence angles, scattering from a Non-Gaussian surface. The model-generated results are shown in solid lines, while the TPMC simulation results are given in dotted lines. The incident direction is depicted with a black arrow.}
    \label{pp_1:fig:ngs_XY_angular_pdfs}
\end{figure}

\begin{figure}[H]
    \centering
    \includegraphics[width=0.99\linewidth]{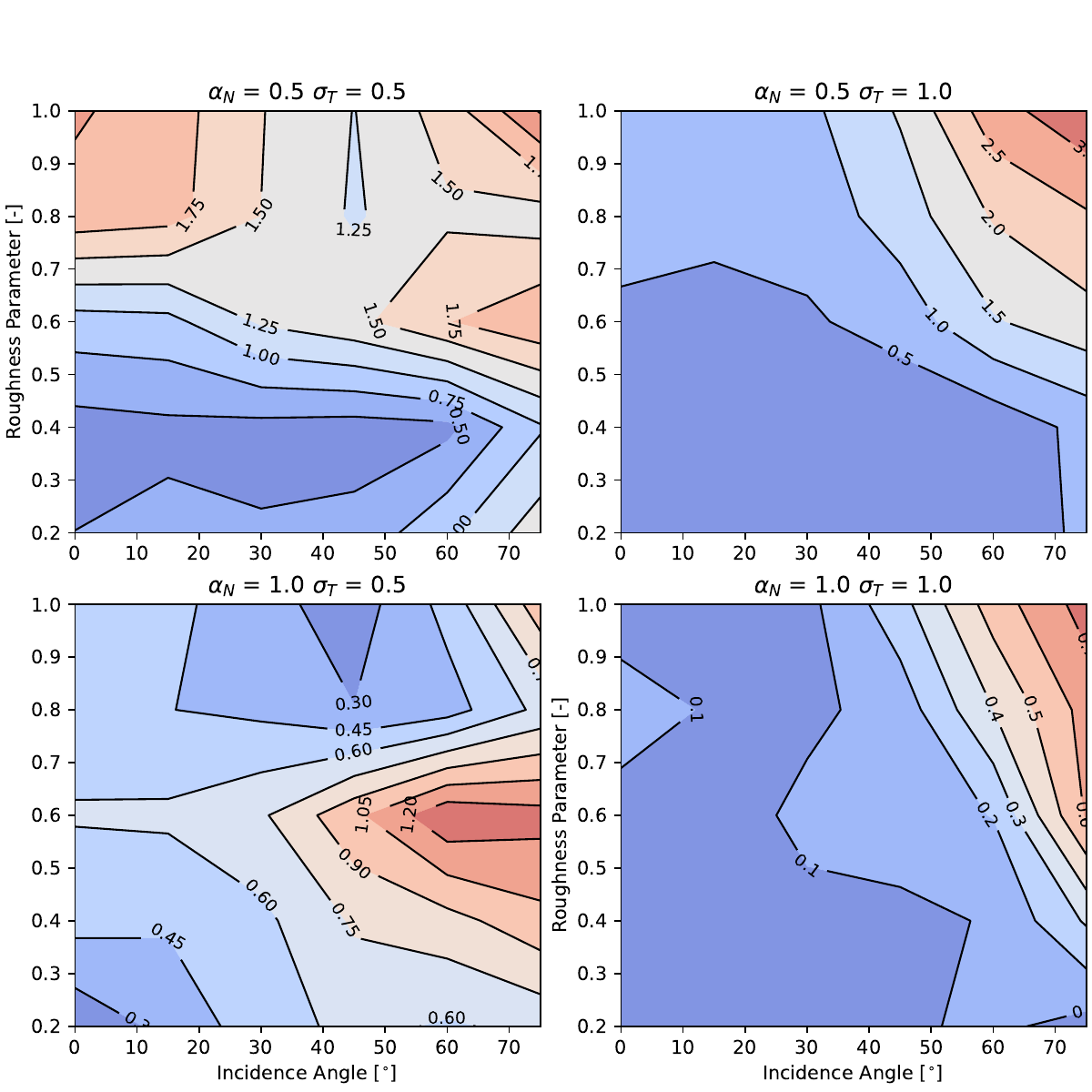}
    \caption{The momentum relative error between the Gaussian Kirchhoff model and the TPMC simulation results is shown as a function of the incidence angle $\theta_i$ and the roughness parameter $\sigma/R$, for four combinations of local CLL parameters.}
    \label{pp_1:fig:gs_error}
\end{figure}

This section presents the angular scattering marginal PDFs in the XY plane for both the Gaussian and poly-Gaussian versions of the Kirchhoff model, in alignment with the analysis detailed in \cref{subsec:Verification}. These PDFs were generated to assess the model's capability in capturing out-of-plane scattering as a result of surface morphology. Visuals of these are given in \cref{pp_1:fig:gs_XY_angular_pdfs} and \cref{pp_1:fig:ngs_XY_angular_pdfs}. Additionally, a relative error contour plot comparing the Kirchhoff and TPMC results for the Gaussian analysis is provided in \cref{pp_1:fig:gs_error}. All three analyses use the same simulation parameters, as specified in \cref{tab:gaussian_analysis} and \cref{tab:polygaussian_analysis}.

Upon examining Figures \ref{pp_1:fig:gs_XY_angular_pdfs} and \ref{pp_1:fig:ngs_XY_angular_pdfs}, an overall excellent agreement is observed between the Kirchhoff PDFs and the TPMC results for both the Gaussian and poly-Gaussian analyses, further confirming the conclusions of \cref{subsec:Verification}. Qualitatively, this agreement surpasses even that of the YZ-plane PDFs presented in \cref{pp_1:fig:gs_angular_pdfs} and \cref{pp_1:fig:ngs_angular_pdfs}, suggesting that the Kirchhoff model can very accurately capture out-of-plane scattering behaviour. The largest discrepancies occur in the poly-Gaussian test case for $(\alpha_N, \sigma_T) = (1.0, 0.0)$, similar to the previous analysis. Specifically, at $\theta_{r_2} = 0^{\circ}$, the PDFs exhibit a local peak that is overestimated by the Kirchhoff model, representing the amount of particles scattering in-plane. Based on this and previous results, it appears that increasing roughness results in larger errors. However, as \cref{pp_1:fig:gs_error} suggests, this is not always the case. In this figure, the relative error between the Kirchhoff model and the TPMC results is plotted for a Gaussian surface as a function of the incidence angle $\theta_i$ and the roughness parameter $\sigma / R$, for different combinations of local parameters. From the four plots, it seems that the error behaviour of the model is highly dependent on the choice of local kernel parameters. Nevertheless, a global trend is observed: the relative error increases with incidence angle for higher roughness levels. That being said, even the largest observed errors do not exceed 2\%, underscoring the accuracy and applicability of the model for Gaussian surfaces. From a phenomenological perspective, all the XY-plane PDFs in \cref{pp_1:fig:gs_XY_angular_pdfs} and \cref{pp_1:fig:ngs_XY_angular_pdfs} display significant backscattering. While in the Gaussian test cases increasing the roughness parameter induces an out-of-plane scattering behaviour similar to DRIA, the poly-Gaussian test case exhibits a more complex behaviour. In particular, the cases with $(\alpha_N, \sigma_T) = (0.0, 0.0)$ and $(\alpha_N, \sigma_T) = (1.0, 0.0)$ show the superposition behaviour discussed in \cref{subsec:Verification}, as they display distinct diffuse and quasi-specular lobes.

\end{appendices}

\bibliographystyle{jasr-model5-names}
\biboptions{authoryear}
\bibliography{refs}

\begin{thebibliography}{83}
\expandafter\ifx\csname natexlab\endcsname\relax\def\natexlab#1{#1}\fi
\ifx\xfnm\relax \def\xfnm[#1]{\unskip,\space#1}\fi

\bibitem[{{Aksenova } \& {Khalidov }(2008)}]{Aksenova2022}
\bibinfo{author}{{Aksenova }},  \& \bibinfo{author}{{Khalidov }} (\bibinfo{year}{2008}).
\newblock \bibinfo{title}{{Aerodynamic Calculation of Rough Surface in Rarefied Gas Flow Applying the Solution of Inverse Problem}}.
\newblock In {\it \bibinfo{booktitle}{{Proceedings of the 9th European Conference for Aerospace Sciences}}\/}.
\newblock \DOIprefix\doi{10.13009/EUCASS2022-4897}.

\bibitem[{Banks et~al.(2004)Banks, Miller \& de~Groh}]{Banks2004}
\bibinfo{author}{Banks, B.}, \bibinfo{author}{Miller, S.},  \& \bibinfo{author}{de~Groh, K.} (\bibinfo{year}{2004}).
\newblock \bibinfo{title}{Low earth orbital atomic oxygen interactions with materials}.
\newblock In {\it \bibinfo{booktitle}{2nd International Energy Conversion Engineering Conference}\/}.
\newblock \bibinfo{publisher}{American Institute of Aeronautics and Astronautics}.
\newblock \DOIprefix\doi{10.2514/6.2004-5638}.

\bibitem[{Barker \& Auerbach(1984)}]{Barker1984}
\bibinfo{author}{Barker, J.},  \& \bibinfo{author}{Auerbach, D.} (\bibinfo{year}{1984}).
\newblock \bibinfo{title}{{Gas—surface interactions and dynamics; Thermal energy atomic and molecular beam studies}}.
\newblock {\it \bibinfo{journal}{Surface Science Reports}\/},  {\it \bibinfo{volume}{4}\/}\bibinfo{issue}{(1–2)}, \bibinfo{pages}{1–99}. \DOIprefix\doi{10.1016/0167-5729(84)90005-0}.

\bibitem[{Beckman et~al.(1987)Beckman, Spizzichino \& Beckmann}]{Beckman1987-kr}
\bibinfo{author}{Beckman, P.}, \bibinfo{author}{Spizzichino, A.},  \& \bibinfo{author}{Beckmann, P.} (\bibinfo{year}{1987}).
\newblock {\it \bibinfo{title}{The scattering of electromagnetic waves from rough surfaces}\/}.
\newblock Radar Library.
\newblock \bibinfo{address}{Norwood, MA}: \bibinfo{publisher}{Artech House}.

\bibitem[{Beckmann(1965)}]{Beckmann1965}
\bibinfo{author}{Beckmann, P.} (\bibinfo{year}{1965}).
\newblock \bibinfo{title}{{Shadowing of random rough surfaces}}.
\newblock {\it \bibinfo{journal}{IEEE Transactions on Antennas and Propagation}\/},  {\it \bibinfo{volume}{13}\/}\bibinfo{issue}{(3)}, \bibinfo{pages}{384–388}. \DOIprefix\doi{10.1109/tap.1965.1138443}.

\bibitem[{Beckmann(1973)}]{Beckmann1973}
\bibinfo{author}{Beckmann, P.} (\bibinfo{year}{1973}).
\newblock \bibinfo{title}{Scattering by non-gaussian surfaces}.
\newblock {\it \bibinfo{journal}{IEEE Transactions on Antennas and Propagation}\/},  {\it \bibinfo{volume}{21}\/}\bibinfo{issue}{(2)}, \bibinfo{pages}{169--175}. \DOIprefix\doi{10.1109/TAP.1973.1140444}.

\bibitem[{Bernstein \& Pilinski(2022)}]{Bernstein2022}
\bibinfo{author}{Bernstein, V.},  \& \bibinfo{author}{Pilinski, M.} (\bibinfo{year}{2022}).
\newblock \bibinfo{title}{Drag coefficient constraints for space weather observations in the upper thermosphere}.
\newblock {\it \bibinfo{journal}{Space Weather}\/},  {\it \bibinfo{volume}{20}\/}\bibinfo{issue}{(5)}. \DOIprefix\doi{10.1029/2021sw002977}.

\bibitem[{{Blender Online Community}(2018)}]{Blender2}
\bibinfo{author}{{Blender Online Community}} (\bibinfo{year}{2018}).
\newblock {\it \bibinfo{title}{{Blender - a 3D modelling and rendering package}}\/}.
\newblock \bibinfo{organization}{Blender Foundation} \bibinfo{address}{Stichting Blender Foundation, Amsterdam}.
\newblock \URLprefix \url{http://www.blender.org}.

\bibitem[{Born \& Oppenheimer(1927)}]{Born1927}
\bibinfo{author}{Born, M.},  \& \bibinfo{author}{Oppenheimer, R.} (\bibinfo{year}{1927}).
\newblock \bibinfo{title}{{Zur Quantentheorie der Molekeln}}.
\newblock {\it \bibinfo{journal}{Annalen der Physik}\/},  {\it \bibinfo{volume}{389}\/}\bibinfo{issue}{(20)}, \bibinfo{pages}{457--484}. \DOIprefix\doi{10.1002/andp.19273892002}.

\bibitem[{Bowman et~al.(2008)Bowman, Tobiska, Marcos, Huang, Lin \& Burke}]{Bowman2008}
\bibinfo{author}{Bowman, B.}, \bibinfo{author}{Tobiska, W.~K.}, \bibinfo{author}{Marcos, F.} et~al. (\bibinfo{year}{2008}).
\newblock \bibinfo{title}{{A New Empirical Thermospheric Density Model {JB2008} Using New Solar and Geomagnetic Indices}}.
\newblock In {\it \bibinfo{booktitle}{{AIAA/AAS Astrodynamics Specialist Conference and Exhibit}}\/}.
\newblock \bibinfo{publisher}{American Institute of Aeronautics and Astronautics}.
\newblock \DOIprefix\doi{10.2514/6.2008-6438}.

\bibitem[{Brown(1980)}]{Brown1980}
\bibinfo{author}{Brown, G.} (\bibinfo{year}{1980}).
\newblock \bibinfo{title}{{Shadowing by non-Gaussian random surfaces}}.
\newblock {\it \bibinfo{journal}{IEEE Transactions on Antennas and Propagation}\/},  {\it \bibinfo{volume}{28}\/}\bibinfo{issue}{(6)}, \bibinfo{pages}{788–790}. \DOIprefix\doi{10.1109/tap.1980.1142437}.

\bibitem[{Bruinsma(2015)}]{Bruinsma2015}
\bibinfo{author}{Bruinsma, S.} (\bibinfo{year}{2015}).
\newblock \bibinfo{title}{The dtm-2013 thermosphere model}.
\newblock {\it \bibinfo{journal}{Journal of Space Weather and Space Climate}\/},  {\it \bibinfo{volume}{5}\/}, \bibinfo{pages}{A1}. \DOIprefix\doi{10.1051/swsc/2015001}.

\bibitem[{Bruinsma et~al.(2023)Bruinsma, Siemes, Emmert \& Mlynczak}]{Bruinsma2023}
\bibinfo{author}{Bruinsma, S.}, \bibinfo{author}{Siemes, C.}, \bibinfo{author}{Emmert, J.~T.} et~al. (\bibinfo{year}{2023}).
\newblock \bibinfo{title}{{Description and comparison of 21st century thermosphere data}}.
\newblock {\it \bibinfo{journal}{Advances in Space Research}\/},  {\it \bibinfo{volume}{72}\/}\bibinfo{issue}{(12)}, \bibinfo{pages}{5476–5489}. \DOIprefix\doi{10.1016/j.asr.2022.09.038}.

\bibitem[{Cercignani \& Lampis(1971)}]{Cercignani1971}
\bibinfo{author}{Cercignani, C.},  \& \bibinfo{author}{Lampis, M.} (\bibinfo{year}{1971}).
\newblock \bibinfo{title}{Kinetic models for gas-surface interactions}.
\newblock {\it \bibinfo{journal}{Transport Theory and Statistical Physics}\/},  {\it \bibinfo{volume}{1}\/}\bibinfo{issue}{(2)}, \bibinfo{pages}{101–114}. \DOIprefix\doi{10.1080/00411457108231440}.

\bibitem[{Cercignani \& Michaelis(2001)}]{Cercignani2001}
\bibinfo{author}{Cercignani, C.},  \& \bibinfo{author}{Michaelis, C.} (\bibinfo{year}{2001}).
\newblock \bibinfo{title}{{{Rarefied Gas Dynamics: From Basic Concepts to Actual Calculations. Cambridge Texts in Applied Mathematics}}}.
\newblock {\it \bibinfo{journal}{Applied Mechanics Reviews}\/},  {\it \bibinfo{volume}{54}\/}\bibinfo{issue}{(5)}, \bibinfo{pages}{B90--B92}. \DOIprefix\doi{10.1115/1.1399679}.

\bibitem[{Chen et~al.(2013)Chen, Madden, Bickmore \& Reches}]{Chen2013}
\bibinfo{author}{Chen, X.}, \bibinfo{author}{Madden, A.~S.}, \bibinfo{author}{Bickmore, B.~R.} et~al. (\bibinfo{year}{2013}).
\newblock \bibinfo{title}{Dynamic weakening by nanoscale smoothing during high-velocity fault slip}.
\newblock {\it \bibinfo{journal}{Geology}\/},  {\it \bibinfo{volume}{41}\/}\bibinfo{issue}{(7)}, \bibinfo{pages}{739–742}. \DOIprefix\doi{10.1130/g34169.1}.

\bibitem[{Chen et~al.(2024)Chen, Gibelli \& Borg}]{Chen2024}
\bibinfo{author}{Chen, Y.}, \bibinfo{author}{Gibelli, L.},  \& \bibinfo{author}{Borg, M.~K.} (\bibinfo{year}{2024}).
\newblock \bibinfo{title}{{Impact of random nanoscale roughness on gas-scattering dynamics}}.
\newblock {\it \bibinfo{journal}{Phys. Rev. E}\/},  {\it \bibinfo{volume}{109}\/}, \bibinfo{pages}{065308}. \DOIprefix\doi{10.1103/PhysRevE.109.065308}.

\bibitem[{Chen et~al.(2023)Chen, Gibelli, Li \& Borg}]{Chen2023}
\bibinfo{author}{Chen, Y.}, \bibinfo{author}{Gibelli, L.}, \bibinfo{author}{Li, J.} et~al. (\bibinfo{year}{2023}).
\newblock \bibinfo{title}{Impact of surface physisorption on gas scattering dynamics}.
\newblock {\it \bibinfo{journal}{Journal of Fluid Mechanics}\/},  {\it \bibinfo{volume}{968}\/}. \DOIprefix\doi{10.1017/jfm.2023.496}.

\bibitem[{Comsa et~al.(1980)Comsa, Fremerey, Lindenau, Messer \& R\"{o}hl}]{Comsa1980}
\bibinfo{author}{Comsa, G.}, \bibinfo{author}{Fremerey, J.~K.}, \bibinfo{author}{Lindenau, B.} et~al. (\bibinfo{year}{1980}).
\newblock \bibinfo{title}{Calibration of a spinning rotor gas friction gauge against a fundamental vacuum pressure standard}.
\newblock {\it \bibinfo{journal}{Journal of Vacuum Science and Technology}\/},  {\it \bibinfo{volume}{17}\/}\bibinfo{issue}{(2)}, \bibinfo{pages}{642–644}. \DOIprefix\doi{10.1116/1.570531}.

\bibitem[{Cook et~al.(1994)Cook, Cross \& Hoffbauer}]{cook1994aiaa}
\bibinfo{author}{Cook, S.}, \bibinfo{author}{Cross, J.},  \& \bibinfo{author}{Hoffbauer, M.} (\bibinfo{year}{1994}).
\newblock \bibinfo{title}{{AIAA--94--2637, Proceedings of the 18th Aerospace Ground Testing Conference}}.
\newblock {\it \bibinfo{journal}{Colorado Springs}\/}, .

\bibitem[{Cook \& Hoffbauer(1997)}]{Cook1997_2}
\bibinfo{author}{Cook, S.~R.},  \& \bibinfo{author}{Hoffbauer, M.~A.} (\bibinfo{year}{1997}).
\newblock \bibinfo{title}{{{Absolute momentum transfer in gas-surface scattering}}}.
\newblock {\it \bibinfo{journal}{Phys. Rev. E}\/},  {\it \bibinfo{volume}{55}\/}, \bibinfo{pages}{R3828--R3831}. \DOIprefix\doi{10.1103/PhysRevE.55.R3828}.

\bibitem[{Cook \& Hoffbauer(1998)}]{Cook1998}
\bibinfo{author}{Cook, S.~R.},  \& \bibinfo{author}{Hoffbauer, M.~A.} (\bibinfo{year}{1998}).
\newblock \bibinfo{title}{Analyzing gas-surface interactions using the reduced force coefficients}.
\newblock {\it \bibinfo{journal}{Phys. Rev. E}\/},  {\it \bibinfo{volume}{58}\/}, \bibinfo{pages}{504--511}. \DOIprefix\doi{10.1103/PhysRevE.58.504}.

\bibitem[{Cutler et~al.(2021)Cutler, Thackeray, Trefonas, Millward, Lee \& Mack}]{Cutler2021}
\bibinfo{author}{Cutler, C.}, \bibinfo{author}{Thackeray, J.~W.}, \bibinfo{author}{Trefonas, P.} et~al. (\bibinfo{year}{2021}).
\newblock \bibinfo{title}{{Pattern roughness analysis using power spectral density: application and impact in photoresist formulation}}.
\newblock {\it \bibinfo{journal}{Journal of Micro/Nanopatterning, Materials, and Metrology}\/},  {\it \bibinfo{volume}{20}\/}\bibinfo{issue}{(01)}. \DOIprefix\doi{10.1117/1.jmm.20.1.010901}.

\bibitem[{Erofeev(1971)}]{Erofeev1971}
\bibinfo{author}{Erofeev, A.~I.} (\bibinfo{year}{1971}).
\newblock \bibinfo{title}{{Effect of roughness on interaction of gas flow with surface of a solid body}}.
\newblock {\it \bibinfo{journal}{Fluid Dynamics}\/},  {\it \bibinfo{volume}{2}\/}\bibinfo{issue}{(6)}, \bibinfo{pages}{57–61}. \DOIprefix\doi{10.1007/bf01013713}.

\bibitem[{Erofeev et~al.(2012)Erofeev, Friedlander, Nikiforov, Nesterov \& Nezhmetdinova}]{Erofeev2012}
\bibinfo{author}{Erofeev, A.~I.}, \bibinfo{author}{Friedlander, O.~G.}, \bibinfo{author}{Nikiforov, A.~P.} et~al. (\bibinfo{year}{2012}).
\newblock \bibinfo{title}{The influence of roughness of the surface on the interchange of momentum between gas flow and solid surface}.
\newblock In {\it \bibinfo{booktitle}{AIP Conference Proceedings}\/}.
\newblock \bibinfo{publisher}{AIP}.
\newblock \DOIprefix\doi{10.1063/1.4769673}.

\bibitem[{Erofeev \& Nikiforov(2014)}]{Erofeev2014-kk}
\bibinfo{author}{Erofeev, A.~I.},  \& \bibinfo{author}{Nikiforov, A.~P.} (\bibinfo{year}{2014}).
\newblock \bibinfo{title}{Angular distribution of free molecular gas flow reflected from a solid body surface}.
\newblock {\it \bibinfo{journal}{TsAGI Sci. J.}\/},  {\it \bibinfo{volume}{45}\/}\bibinfo{issue}{(8)}, \bibinfo{pages}{927--948}.

\bibitem[{Gong et~al.(2016)Gong, Misture, Gao \& Mellott}]{Gong2016}
\bibinfo{author}{Gong, Y.}, \bibinfo{author}{Misture, S.~T.}, \bibinfo{author}{Gao, P.} et~al. (\bibinfo{year}{2016}).
\newblock \bibinfo{title}{{Surface Roughness Measurements Using Power Spectrum Density Analysis with Enhanced Spatial Correlation Length}}.
\newblock {\it \bibinfo{journal}{The Journal of Physical Chemistry C}\/},  {\it \bibinfo{volume}{120}\/}\bibinfo{issue}{(39)}, \bibinfo{pages}{22358–22364}. \DOIprefix\doi{10.1021/acs.jpcc.6b06635}.

\bibitem[{Goodman(1965)}]{Goodman1965}
\bibinfo{author}{Goodman, F.} (\bibinfo{year}{1965}).
\newblock \bibinfo{title}{{On the theory of accommodation coefficients—IV. Simple distribution function theory of gas-solid interaction systems}}.
\newblock {\it \bibinfo{journal}{Journal of Physics and Chemistry of Solids}\/},  {\it \bibinfo{volume}{26}\/}\bibinfo{issue}{(1)}, \bibinfo{pages}{85–105}. \DOIprefix\doi{10.1016/0022-3697(65)90077-6}.

\bibitem[{Healy(1967)}]{Healy1967}
\bibinfo{author}{Healy, T.~J.} (\bibinfo{year}{1967}).
\newblock \bibinfo{title}{The scattering of particles from rough surfaces}.
\newblock In {\it \bibinfo{booktitle}{Fundamentals of Gas–Surface Interactions}\/} (p. \bibinfo{pages}{435–447}).
\newblock \bibinfo{publisher}{Elsevier}.
\newblock \DOIprefix\doi{10.1016/b978-1-4832-2901-0.50030-5}.

\bibitem[{Hellwig et~al.(2019)Hellwig, K\"{o}ppen, Hiller, Koslowski, Litnovsky, Schmid, Schwab \& De~Souza}]{Hellwig2019}
\bibinfo{author}{Hellwig, M.}, \bibinfo{author}{K\"{o}ppen, M.}, \bibinfo{author}{Hiller, A.} et~al. (\bibinfo{year}{2019}).
\newblock \bibinfo{title}{{Impact of Surface Roughness on Ion-Surface Interactions Studied with Energetic Carbon Ions 13C+ on Tungsten Surfaces}}.
\newblock {\it \bibinfo{journal}{Condensed Matter}\/},  {\it \bibinfo{volume}{4}\/}\bibinfo{issue}{(1)}, \bibinfo{pages}{29}. \DOIprefix\doi{10.3390/condmat4010029}.

\bibitem[{Kac(1939)}]{Kac1939}
\bibinfo{author}{Kac, M.} (\bibinfo{year}{1939}).
\newblock \bibinfo{title}{{On a Characterization of the Normal Distribution}}.
\newblock {\it \bibinfo{journal}{American Journal of Mathematics}\/},  {\it \bibinfo{volume}{61}\/}\bibinfo{issue}{(3)}, \bibinfo{pages}{726}. \DOIprefix\doi{10.2307/2371328}.

\bibitem[{Karan \& Mallik(2008)}]{Karan2008}
\bibinfo{author}{Karan, S.},  \& \bibinfo{author}{Mallik, B.} (\bibinfo{year}{2008}).
\newblock \bibinfo{title}{{Power spectral density analysis and photoconducting behavior in copper(ii) phthalocyanine nanostructured thin films}}.
\newblock {\it \bibinfo{journal}{Physical Chemistry Chemical Physics}\/},  {\it \bibinfo{volume}{10}\/}\bibinfo{issue}{(45)}, \bibinfo{pages}{6751}. \DOIprefix\doi{10.1039/b809648a}.

\bibitem[{Kleyn(2003)}]{Kleyn2003}
\bibinfo{author}{Kleyn, A.~W.} (\bibinfo{year}{2003}).
\newblock \bibinfo{title}{{Molecular beams and chemical dynamics at surfaces}}.
\newblock {\it \bibinfo{journal}{Chemical Society Reviews}\/},  {\it \bibinfo{volume}{32}\/}\bibinfo{issue}{(2)}, \bibinfo{pages}{87–95}. \DOIprefix\doi{10.1039/b105760j}.

\bibitem[{Knechtel \& Pitts(1973)}]{knechtel1973normal}
\bibinfo{author}{Knechtel, E.~D.},  \& \bibinfo{author}{Pitts, W.~C.} (\bibinfo{year}{1973}).
\newblock \bibinfo{title}{Normal and tangential momentum accommodation for earth satellite conditions}.
\newblock {\it \bibinfo{journal}{Astronautica Acta}\/},  {\it \bibinfo{volume}{18}\/}\bibinfo{issue}{(NAS 1.15: 112784)}.

\bibitem[{Koch(1999)}]{Koch1997}
\bibinfo{author}{Koch, K.-R.} (\bibinfo{year}{1999}).
\newblock {\it \bibinfo{title}{{Parameter Estimation and Hypothesis Testing in Linear Models}}\/}.
\newblock \bibinfo{address}{Berlin}: \bibinfo{publisher}{Springer}.
\newblock \DOIprefix\doi{10.1007/978-3-662-03976-2}.

\bibitem[{Liang et~al.(2018)Liang, Li \& Ye}]{Liang2018}
\bibinfo{author}{Liang, T.}, \bibinfo{author}{Li, Q.},  \& \bibinfo{author}{Ye, W.} (\bibinfo{year}{2018}).
\newblock \bibinfo{title}{A physical-based gas–surface interaction model for rarefied gas flow simulation}.
\newblock {\it \bibinfo{journal}{Journal of Computational Physics}\/},  {\it \bibinfo{volume}{352}\/}, \bibinfo{pages}{105–122}. \DOIprefix\doi{10.1016/j.jcp.2017.08.061}.

\bibitem[{Liang et~al.(2021)Liang, Zhang \& Li}]{Liang2021}
\bibinfo{author}{Liang, T.}, \bibinfo{author}{Zhang, J.},  \& \bibinfo{author}{Li, Q.} (\bibinfo{year}{2021}).
\newblock \bibinfo{title}{A parameter-free physical model for gas–surface interaction}.
\newblock {\it \bibinfo{journal}{Physics of Fluids}\/},  {\it \bibinfo{volume}{33}\/}\bibinfo{issue}{(8)}. \DOIprefix\doi{10.1063/5.0059029}.

\bibitem[{Liao et~al.(2018)Liao, Grenier, To, de~Lara-Castells \& Léonard}]{Liao2018}
\bibinfo{author}{Liao, M.}, \bibinfo{author}{Grenier, R.}, \bibinfo{author}{To, Q.-D.} et~al. (\bibinfo{year}{2018}).
\newblock \bibinfo{title}{{Helium and Argon Interactions with Gold Surfaces: Ab Initio-Assisted Determination of the He–Au Pairwise Potential and Its Application to Accommodation Coefficient Determination}}.
\newblock {\it \bibinfo{journal}{The Journal of Physical Chemistry C}\/},  {\it \bibinfo{volume}{122}\/}\bibinfo{issue}{(26)}, \bibinfo{pages}{14606–14614}. \DOIprefix\doi{10.1021/acs.jpcc.8b03555}.

\bibitem[{Litvak \& Malyugin(2012)}]{Litvak2012}
\bibinfo{author}{Litvak, M.~Y.},  \& \bibinfo{author}{Malyugin, V.~I.} (\bibinfo{year}{2012}).
\newblock \bibinfo{title}{{Poly-Gaussian models of a non-Gaussian randomly rough surface}}.
\newblock {\it \bibinfo{journal}{Technical Physics}\/},  {\it \bibinfo{volume}{57}\/}\bibinfo{issue}{(4)}, \bibinfo{pages}{524–533}. \DOIprefix\doi{10.1134/s1063784212040172}.

\bibitem[{Liu et~al.(1979)Liu, Sharma \& Knuth}]{Liu1979}
\bibinfo{author}{Liu, S.-M.}, \bibinfo{author}{Sharma, F.~K.},  \& \bibinfo{author}{Knuth, E.~L.} (\bibinfo{year}{1979}).
\newblock \bibinfo{title}{{Satellite drag coefficients calculated from measured distributions of reflected helium atoms}}.
\newblock {\it \bibinfo{journal}{AIAA Journal}\/},  {\it \bibinfo{volume}{17}\/}\bibinfo{issue}{(12)}, \bibinfo{pages}{1314–1319}. \DOIprefix\doi{10.2514/3.7629}.

\bibitem[{Livadiotti et~al.(2020)Livadiotti, Crisp, Roberts, Worrall, Oiko, Edmondson, Haigh, Huyton, Smith, Sinpetru, Holmes, Becedas, Domínguez, Cañas, Christensen, Mølgaard, Nielsen, Bisgaard, Chan, Herdrich, Romano, Fasoulas, Traub, Garcia-Almiñana, Rodriguez-Donaire, Sureda, Kataria, Belkouchi, Conte, Perez, Villain \& Outlaw}]{Livadiotti2020}
\bibinfo{author}{Livadiotti, S.}, \bibinfo{author}{Crisp, N.~H.}, \bibinfo{author}{Roberts, P.~C.} et~al. (\bibinfo{year}{2020}).
\newblock \bibinfo{title}{A review of gas-surface interaction models for orbital aerodynamics applications}.
\newblock {\it \bibinfo{journal}{Progress in Aerospace Sciences}\/},  {\it \bibinfo{volume}{119}\/}, \bibinfo{pages}{100675}. \DOIprefix\doi{10.1016/j.paerosci.2020.100675}.

\bibitem[{Logan \& Keck(1968)}]{Logan1968}
\bibinfo{author}{Logan, R.~M.},  \& \bibinfo{author}{Keck, J.~C.} (\bibinfo{year}{1968}).
\newblock \bibinfo{title}{{Classical Theory for the Interaction of Gas Atoms with Solid Surfaces}}.
\newblock {\it \bibinfo{journal}{The Journal of Chemical Physics}\/},  {\it \bibinfo{volume}{49}\/}\bibinfo{issue}{(2)}, \bibinfo{pages}{860–876}. \DOIprefix\doi{10.1063/1.1670153}.

\bibitem[{Logan \& Stickney(1966)}]{Logan1966}
\bibinfo{author}{Logan, R.~M.},  \& \bibinfo{author}{Stickney, R.~E.} (\bibinfo{year}{1966}).
\newblock \bibinfo{title}{{Simple Classical Model for the Scattering of Gas Atoms from a Solid Surface}}.
\newblock {\it \bibinfo{journal}{The Journal of Chemical Physics}\/},  {\it \bibinfo{volume}{44}\/}\bibinfo{issue}{(1)}, \bibinfo{pages}{195–201}. \DOIprefix\doi{10.1063/1.1726446}.

\bibitem[{Lord(1995)}]{Lord1995}
\bibinfo{author}{Lord, R.~G.} (\bibinfo{year}{1995}).
\newblock \bibinfo{title}{{Some further extensions of the Cercignani–Lampis gas–surface interaction model}}.
\newblock {\it \bibinfo{journal}{Physics of Fluids}\/},  {\it \bibinfo{volume}{7}\/}\bibinfo{issue}{(5)}, \bibinfo{pages}{1159–1161}. \DOIprefix\doi{10.1063/1.868557}.

\bibitem[{MacKay(2003)}]{MacKay2003-jc}
\bibinfo{author}{MacKay, D. J.~C.} (\bibinfo{year}{2003}).
\newblock {\it \bibinfo{title}{Information theory, inference and learning algorithms}\/}.
\newblock \bibinfo{address}{Cambridge, England}: \bibinfo{publisher}{Cambridge University Press}.

\bibitem[{March et~al.(2019)March, Doornbos \& Visser}]{March2019}
\bibinfo{author}{March, G.}, \bibinfo{author}{Doornbos, E.},  \& \bibinfo{author}{Visser, P.} (\bibinfo{year}{2019}).
\newblock \bibinfo{title}{{High-fidelity geometry models for improving the consistency of CHAMP, GRACE, GOCE and Swarm thermospheric density data sets}}.
\newblock {\it \bibinfo{journal}{Advances in Space Research}\/},  {\it \bibinfo{volume}{63}\/}\bibinfo{issue}{(1)}, \bibinfo{pages}{213–238}. \DOIprefix\doi{10.1016/j.asr.2018.07.009}.

\bibitem[{March et~al.(2021)March, van~den IJssel, Siemes, Visser, Doornbos \& Pilinski}]{March2021}
\bibinfo{author}{March, G.}, \bibinfo{author}{van~den IJssel, J.}, \bibinfo{author}{Siemes, C.} et~al. (\bibinfo{year}{2021}).
\newblock \bibinfo{title}{Gas-surface interactions modelling influence on satellite aerodynamics and thermosphere mass density}.
\newblock {\it \bibinfo{journal}{Journal of Space Weather and Space Climate}\/},  {\it \bibinfo{volume}{11}\/}, \bibinfo{pages}{54}. \DOIprefix\doi{10.1051/swsc/2021035}.

\bibitem[{Mateljevic et~al.(2009)Mateljevic, Kerwin, Roy, Schmidt \& Tully}]{Mateljevic2009}
\bibinfo{author}{Mateljevic, N.}, \bibinfo{author}{Kerwin, J.}, \bibinfo{author}{Roy, S.} et~al. (\bibinfo{year}{2009}).
\newblock \bibinfo{title}{Accommodation of gases at rough surfaces}.
\newblock {\it \bibinfo{journal}{The Journal of Physical Chemistry C}\/},  {\it \bibinfo{volume}{113}\/}\bibinfo{issue}{(6)}, \bibinfo{pages}{2360–2367}. \DOIprefix\doi{10.1021/jp8077634}.

\bibitem[{Maxwell(1879)}]{Maxwell1879}
\bibinfo{author}{Maxwell, J.~C.} (\bibinfo{year}{1879}).
\newblock \bibinfo{title}{On stresses in rarified gases arising from inequalities of temperature}.
\newblock {\it \bibinfo{journal}{Philosophical Transactions of the Royal Society of London}\/},  {\it \bibinfo{volume}{170}\/}, \bibinfo{pages}{231–256}. \DOIprefix\doi{10.1098/rstl.1879.0067}.

\bibitem[{McLaughlin et~al.(2011)McLaughlin, Mance \& Lichtenberg}]{McLaughlin2011}
\bibinfo{author}{McLaughlin, C.~A.}, \bibinfo{author}{Mance, S.},  \& \bibinfo{author}{Lichtenberg, T.} (\bibinfo{year}{2011}).
\newblock \bibinfo{title}{{Drag Coefficient Estimation in Orbit Determination}}.
\newblock {\it \bibinfo{journal}{The Journal of the Astronautical Sciences}\/},  {\it \bibinfo{volume}{58}\/}\bibinfo{issue}{(3)}, \bibinfo{pages}{513–530}. \DOIprefix\doi{10.1007/bf03321183}.

\bibitem[{Mehta \& Linares(2018)}]{Mehta2018}
\bibinfo{author}{Mehta, P.~M.},  \& \bibinfo{author}{Linares, R.} (\bibinfo{year}{2018}).
\newblock \bibinfo{title}{{A New Transformative Framework for Data Assimilation and Calibration of Physical Ionosphere‐Thermosphere Models}}.
\newblock {\it \bibinfo{journal}{Space Weather}\/},  {\it \bibinfo{volume}{16}\/}\bibinfo{issue}{(8)}, \bibinfo{pages}{1086–1100}. \DOIprefix\doi{10.1029/2018sw001875}.

\bibitem[{Mehta et~al.(2023)Mehta, Paul, Crisp, Sheridan, Siemes, March \& Bruinsma}]{Mehta2023}
\bibinfo{author}{Mehta, P.~M.}, \bibinfo{author}{Paul, S.~N.}, \bibinfo{author}{Crisp, N.~H.} et~al. (\bibinfo{year}{2023}).
\newblock \bibinfo{title}{Satellite drag coefficient modeling for thermosphere science and mission operations}.
\newblock {\it \bibinfo{journal}{Advances in Space Research}\/},  {\it \bibinfo{volume}{72}\/}\bibinfo{issue}{(12)}, \bibinfo{pages}{5443–5459}. \DOIprefix\doi{10.1016/j.asr.2022.05.064}.

\bibitem[{Mehta et~al.(2014)Mehta, Walker, McLaughlin \& Koller}]{Mehta2014}
\bibinfo{author}{Mehta, P.~M.}, \bibinfo{author}{Walker, A.}, \bibinfo{author}{McLaughlin, C.~A.} et~al. (\bibinfo{year}{2014}).
\newblock \bibinfo{title}{Comparing physical drag coefficients computed using different gas–surface interaction models}.
\newblock {\it \bibinfo{journal}{Journal of Spacecraft and Rockets}\/},  {\it \bibinfo{volume}{51}\/}\bibinfo{issue}{(3)}, \bibinfo{pages}{873–883}. \DOIprefix\doi{10.2514/1.a32566}.

\bibitem[{Mehta et~al.(2017)Mehta, Walker, Sutton \& Godinez}]{Mehta2017}
\bibinfo{author}{Mehta, P.~M.}, \bibinfo{author}{Walker, A.~C.}, \bibinfo{author}{Sutton, E.~K.} et~al. (\bibinfo{year}{2017}).
\newblock \bibinfo{title}{{New density estimates derived using accelerometers on board the CHAMP and GRACE satellites}}.
\newblock {\it \bibinfo{journal}{Space Weather}\/},  {\it \bibinfo{volume}{15}\/}\bibinfo{issue}{(4)}, \bibinfo{pages}{558–576}. \DOIprefix\doi{10.1002/2016sw001562}.

\bibitem[{Metropolis et~al.(1953)Metropolis, Rosenbluth, Rosenbluth, Teller \& Teller}]{Metropolis1953}
\bibinfo{author}{Metropolis, N.}, \bibinfo{author}{Rosenbluth, A.~W.}, \bibinfo{author}{Rosenbluth, M.~N.} et~al. (\bibinfo{year}{1953}).
\newblock \bibinfo{title}{{Equation of State Calculations by Fast Computing Machines}}.
\newblock {\it \bibinfo{journal}{The Journal of Chemical Physics}\/},  {\it \bibinfo{volume}{21}\/}\bibinfo{issue}{(6)}, \bibinfo{pages}{1087–1092}. \DOIprefix\doi{10.1063/1.1699114}.

\bibitem[{Moe \& Moe(2005)}]{Moe2005}
\bibinfo{author}{Moe, K.},  \& \bibinfo{author}{Moe, M.~M.} (\bibinfo{year}{2005}).
\newblock \bibinfo{title}{{Gas–surface interactions and satellite drag coefficients}}.
\newblock {\it \bibinfo{journal}{Planetary and Space Science}\/},  {\it \bibinfo{volume}{53}\/}\bibinfo{issue}{(8)}, \bibinfo{pages}{793–801}. \DOIprefix\doi{10.1016/j.pss.2005.03.005}.

\bibitem[{Moe et~al.(1972)Moe, Moe \& Yelaca}]{Moe1972}
\bibinfo{author}{Moe, K.}, \bibinfo{author}{Moe, M.~M.},  \& \bibinfo{author}{Yelaca, N.~W.} (\bibinfo{year}{1972}).
\newblock \bibinfo{title}{{Effect of surface heterogeneity on the adsorptive behavior of orbiting pressure gages}}.
\newblock {\it \bibinfo{journal}{Journal of Geophysical Research}\/},  {\it \bibinfo{volume}{77}\/}\bibinfo{issue}{(22)}, \bibinfo{pages}{4242–4247}. \DOIprefix\doi{10.1029/ja077i022p04242}.

\bibitem[{Moe et~al.(1993)Moe, Wallace \& Moe}]{Moe1993}
\bibinfo{author}{Moe, M.~M.}, \bibinfo{author}{Wallace, S.~D.},  \& \bibinfo{author}{Moe, K.} (\bibinfo{year}{1993}).
\newblock \bibinfo{title}{{Refinements in determining satellite drag coefficients - Method for resolving density discrepancies}}.
\newblock {\it \bibinfo{journal}{Journal of Guidance, Control, and Dynamics}\/},  {\it \bibinfo{volume}{16}\/}\bibinfo{issue}{(3)}, \bibinfo{pages}{441–445}. \DOIprefix\doi{10.2514/3.21029}.

\bibitem[{Murray et~al.(2015)Murray, Marshall, Woodburn \& Minton}]{Murray2015}
\bibinfo{author}{Murray, V.~J.}, \bibinfo{author}{Marshall, B.~C.}, \bibinfo{author}{Woodburn, P.~J.} et~al. (\bibinfo{year}{2015}).
\newblock \bibinfo{title}{{Inelastic and Reactive Scattering Dynamics of Hyperthermal O and O2 on Hot Vitreous Carbon Surfaces}}.
\newblock {\it \bibinfo{journal}{The Journal of Physical Chemistry C}\/},  {\it \bibinfo{volume}{119}\/}\bibinfo{issue}{(26)}, \bibinfo{pages}{14780–14796}. \DOIprefix\doi{10.1021/acs.jpcc.5b00924}.

\bibitem[{Murray et~al.(2017)Murray, Pilinski, Smoll, Qian, Minton, Madzunkov \& Darrach}]{Murray2017}
\bibinfo{author}{Murray, V.~J.}, \bibinfo{author}{Pilinski, M.~D.}, \bibinfo{author}{Smoll, E.~J.} et~al. (\bibinfo{year}{2017}).
\newblock \bibinfo{title}{{Gas–Surface Scattering Dynamics Applied to Concentration of Gases for Mass Spectrometry in Tenuous Atmospheres}}.
\newblock {\it \bibinfo{journal}{The Journal of Physical Chemistry C}\/},  {\it \bibinfo{volume}{121}\/}\bibinfo{issue}{(14)}, \bibinfo{pages}{7903––7922}. \DOIprefix\doi{10.1021/acs.jpcc.7b00456}.

\bibitem[{Mwema et~al.(2018)Mwema, Oladijo, Sathiaraj \& Akinlabi}]{Mwema2018}
\bibinfo{author}{Mwema, F.~M.}, \bibinfo{author}{Oladijo, O.~P.}, \bibinfo{author}{Sathiaraj, T.~S.} et~al. (\bibinfo{year}{2018}).
\newblock \bibinfo{title}{Atomic force microscopy analysis of surface topography of pure thin aluminum films}.
\newblock {\it \bibinfo{journal}{Materials Research Express}\/},  {\it \bibinfo{volume}{5}\/}\bibinfo{issue}{(4)}, \bibinfo{pages}{046416}. \DOIprefix\doi{10.1088/2053-1591/aabe1b}.

\bibitem[{Ozhgibesov et~al.(2013)Ozhgibesov, Leu, Cheng \& Utkin}]{Ozhgibesov2013}
\bibinfo{author}{Ozhgibesov, M.}, \bibinfo{author}{Leu, T.}, \bibinfo{author}{Cheng, C.} et~al. (\bibinfo{year}{2013}).
\newblock \bibinfo{title}{Studies on argon collisions with smooth and rough tungsten surfaces}.
\newblock {\it \bibinfo{journal}{Journal of Molecular Graphics and Modelling}\/},  {\it \bibinfo{volume}{45}\/}, \bibinfo{pages}{45–49}. \DOIprefix\doi{10.1016/j.jmgm.2013.08.010}.

\bibitem[{Padilla \& Boyd(2007)}]{Padilla2007}
\bibinfo{author}{Padilla, J.},  \& \bibinfo{author}{Boyd, I.} (\bibinfo{year}{2007}).
\newblock \bibinfo{title}{{Assessment of Gas-Surface Interaction Models in DSMC Analysis of Rarefied Hypersonic Flow}}.
\newblock In {\it \bibinfo{booktitle}{39th AIAA Thermophysics Conference}\/}.
\newblock \bibinfo{publisher}{American Institute of Aeronautics and Astronautics}.
\newblock \DOIprefix\doi{10.2514/6.2007-3891}.

\bibitem[{Pardini et~al.(2010)Pardini, Anselmo, Moe \& Moe}]{Pardini2010}
\bibinfo{author}{Pardini, C.}, \bibinfo{author}{Anselmo, L.}, \bibinfo{author}{Moe, K.} et~al. (\bibinfo{year}{2010}).
\newblock \bibinfo{title}{{Drag and energy accommodation coefficients during sunspot maximum}}.
\newblock {\it \bibinfo{journal}{Advances in Space Research}\/},  {\it \bibinfo{volume}{45}\/}\bibinfo{issue}{(5)}, \bibinfo{pages}{638–650}. \DOIprefix\doi{10.1016/j.asr.2009.08.034}.

\bibitem[{Pardini et~al.(2006)Pardini, Tobiska \& Anselmo}]{Pardini2006}
\bibinfo{author}{Pardini, C.}, \bibinfo{author}{Tobiska, W.~K.},  \& \bibinfo{author}{Anselmo, L.} (\bibinfo{year}{2006}).
\newblock \bibinfo{title}{Analysis of the orbital decay of spherical satellites using different solar flux proxies and atmospheric density models}.
\newblock {\it \bibinfo{journal}{Advances in Space Research}\/},  {\it \bibinfo{volume}{37}\/}\bibinfo{issue}{(2)}, \bibinfo{pages}{392–400}. \DOIprefix\doi{10.1016/j.asr.2004.10.009}.

\bibitem[{Picone et~al.(2002)Picone, Hedin, Drob \& Aikin}]{Picone2002}
\bibinfo{author}{Picone, J.~M.}, \bibinfo{author}{Hedin, A.~E.}, \bibinfo{author}{Drob, D.~P.} et~al. (\bibinfo{year}{2002}).
\newblock \bibinfo{title}{Nrlmsise‐00 empirical model of the atmosphere: Statistical comparisons and scientific issues}.
\newblock {\it \bibinfo{journal}{Journal of Geophysical Research: Space Physics}\/},  {\it \bibinfo{volume}{107}\/}\bibinfo{issue}{(A12)}. \DOIprefix\doi{10.1029/2002ja009430}.

\bibitem[{Pilinski et~al.(2013)Pilinski, Argrow, Palo \& Bowman}]{Pilinski2013}
\bibinfo{author}{Pilinski, M.~D.}, \bibinfo{author}{Argrow, B.~M.}, \bibinfo{author}{Palo, S.~E.} et~al. (\bibinfo{year}{2013}).
\newblock \bibinfo{title}{{Semi-Empirical Satellite Accommodation Model for Spherical and Randomly Tumbling Objects}}.
\newblock {\it \bibinfo{journal}{Journal of Spacecraft and Rockets}\/},  {\it \bibinfo{volume}{50}\/}\bibinfo{issue}{(3)}, \bibinfo{pages}{556–571}. \DOIprefix\doi{10.2514/1.a32348}.

\bibitem[{Pollak \& Tatchen(2009)}]{Pollak2009}
\bibinfo{author}{Pollak, E.},  \& \bibinfo{author}{Tatchen, J.} (\bibinfo{year}{2009}).
\newblock \bibinfo{title}{{Rainbow scattering of argon from $2H\text{-W}(100)$}}.
\newblock {\it \bibinfo{journal}{Phys. Rev. B}\/},  {\it \bibinfo{volume}{80}\/}, \bibinfo{pages}{115404}. \DOIprefix\doi{10.1103/PhysRevB.80.115404}.

\bibitem[{Rettner et~al.(1991)Rettner, Barker \& Bethune}]{Rettner1991}
\bibinfo{author}{Rettner, C.~T.}, \bibinfo{author}{Barker, J.~A.},  \& \bibinfo{author}{Bethune, D.~S.} (\bibinfo{year}{1991}).
\newblock \bibinfo{title}{{Angular and velocity distributions characteristic of the transition between the thermal and structure regimes of gas-surface scattering}}.
\newblock {\it \bibinfo{journal}{Physical Review Letters}\/},  {\it \bibinfo{volume}{67}\/}\bibinfo{issue}{(16)}, \bibinfo{pages}{2183–2186}. \DOIprefix\doi{10.1103/physrevlett.67.2183}.

\bibitem[{Roman et~al.(2023)Roman, Knight, Moon, Lane, Greaves, Costen \& McKendrick}]{Roman2023}
\bibinfo{author}{Roman, M.~J.}, \bibinfo{author}{Knight, A.~G.}, \bibinfo{author}{Moon, D.~R.} et~al. (\bibinfo{year}{2023}).
\newblock \bibinfo{title}{{Inelastic scattering of OH from a liquid PFPE surface: Resolution of correlated speed and angular distributions}}.
\newblock {\it \bibinfo{journal}{The Journal of Chemical Physics}\/},  {\it \bibinfo{volume}{158}\/}\bibinfo{issue}{(24)}. \DOIprefix\doi{10.1063/5.0153314}.

\bibitem[{Sazhin et~al.(2001)Sazhin, Borisov \& Sharipov}]{Sazhin2001}
\bibinfo{author}{Sazhin, O.~V.}, \bibinfo{author}{Borisov, S.~F.},  \& \bibinfo{author}{Sharipov, F.} (\bibinfo{year}{2001}).
\newblock \bibinfo{title}{{Accommodation coefficient of tangential momentum on atomically clean and contaminated surfaces}}.
\newblock {\it \bibinfo{journal}{Journal of Vacuum Science \& Technology A: Vacuum, Surfaces, and Films}\/},  {\it \bibinfo{volume}{19}\/}\bibinfo{issue}{(5)}, \bibinfo{pages}{2499–2503}. \DOIprefix\doi{10.1116/1.1388622}.

\bibitem[{Sentman(1961)}]{Sentman1961FREEMF}
\bibinfo{author}{Sentman, L.~H.} (\bibinfo{year}{1961}).
\newblock \bibinfo{title}{{FREE MOLECULE FLOW THEORY AND ITS APPLICATION TO THE DETERMINATION OF AERODYNAMIC FORCES}}.
\newblock \URLprefix \url{https://api.semanticscholar.org/CorpusID:92321666}.

\bibitem[{Shoda et~al.(2022)Shoda, Kano, Jotaki, Ezaki, Itatani, Ozawa, Yamashita, Nishiyama, Yokota \& Tagawa}]{Shoda2022}
\bibinfo{author}{Shoda, K.}, \bibinfo{author}{Kano, N.}, \bibinfo{author}{Jotaki, Y.} et~al. (\bibinfo{year}{2022}).
\newblock \bibinfo{title}{{Anisotropic molecular scattering at microstructured surface for rarefied gas compression inside air breathing ion engine}}.
\newblock {\it \bibinfo{journal}{CEAS Space Journal}\/},  {\it \bibinfo{volume}{15}\/}\bibinfo{issue}{(3)}, \bibinfo{pages}{403–411}. \DOIprefix\doi{10.1007/s12567-022-00430-7}.

\bibitem[{Shu et~al.(2023)Shu, Wu, Zhong, Yan \& Huang}]{Shu2023}
\bibinfo{author}{Shu, C.}, \bibinfo{author}{Wu, X.}, \bibinfo{author}{Zhong, M.} et~al. (\bibinfo{year}{2023}).
\newblock \bibinfo{title}{The atomic oxygen resistant study of a transparent polyimide film containing phosphorus and fluorine}.
\newblock {\it \bibinfo{journal}{Applied Surface Science}\/},  {\it \bibinfo{volume}{631}\/}, \bibinfo{pages}{157562}. \DOIprefix\doi{10.1016/j.apsusc.2023.157562}.

\bibitem[{Siemes et~al.(2023)Siemes, Borries, Bruinsma, Fernandez-Gomez, Hładczuk, den IJssel, Kodikara, Vielberg \& Visser}]{Siemes2023}
\bibinfo{author}{Siemes, C.}, \bibinfo{author}{Borries, C.}, \bibinfo{author}{Bruinsma, S.} et~al. (\bibinfo{year}{2023}).
\newblock \bibinfo{title}{New thermosphere neutral mass density and crosswind datasets from champ, grace, and grace-fo}.
\newblock {\it \bibinfo{journal}{Journal of Space Weather and Space Climate}\/},  {\it \bibinfo{volume}{13}\/}, \bibinfo{pages}{16}. \DOIprefix\doi{10.1051/swsc/2023014}.

\bibitem[{Siemes et~al.(2024)Siemes, {den IJssel} \& Visser}]{SIEMES2024}
\bibinfo{author}{Siemes, C.}, \bibinfo{author}{{den IJssel}, J.},  \& \bibinfo{author}{Visser, P.} (\bibinfo{year}{2024}).
\newblock \bibinfo{title}{Uncertainty of thermosphere mass density observations derived from accelerometer and gnss tracking data}.
\newblock {\it \bibinfo{journal}{Advances in Space Research}\/}, . \DOIprefix\doi{https://doi.org/10.1016/j.asr.2024.02.057}.

\bibitem[{Smith(1967)}]{Smith1967}
\bibinfo{author}{Smith, B.} (\bibinfo{year}{1967}).
\newblock \bibinfo{title}{{Geometrical shadowing of a random rough surface}}.
\newblock {\it \bibinfo{journal}{IEEE Transactions on Antennas and Propagation}\/},  {\it \bibinfo{volume}{15}\/}\bibinfo{issue}{(5)}, \bibinfo{pages}{668–671}. \DOIprefix\doi{10.1109/tap.1967.1138991}.

\bibitem[{Tully(1990)}]{Tully1990}
\bibinfo{author}{Tully, J.~C.} (\bibinfo{year}{1990}).
\newblock \bibinfo{title}{{Washboard model of gas–surface scattering}}.
\newblock {\it \bibinfo{journal}{The Journal of Chemical Physics}\/},  {\it \bibinfo{volume}{92}\/}\bibinfo{issue}{(1)}, \bibinfo{pages}{680–686}. \DOIprefix\doi{10.1063/1.458421}.

\bibitem[{Vallado(2007)}]{Vallado2007}
\bibinfo{author}{Vallado, D.~A.} (\bibinfo{year}{2007}).
\newblock {\it \bibinfo{title}{Fundamentals of Astrodynamics and Applications}\/}.
\newblock (\bibinfo{edition}{2nd} ed.).
\newblock \bibinfo{publisher}{Springer-Netherlands}.

\bibitem[{Walker et~al.(2014{\natexlab{a}})Walker, Mehta \& Koller}]{Walker2014}
\bibinfo{author}{Walker, A.}, \bibinfo{author}{Mehta, P.},  \& \bibinfo{author}{Koller, J.} (\bibinfo{year}{2014}{\natexlab{a}}).
\newblock \bibinfo{title}{{Drag Coefficient Model Using the Cercignani–Lampis–Lord Gas–Surface Interaction Model}}.
\newblock {\it \bibinfo{journal}{Journal of Spacecraft and Rockets}\/},  {\it \bibinfo{volume}{51}\/}\bibinfo{issue}{(5)}, \bibinfo{pages}{1544–1563}. \DOIprefix\doi{10.2514/1.a32677}.

\bibitem[{Walker et~al.(2014{\natexlab{b}})Walker, Mehta \& Koller}]{Mehta2014_2}
\bibinfo{author}{Walker, A.}, \bibinfo{author}{Mehta, P.},  \& \bibinfo{author}{Koller, J.} (\bibinfo{year}{2014}{\natexlab{b}}).
\newblock \bibinfo{title}{The effect of different adsorption models on satellite drag coefficients}.
\newblock In {\it \bibinfo{booktitle}{Astrodynamics 2013 - Advances in the Astronautical Sciences}\/} Advances in the Astronautical Sciences (pp. \bibinfo{pages}{675--686}).
\newblock \bibinfo{publisher}{Univelt Inc.}
\newblock \bibinfo{note}{2013 AAS/AIAA Astrodynamics Specialist Conference, Astrodynamics 2013 ; Conference date: 11-08-2013 Through 15-08-2013}.

\bibitem[{Wiener(1930)}]{Wiener1930}
\bibinfo{author}{Wiener, N.} (\bibinfo{year}{1930}).
\newblock \bibinfo{title}{{Generalized harmonic analysis}}.
\newblock {\it \bibinfo{journal}{Acta Mathematica}\/},  {\it \bibinfo{volume}{55}\/}\bibinfo{issue}{(0)}, \bibinfo{pages}{117–258}. \DOIprefix\doi{10.1007/bf02546511}.

\bibitem[{Xu et~al.(2023)Xu, Murray, Pilinski, Schwartzentruber, Poovathingal \& Minton}]{Xu2023}
\bibinfo{author}{Xu, C.}, \bibinfo{author}{Murray, V.~J.}, \bibinfo{author}{Pilinski, M.~D.} et~al. (\bibinfo{year}{2023}).
\newblock \bibinfo{title}{{Gas concentration in rarefied flows: Experiments and modeling}}.
\newblock {\it \bibinfo{journal}{Aerospace Science and Technology}\/},  {\it \bibinfo{volume}{142}\/}, \bibinfo{pages}{108568}. \DOIprefix\doi{10.1016/j.ast.2023.108568}.

\end{thebibliography}

\end{document}